\documentclass[12pt]{article}
\usepackage{comment}
\usepackage{textcomp}
\usepackage{chetnew}
\usepackage{amssymb,amsmath,amsfonts,mathrsfs,braket}
\usepackage[utf8]{inputenc}
\usepackage{graphicx,longtable,afterpage}
\usepackage{cleveref}

\usepackage{cancel}

\usepackage[backend=biber,style=numeric-comp,sorting=none]{biblatex}
\usepackage{floatrow}
\newfloatcommand{capbtabbox}{table}[][\FBwidth]

\def\one{{\,\hbox{1\kern-.8mm l}}}

\newcommand{\CC}{\mathcal{C}}

\allowdisplaybreaks

\def\makeatletter{\catcode`\@=11}
\makeatletter
\def\mathbox#1{\hbox{$\m@th#1$}}%
\def\math@ccstyles#1#2#3#4#5#6#7{{\leavevmode
      \setbox0\mathbox{#6#7}%
      \setbox2\mathbox{#4#5}%
      \dimen@ #3%
      \baselineskip\z@\lineskiplimit#1\lineskip\z@
      \vbox{\ialign{##\crcr
             \hfil \kern #2\box2 \hfil\crcr
             \noalign{\kern\dimen@}%
             \hfil\box0\hfil\crcr}}}}
\def\mathaccstyles{\math@ccstyles\maxdimen}
\def\maththroughstyles{\math@ccstyles{-\maxdimen}}
\def\unity%
 {\maththroughstyles{.45\ht0}\z@\displaystyle {\mathchar"006C}\displaystyle 1}

\def\AA{{\cal A}}
\def\BB{{\cal B}}
\def\CC{{\cal C}}
\def\DD{{\cal D}}
\def\EE{{\cal E}}
\def\FF{{\cal F}}
\def\GG{{\cal G}} 
\def\HH{{\cal H}}

\def\LL{{\cal L}}

\def\NN{{\cal N}}
\def\OO{{\cal O}}

\def\TT{{\cal T}}

\def\VV{{\cal V}}
\def\WW{{\cal W}}
\def\XX{{\cal X}}

\def\beq{\begin{equation}}
\def\eeq{\end{equation}}
\newcommand{\bea}{\begin{eqnarray}}
\newcommand{\eea}{\end{eqnarray}}
\def\bal{\begin{align}}
\def\eal{\end{align}}

\preprint{\hfill QMUL-PH-26-02}

\title{\vspace{-1.cm} 
Reduced superblocks at next-to-next-to-extremality for all maximally supersymmetric CFTs} 

\author{
 Mitchell~Woolley}

\affiliation{
 Centre for Theoretical Physics, Department of Physics and Astronomy\\ Queen Mary University of London, London E1 4NS, UK \vspace{0.3cm} $ $ \\
\vspace{0.3cm} $ $

\vspace{0.3cm}
{\tt \small
mitchell.woolley@qmul.ac.uk}}

\abstract{We consider mixed four-point correlators of 1/2-BPS operators $\mathcal{O}_{k}$ in the maximally supersymmetric CFTs, i.e. the 3d $\mathcal{N}=8$, 4d $\mathcal{N}=4$, and 6d $\mathcal{N}=(2,0)$ theories. In \cite{Dolan:2004mu}, Dolan, Gallot, and Sokatchev demonstrated that four-point correlators of identical $\mathcal{O}_{k}$ in these SCFTs can be expressed in terms of unconstrained ``reduced correlators" $\mathcal{T}^{\{k_i\}}_{I,J}(U,V)$, $h^{\{k_i\}}_I(z)$ acted on by a $2(\varepsilon-1)$nd order differential operator $\Delta_\varepsilon$, which is non-local in odd dimensions $d=2(\varepsilon+1)$. We generalize this construction to mixed correlators $\langle \mathcal{O}_{k_1}\mathcal{O}_{k_2}\mathcal{O}_{k_3}\mathcal{O}_{k_1+k_2+k_3-2\mathcal{E}}\rangle$ up to extremality $\mathcal{E}=2$. To construct superconformal blocks, we generalize the R-symmetry channel equations and use Jack polynomial expansions to recursively generate the full spectrum of conformal blocks in a superblock from a single channel. We observe that for each $\varepsilon$, this channel equation can be inverted to expand the reduced correlators in ``reduced superblocks" involving blocks with shifted external kinematics. These reduced blocks reproduce what is known in 4d, generalize the known $\langle \mathcal{O}_{2}\mathcal{O}_{2}\mathcal{O}_{k}\mathcal{O}_{k}\rangle$ case in 6d, and offer a novel result in 3d.}

\date{\today}

\begin{document}

\maketitle

\hypersetup{pageanchor=true}

\setcounter{tocdepth}{2}

\toc

\newpage

\section{Introduction}
\label{intro}
The conformal bootstrap program in dimensions $d>2$ initiated in \cite{Rattazzi:2010yc} generated a surge of constraints on the non-perturbative dynamics of conformal field theories (CFTs). In its original formulation, this method imposed crossing symmetry on four-point correlation functions of scalar conformal primaries to constrain scaling dimensions $\Delta_i$ and three-point coefficients $\lambda_{ijk}$ of the operators $\OO_i$ in these theories. 

A broader aim of the conformal bootstrap program is to rigorously carve out the space of consistent CFTs. Superconformal field theories (SCFTs) occupy a special place in this space and supersymmetry gives rise to protected subsectors whose CFT data $\{\Delta_i,\lambda_{ijk}\}$ can be determined exactly either using the superconformal bootstrap directly or other analytical methods enabled by supersymmetry like localization or integrability. The most distinguished among such theories are the interacting theories with the maximal amount supersymmetry, i.e. sixteen real Poincaré supercharges. These are
\begin{itemize}
    \item the 6d $\NN=(2,0)$ theories with superconformal algebra $\mathfrak{osp}(8^*|4)$
     \item the 4d $\NN=4$ theories with superconformal algebra $\mathfrak{psu}(2,2|4)$
    \item the 3d $\NN=8$ theories with superconformal algebra $\mathfrak{osp}(8|4)$
\end{itemize}
These theories are not only of interest as quantum field theories - they are also holographically dual to quantum gravity on maximally superconformal backgrounds, namely M-theory on $AdS_4\times S^7$, type IIB string theory on $AdS_5\times S^5$, and M-theory on $AdS_7\times S^4$\footnote{We note that these backgrounds can be modified to include a supersymmetry-preserving $\mathbb{Z}_2$ action. The SCFTs dual to these configurations have modified spectra, as described in \cite{Witten:1998xy,Aharony:1998rm}.}.

A basic technical ingredient in any conformal bootstrap study is the decomposition of four-point correlators into conformal blocks, encoding the exchange of a conformal primary and all of its conformal descendants. Supersymmetry refines this structure by relating conformal primaries and packaging them into superconformal multiplets wherein all CFT data is related to that of the superconformal primary operator. The superconformal Ward identities allow one to decompose four-point functions in a superconformal block (superblock) decomposition which packages together the superconformal primary and its conformal descendants, together with superconformal descendant primaries and their conformal descendant operators.  

An implementation of this structure involves determining 1) the spectrum of conformal families in a given superconformal multiplet and 2) the proportionality constants that relate the three-point coefficients $\lambda_{ijk}$ of superconformal descendants to that of the superconformal primary. The former can be accomplished using the Racah-Speiser algorithm described e.g. in \cite{Cordova:2016emh,Buican:2016hpb} and implemented in Appendix B of \cite{Hayling:2020mbp}. This spectrum can be used to construct a superblock ansatz and the aforementioned proportionality constants can be determined by imposing the superconformal Ward identities order-by-order in a small conformal cross-ratio expansion, following the approach introduced in \cite{Chester:2014fya}. This strategy is necessarily case-by-case and the aim of this paper will be to provide an alternative derivation of the spectra of superconformal descendants and their corresponding proportionality constants for a general class of kinematic configurations, using only the superconformal Ward identities and a convenient basis of solutions.

The authors of \cite{Dolan:2004mu} expressed such solutions in terms of a differential operator $\Delta_{\varepsilon}$ acting on a set of unconstrained functions $\TT^{\{k_i\}}_{I,J}(z,\bar{z})$ and $h^{\{k_i\}}_{I}(z)$ that we will call reduced correlators. These simpler functions encode all of the non-trivial dynamics of the full four-point function and provide an economical implementation of the superconformal bootstrap. In 4d, the operator $\Delta_{1}$ is a multiplicative factor and it has long been known how to work directly with the reduced correlators in a reduced superblock expansion. In 6d, $\Delta_{2}$ is a second order partial differential operator with a non-trivial kernel, which obfuscates the reduced correlator function formalism. The situation in odd dimensions is even worse, where $\Delta_{\varepsilon}$ becomes a non-local operator defined only in terms of its action on the functional space spanned by the so-called Jack polynomials introduced in \cite{Jack:1970}. A result of this work will be to overcome these obstacles and extract reduced block decompositions of $\TT^{\{k_i\}}_{I,J}(z,\bar{z})$ and $h^{\{k_i\}}_{I}(z)$ for a class of next-to-next-to-extremal four-point functions $\langle\OO_{k_1}\OO_{k_2}\OO_{k_3}\OO_{k_1+k_2+k_3-4}\rangle$ in each number of dimensions $d=2(\varepsilon+1)=3,4,6$.

The half-maximally supersymmetric analogue of these calculations were initiated in \cite{Dolan:2004mu} and pursued further in \cite{Baume:2019aid,Bobev:2017jhk,Chang:2017xmr} for theories in $d=3,4,5,6$ with eight real Poincaré supercharges and R-symmetry algebra $\mathfrak{su}(2)_R$. While the existence of fewer supercharges means that superblocks are smaller and less complicated, the loss of an R-symmetry cross-ratio makes the extraction of reduced correlators more delicate. Despite this, reduced correlator and their superblock decompositions have been derived in position space in 4d in \cite{Dolan:2001tt,Beem:2014zpa} and in 6d in \cite{Chang:2017xmr}. More recently, reduced correlators of holographic half-maximally supersymmetric CFTs in $3\leq d\leq6$ were found in Mellin space in \cite{Chester:2025jxg}. Alternative formulations of superblocks in a diverse range of dimensions and amounts of supersymmetry have been developed, e.g. in \cite{Bobev:2015jxa,Buric:2019rms,Aprile:2021pwd}.

We organize the remainder of this work as follows. In Section \ref{setup}, we begin by reviewing the mixed four-point functions of 1/2-BPS operators that are central to this work. After setting conventions and specifying the class of configurations $\langle\OO_{k_1}\OO_{k_2}\OO_{k_3}\OO_{k_1+k_2+k_3-2\EE}\rangle$ with $\EE=1,2$ that we will consider, we will define notation to collectively refer to the R-symmetry representations and supermultiplets appearing in such correlators in the 3d $\NN=8$, 4d $\NN=4$, and 6d $\NN=(2,0)$ SCFTs. With this, we review the superconformal Ward identities and the superconformal block decompositions of their solutions. To explicitly construct such superconformal blocks, we describe another method for satisfying the superconformal Ward identities introduced in \cite{Dolan:2004mu}. In Section \ref{E1}, we warm up to this perspective in the simple example of extremality $\EE=1$ correlators $\langle\OO_{k_1}\OO_{k_2}\OO_{k_3}\OO_{k_1+k_2+k_3-2}\rangle$, which turn out to be encoded by the single-variable reduced correlators $h^{\{k_i\}}_1$.

In a generalization to mixed correlators with $\EE=2$, Section \ref{E2} presents a system of six R-symmetry channel equations relating superblocks to a combination of $\Delta_\varepsilon$ acting on a two-variable reduced correlator $\TT^{\{k_i\}}$ and a number of single-variable functions $h^{\{k_i\}}_i$. We then describe how these operator equations can be solved to obtain explicit superblocks for $\langle\OO_{k_1}\OO_{k_2}\OO_{k_3}\OO_{k_1+k_2+k_3-4}\rangle$. In Section \ref{reducedblocks} we demonstrate how one can invert a particularly simple channel equation to obtain an expansion of $\TT^{\{k_i\}}$ in ``reduced" blocks with shifted external kinematics $(\tilde{\Delta}_{12},\tilde{\Delta}_{34})=\left(\Delta_{12},\Delta_{34}-2(\varepsilon-1)\right)$. Focusing on 6d and 3d, we describe the $\varepsilon$-dependent subtleties of these reduced block decompositions and point out an apparent paradox concerning single-variable reduced correlators in 3d.

Several technicalities are relegated to Appendices and an ancilliary \texttt{Mathematica} notebook. In Appendix \ref{superblockconventions}, we set our conformal block and R-symmetry polynomial conventions, while in Appendix \ref{jackpolys}, we define and list useful properties obeyed by the Jack polynomials that can be used to express and deduce relations between conformal blocks. In Appendices \ref{hrecurrencerelations} and \ref{Trecurrencerelations}, we present recursion relations obeyed by conformal blocks $G^{\Delta_{12},\Delta_{34}}_{\Delta,\ell}$ in the presence of the operator $\Delta_\varepsilon$ which we use to solve the system of R-symmetry channel differential equations in \eqref{eq:AT1}-\eqref{eq:AE2h6}. In Appendix \ref{6dsupermultipletsinE2}, we give an 6d $\NN=(2,0)$ example of the outcome of the Racah-Speiser algorithm for enumerating superconformal descendant primaries in the supermultiplets exchanged in $\langle\OO_{k_1}\OO_{k_2}\OO_{k_3}\OO_{k_1+k_2+k_3-4}\rangle$. Finally, we attach a \texttt{Mathematica} notebook with an implementation of the functions and recursion relations in Appendices \ref{hrecurrencerelations} and \ref{Trecurrencerelations} and R-symmetry channel equations in \eqref{eq:AT1}-\eqref{eq:AE2h6}.

\section{Four-point functions of 1/2-BPS operators}
\label{setup}
In this section, we define the class of four-point functions we will consider, set conventions, and write down the crossing equations obeyed by these mixed correlators. We will also outline structural similarities in the R-symmetry representation theory content of such correlators in the 3d $\NN=8$, 4d $\NN=4$, and 6d $\NN=(2,0)$ SCFTs and define a notation that unifies this. We go on to describe the superconformal Ward identities and two important decompositions of their solutions. The first is a superblock decomposition, which organizes operators into superblocks, wherein in all conformal primary operators have three-point coefficients proportional to that of the superconformal primary, with proportionality constants \textit{a priori} unknown. The second decomposition is that of \cite{Dolan:2004mu}, which in later sections we will use to solve for the proportionality constants in the superblock decomposition.

\subsection{Kinematics and crossing equations}
We will be concerned with four-point functions of 1/2-BPS superconformal primary operators $\OO_k$. These are rank-$k$ symmetric, traceless representations of the appropriate $\mathfrak{so}(\texttt{d})_R$ R-symmetry group with protected conformal dimension $\Delta=\varepsilon k$\footnote{Beyond this representation theory information, we will not make any specialization to a particular realization of these 1/2-BPS operators. For instance, we do not require these to be the ``single-trace" operators $S_k$ frequently studied as a starting point in the superconformal bootstrap.}. As is standard in the superconformal bootstrap, we will contract $\OO_{I_1\dots I_k}(x)$ with auxiliary complex $SO(\texttt{d})_R$ polarization vectors $t^I$:
\begin{align}
    \OO_k(x,t)\equiv \OO_{I_1\dots I_k}(x)t^{I_1}\cdots t^{I_k}.
\end{align}
This transfers $\mathfrak{so}(\texttt{d})_R$ representation theoretical information into polynomials of products of $t^{I_i}$, where tracelessness is encoded by the null condition $t^It_I=0$. Superconformal symmetry then allow us to factorize four-point functions of 1/2-BPS operators into a kinematic factor and a dynamical function $\mathcal{G}^{k_1k_2k_3k_4}$. Following \cite{Agmon:2019imm, Bissi:2020jve}, we extract a kinematical factor by writing our four-point function as\footnote{Our kinematic factor does not vary between ``Case I" and ``Case II" kinematics, as defined in e.g. \cite{Alday:2020dtb,Behan:2021pzk}. Instead, we will later compensate for negative powers of $t_{14}$ in ``Case II" by including appropriate powers of $\tau$ in $\mathcal{G}^{k_1k_2k_3k_4}$, following the 4d $\mathcal{N}=4$ analogy in \cite{Nirschl:2004pa}.} 
\begin{align}
    \langle \OO_{k_1}(x_1,t_1)\OO_{k_2}(x_2,t_2)\OO_{k_3}(x_3,t_3)\OO_{k_4}(x_4,t_4) \rangle =&\; \nonumber \\
    \left(\frac{t_{12}}{x^{2\varepsilon}_{12}}\right)^{\frac{k_1+k_2}{2}}
    \left(\frac{t_{34}}{x^{2\varepsilon}_{34}}\right)^{\frac{k_3+k_4}{2}}
    \left(\frac{t_{14}}{t_{24}}\right)^{\frac{k_{12}}{2}}&\;
    \left(\frac{t_{12}t_{34}}{t_{14}t_{24}}\right)^{\frac{k_{34}}{2}}
    \left(\frac{x_{14}^{2\varepsilon}x_{24}^{2\varepsilon}}{x_{12}^{2\varepsilon}x_{34}^{2\varepsilon}}\right)^{\frac{k_{34}}{2}}\mathcal{G}^{k_1k_2k_3k_4}(U,V;\sigma,\tau),
\label{eq:full4pt}
\end{align}
 where we consider operator orderings with $k_1,k_2,k_3\leq k_4$ and we define $x_{ij}=x_i-x_j$, $t_{ij}=t_i\cdot t_j$, and $k_{ij}=k_i-k_j$. The dynamical function $\mathcal{G}^{k_1k_2k_3k_4}$\footnote{To avoid clutter, we suppress the dependence of $\mathcal{G}^{k_1k_2k_3k_4}$ on $\varepsilon$, which will be made clear by the context.} depends on conformal cross-ratios $U$ and $V$ and $\mathfrak{so}(\texttt{d})_R$ cross-ratios $\sigma$ and $\tau$ defined by
 \begin{align}
    U=\frac{x_{12}^2x_{34}^2}{x_{13}^2x_{24}^2},\;\;\;\;V=\frac{x_{14}^2x_{23}^2}{x_{13}^2x_{24}^2},\;\;\;\;\sigma=\frac{t_{13}t_{24}}{t_{12}t_{34}},\;\;\;\;\tau=\frac{t_{14}t_{23}}{t_{12}t_{34}}.
\end{align}
 This organization implies that $\mathcal{G}^{k_1k_2k_3k_4}$ is a polynomial in $\sigma, \tau$ of degree min$\{k_i\}$, modulo a factor $\tau^{\frac{1}{2}\left(k_2+k_3-k_1-k_4\right)}$ that we include for kinematical configurations satisfying $k_1+k_4<k_2+k_3$. Permuting the operators $S_{k_i}(x_i,t_i)$ in  full correlator \eqref{eq:full4pt} imposes crossing relations between different channels. Swapping $1\leftrightarrow2$ gives the relation
\begin{align}
\mathcal{G}^{k_1k_2k_3k_4}(U,V;\sigma,\tau)=\mathcal{G}^{k_2k_1k_3k_4}\left(\frac{U}{V},\frac{1}{V};\tau,\sigma\right),
\label{eq:12cross}
\end{align}
while swapping $1\leftrightarrow3$ gives
\begin{align}
\mathcal{G}^{k_1k_2k_3k_4}(U,V;\sigma,\tau)=\left(\frac{U^{\varepsilon}}{V^\varepsilon}\tau\right)^{\frac{k_1+k_2+k_3-k_4}{2}}\mathcal{G}^{k_3k_2k_1k_4}\left(V,U;\frac{\sigma}{\tau},\frac{1}{\tau}\right),
\label{eq:13cross}
\end{align}
and finally, swapping $2\leftrightarrow3$
\begin{align}
\mathcal{G}^{k_1k_2k_3k_4}(U,V;\sigma,\tau)=\left(U^\varepsilon\sigma\right)^{\frac{k_1+k_2+k_3-k_4}{2}}\mathcal{G}^{k_1k_3k_2k_4}\left(\frac{1}{U},\frac{V}{U};\frac{1}{\sigma},\frac{\tau}{\sigma}\right).
\label{eq:23cross}
\end{align}
While the condition derived by swapping $1\leftrightarrow2$ is manifestly satisfied term-by-term in the block expansions appearing below, the remaining crossing relations \eqref{eq:13cross} and \eqref{eq:23cross} imply non-trivial constraints on four-point functions and are the foundation of the conformal bootstrap.


\subsection{Superconformal representation theory and superblock decompositions}
The bosonic subalgebra of the superconformal algebra allows us to decompose $\mathcal{G}^{k_1k_2k_3k_4}$ into a basis of exchanged irreducible representations (irreps) of the R-symmetry algebra $\mathfrak{so}(\texttt{d})_R$. For the R-symmetry algebras relevant to $\varepsilon=\frac{1}{2},1,2$ and assuming $k_1\leq k_2$, the allowed irrep exchanges in the OPE $S_{k_1}\times S_{k_2}$ are given by 
\begin{align}
    \mathfrak{so}(8)_R:\hspace{1cm} [0,0,k_1,0]\otimes[0,0,k_2,0]=&\;\bigoplus_{m=0}^{k_1} \bigoplus_{n=0}^{k_1-m}\left[0,n,k_1+k_2-2(m+n),0\right],
    \label{eq:so8tp}
    \\ 
    \mathfrak{so}(6)_R:\hspace{1cm} [0,k_1,0]\otimes [0,k_2,0]=&\;\bigoplus_{m=0}^{k_1} \bigoplus_{n=0}^{k_1-m}\left[n,k_1+k_1-2(m+n),n\right],
    \label{eq:so6tp}
    \\
    \mathfrak{so}(5)_R:\hspace{1cm} [k_1,0]\otimes [k_2,0]=&\;\bigoplus_{m=0}^{k_1} \bigoplus_{n=0}^{k_1-m} [k_1 + k_2-2(m+n), 2 n].
    \label{eq:so5tp}
\end{align}
It will be convenient to uniformize these selection rules by making the definitions
\begin{align}
    \mathfrak{so}(5)_R: \left[d_1,d_2\right],\hspace{1cm}\mathfrak{so}(6)_R: \left[\frac{d_2}{2},d_1,\frac{d_2}{2}\right],\hspace{1cm}\mathfrak{so}(8)_R: \left[0,\frac{d_2}{2},d_1,0\right].
\label{eq:uniformquantumnumbers}
\end{align}
This allows us to use \eqref{eq:so5tp} to collectively encode $\mathfrak{so}(\texttt{d})_R$ tensor products.

Focusing on the $s$-channel, the admissible $\mathfrak{so}(\texttt{d})_R$ 
irreps exchanged in $\mathcal{G}^{k_1k_2k_3k_4}$ are captured by $([k_1,0]\otimes [k_2,0])\cap([k_3,0]\otimes [k_4,0])$. Temporarily taking $k_1\leq k_2$ again, we can decompose $\mathcal{G}^{k_1k_2k_3k_4}\equiv\mathcal{G}^{\{k_i\}}$ into these irreps by writing 
\begin{align}
    \mathcal{G}^{\{k_i\}}(U,V;\sigma,\tau)=\tau^{\delta}\sum_{m=\frac{k_2-k_1}{2}}^{\frac{k_1+k_2}{2}} \sum_{n=\frac{k_2-k_1}{2}}^{m} Y_{mn}^{\frac{|k_{12}-k_{34}|}{2},\frac{|k_{23}-k_{14}|}{2}}(\sigma,\tau) A_{mn}^{k_{12},k_{34}}(U,V).
\label{eq:Adecomp}
\end{align}
The functions $Y_{mn}^{a,b}(\sigma,\tau)$ are eigenfunctions of the $\mathfrak{so}(\texttt{d})_R$ quadratic Casimir associated with exchanged irreps $[2n,2(m-n)]$. They are polynomials in $\sigma$ and $\tau$ of degree $m-\frac{a+b}{2}$ and can be computed as reviewed in Appendix \ref{superblockconventions}. The exponent $\delta$ is 
\begin{align}
    \delta=
    \begin{cases}
        0 \;\;\;& k_1+k_4\geq k_2+k_3,\\ \frac{k_2+k_3-k_1-k_4}{2} \;\;\;& k_1+k_4 < k_2+k_3,
    \end{cases}
\label{eq:delta}
\end{align}
and serves to cancel negative powers of $t_{14}$ in the kinematic factor, as described in \cite{Nirschl:2004pa}. To avoid unnecessarily many conditional expressions and without loss of generality, we will restrict to the class of configurations where $k_4=k_1+k_2+k_3-2\EE$ where the extremality $\EE$ is an integer obeying $k_i\geq\EE\geq0$ which quantifies the complexity of the R-symmetry representations exchanged in $\GG^{\{k_i\}}$. We do not impose any ordering among $k_1,k_2,k_3$ and these configurations are sufficient to consider all of the channels appearing in the crossing equations \eqref{eq:12cross}, \eqref{eq:13cross}, and \eqref{eq:23cross}. For this class of configurations, we always have that $|k_{12}|\leq|k_{34}|$ and $\delta=0$. We can then take the $s$-channel OPEs $\OO_{k_1}\times \OO_{k_2}$ and $\OO_{k_3}\times \OO_{k_4}$ and expand the R-symmetry channel functions $A_{mn}^{k_{12},k_{34}}$ in the $2(\varepsilon+1)$-dimensional bosonic conformal blocks $G_{\Delta,\ell}^{\Delta_{12},\Delta_{34}}(U,V)$ defined in Appendix \ref{superblockconventions} as 
\begin{align}
    A_{mn}^{k_{12},k_{34}}(U,V)=U^{\varepsilon\frac{k_{34}}{2}}\sum_{\Delta,\ell}\lambda_{k_1k_2\mathcal{O}_{\Delta,\ell,mn}}\lambda_{k_3k_4\mathcal{O}_{\Delta,\ell,mn}}G_{\Delta,\ell}^{\Delta_{12},\Delta_{34}}(U,V).
\label{eq:Ablockdecomp}
\end{align}

The fermionic subalgebra of each superconformal algebra $d=2(\varepsilon+1)$ dimensions imposes additional constraints which are captured by the superconformal Ward identities \cite{Dolan:2004mu}
\begin{align}
    \left(z\partial_z-\varepsilon\alpha\partial_\alpha\right)\mathcal{G}^{\{k_i\}}\left(z,\bar z;\alpha,\bar \alpha\right)\bigg|_{\alpha\rightarrow z^{-1}}=\left(\bar{z}\partial_{\bar{z}}-\varepsilon\bar{\alpha}\partial_{\bar{\alpha}}\right)\mathcal{G}^{\{k_i\}}\left(z,\bar z;\alpha,\bar \alpha\right)\bigg|_{\bar{\alpha}\rightarrow \bar{z}^{-1}}=0,
\label{eq:scwi}
\end{align}
where the variables $z,\bar{z}$ and $\alpha,\bar{\alpha}$ are related to conformal and R-symmetry cross-ratios by 
\begin{align}
    U=z \bar{z},\;\;\;\;\;\;V=(1-z)(1- \bar{z}),\;\;\;\;\;\;\;\;\sigma=\alpha \bar{\alpha},\;\;\;\;\;\;\tau=(1-\alpha) (1-\bar{\alpha}).
\end{align}
\par 
A consequence of these Ward identities is that, in the same way conformal symmetry allows us to decompose $\mathcal{G}^{\{k_i\}}$ in a sum of conformal blocks $G_{\Delta,\ell}^{\Delta_{12},\Delta_{34}}(U,V)$, i.e. in a basis of functions that individually satisfy the $\mathfrak{so}(d,2)$ conformal Casimir equation, superconformal symmetry allows us to repackage $\mathcal{G}^{\{k_i\}}$ into an expansion in superblocks $\mathfrak{G}_\mathcal{X}^{\Delta_{12},\Delta_{34}}(U,V;\sigma,\tau)$, i.e. a basis of functions that individually satisfy \eqref{eq:scwi}. So we can write  
\begin{align}
\mathcal{G}^{\{k_i\}}(U,V;\sigma,\tau)=\sum_{\mathcal{X}\in (\OO_1\times \OO_2)\cap(\OO_3\times \OO_4)}\lambda_{k_1 k_2 \mathcal{X}}\lambda_{k_3 k_4 \mathcal{X}}\mathfrak{G}_\mathcal{X}^{\Delta_{12},\Delta_{34}}(U,V;\sigma,\tau),
\label{eq:superblockdecomp}
\end{align}
where for $k_1\leq k_2$ superblocks  then take the form
\begin{align}
    \mathfrak{G}_\mathcal{X}^{\Delta_{12},\Delta_{34}}(U,V;\sigma,\tau)=&\;\tau^{\delta}\sum_{m=\frac{k_2-k_1}{2}}^{\frac{k_1+k_2}{2}}\sum_{n=\frac{k_2-k_1}{2}}^{m} Y_{mn}^{a,b}(\sigma,\tau)\sum_{\mathcal{O}\in\mathcal{X}} \mathcal{C}^{\mathcal{X},\Delta_{12},\Delta_{34}}_{\mathcal{O}_{\Delta,\ell,mn}}\;U^{\varepsilon\frac{k_{34}}{2}}\;G_{\Delta,\ell}^{\Delta_{12},\Delta_{34}}(U,V)\nonumber \\
    \equiv&\;\tau^{\delta}\sum_{m=\frac{k_2-k_1}{2}}^{\frac{k_1+k_2}{2}}\sum_{n=\frac{k_2-k_1}{2}}^{m} Y_{mn}^{a,b}(\sigma,\tau)A_{mn;\Delta',\ell'}^{k_{12},k_{34}}\left(U,V\right),
\label{eq:superblock}
\end{align}
where we abbreviated $\left(a,b\right)=\left(\frac{|k_{12}-k_{34}|}{2},\frac{|k_{23}-k_{14}|}{2}\right)$ and we defined $A_{mn;\Delta,\ell}^{k_{12},k_{34}}$ to be the channel contribution coming from the supermultiplet $\XX$ with (a unique) primary $(\Delta',\ell')$ transforming in the highest R-symmetry irrep $\left[\frac{k_1+k_2}{2},\frac{k_1+k_2}{2}\right]$.

In passing from the block decomposition in \eqref{eq:Adecomp} and \eqref{eq:Ablockdecomp} to the superblock decomposition in \eqref{eq:superblockdecomp}, supersymmetry has identified all products of three-point coefficients of superdescendent primaries $\mathcal{O}_{\Delta,\ell,mn}$ in a supermultiplet to be proportional to those of the superconformal primary, which we also denote $\mathcal{X}$. The proportionality constants $\mathcal{C}^{\mathcal{X},\Delta_{12},\Delta_{34}}_{\mathcal{O}_{\Delta,\ell,mn}}=\frac{\lambda_{k_1 k_2 \mathcal{O}_{\Delta,\ell,mn}}\lambda_{k_3 k_4 \mathcal{O}_{\Delta,\ell,mn}}}{\lambda_{k_1 k_2 \mathcal{X}}\lambda_{k_3 k_4 \mathcal{X}}}$ weight different primaries in a supermultiplet. One way to compute these is to follow the strategy in \cite{Chester:2014fya,Agmon:2017xes,Alday:2020tgi} by expanding the ansatz \eqref{eq:superblock} around $z,\bar{z}\ll1$ (e.g. using Jack polynomials) and imposing \eqref{eq:scwi} order-by-order. To actually generate the list of superdescendent primaries $\mathcal{O}_{\Delta,\ell,mn}$ in a given supermultiplet $\XX$, we used the Python implementation \cite{Hayling:2020mbp} of the Racah-Speiser algorithm \cite{Cordova:2016emh,Buican:2016hpb}, taking into account the R-symmetry selection rules and the $\ell$ selection rules discussed in the following Subsection. One of the results of this paper will be to provide an alternative derivation of the spectra of superconformal descendant primaries and their corresponding proportionality constants $\mathcal{C}^{\mathcal{X},\Delta_{12},\Delta_{34}}_{\mathcal{O}_{\Delta,\ell,mn}}$ in mixed correlators $\langle\OO_{k_1}\OO_{k_2}\OO_{k_3}\OO_{k_1+k_2+k_3-4}\rangle$.

The superconformal multiplets $\mathcal{X}$ that are allowed to appear in a given OPE $\OO_{k_1}\times \OO_{k_2}$ are dictated by the selection rules for 1/2-BPS operators worked out for the maximally supersymmetric CFTs in several studies, e.g. \cite{Ferrara:2001uj,Eden:2001ec,Nirschl:2004pa,Heslop:2004du}. It will be instructive to quote these superselection rules for each theory, using the language adopted in the numerical bootstrap studies in \cite{Chester:2014fya,Beem:2016wfs,Beem:2015aoa}\footnote{We slightly adapt this notation to make quantum numbers $(\Delta,\ell)$ appear explicitly.}\footnote{We refer the reader to \cite{Cordova:2016emh}, whose authors introduced an alternative notation for uniformly referring to superconformal multiplets in diverse dimensions.}. Taking $k_1\leq k_2$, we have \\\\
\noindent
\underline{6d $\NN=(2,0)$:}
\begin{align}
    \mathcal{D}[k_1,0]_{2k_1,0}\otimes\mathcal{D}[k_2,0]_{2k_2,0}=&\; \bigoplus_{m=0}^{k_1}\bigoplus_{\substack{n=0}}^{k_1-m}\mathcal{D}[k_1+k_2-2(m+n),2n]_{2(k_1+k_2-2m),0} \nonumber \\
    \oplus&\;\bigoplus_{m=1}^{k_1}\bigoplus_{\substack{n=0}}^{k_1-m}\bigoplus_{\ell=0}^{\infty}\mathcal{B}[k_1+k_2-2(m+n),2n]_{4+\ell+2(k_1+k_2-2m),\ell} 
    \label{eq:6dsuperselect}
    \\ \oplus&\;\bigoplus_{m=2}^{k_1}\bigoplus_{\substack{n=0}}^{k_1-m}\bigoplus_{\ell=0}^{\infty}\bigoplus_{\Delta}\mathcal{L}[k_1+k_2-2(m+n),2n]_{\Delta>6+\ell+2(k_1+k_2-2m),\ell}. \nonumber
\end{align}
\noindent
\underline{4d $\NN=4$:}
\begin{align}
    \mathcal{B}_{[0,k_1,0],k_1,0}\otimes\mathcal{B}_{[0,k_2,0],k_2,0}=&\; \bigoplus_{m=0}^{k_1}\bigoplus_{\substack{n=0}}^{k_1-m}\mathcal{B}_{[n,k_1+k_2-2(m+n),n],k_1+k_2-2m,0} \nonumber \\
    \oplus&\;\bigoplus_{m=1}^{k_1}\bigoplus_{\substack{n=0}}^{k_1-m}\bigoplus_{\ell=0}^{\infty}\mathcal{C}_{[n,k_1+k_2-
    2(m+n),n],2+\ell+k_1+k_2-2m,\ell}
    \label{eq:4dsuperselect}
    \\ \oplus&\;\bigoplus_{m=2}^{k_1}\bigoplus_{\substack{n=0}}^{k_1-m}\bigoplus_{\ell=0}^{\infty}\bigoplus_{\Delta}\mathcal{A}_{[n,k_1+k_2-2(m+n),n],\Delta\geq2+\ell+k_1+k_2-2m,\ell}. \nonumber
\end{align}
\noindent
\underline{3d $\NN=8$:}
\begin{align}
    \left(B,+\right)_{\frac{k_1}{2},0}^{[0,0,k_1,0]}\otimes\left(B,+\right)_{\frac{k_2}{2},0}^{[0,0,k_2,0]}=& \bigoplus_{m=0}^{k_1}\bigoplus_{\substack{n=0}}^{k_1-m}\left(B,i\right)_{\frac{k_1+k_2-2m}{2},0}^{\left[0,n,k_1+k_2-2(m+n),0\right]}\nonumber \\
    \oplus&\;\bigoplus_{m=1}^{k_1}\bigoplus_{\substack{n=0}}^{k_1-m}\bigoplus_{\ell=0}^{\infty}(A,i)^{[0,n,k_1+k_2-2(m+n),0]}_{1+\ell+\frac{k_1+k_2-2m}{2},\ell}
    \label{eq:3dsuperselect}
    \\ \oplus&\;\bigoplus_{m=2}^{k_1}\bigoplus_{\substack{n=0}}^{k_1-m}\bigoplus_{\ell=0}^{\infty}\bigoplus_{\Delta}\left(A,0\right)^{[0,n,k_1+k_2-2(m+n),0]}_{\Delta\geq1+\ell+\frac{k_1+k_2-2m}{2},\ell}, \nonumber
\end{align}
where 
\begin{align}
    (B,i)^{[0,d_2,d_3,0]}_{\Delta,\ell}=&\;
    \begin{cases}
        (B,+)^{[0,0,d_3,0]}_{\frac{d_3}{2},0} \text{ for } d_2=0\\
        (B,2)^{[0,d_2,d_3,0]}_{\frac{d_2+d_3}{2},0} \text{ for } d_2>0
    \end{cases},\\
    (A,i)^{[0,d_2,d_3,0]}_{\Delta,\ell}=&\;
    \begin{cases}
        (A,+)^{[0,0,d_3,0]}_{1+\ell+\frac{d_3}{2},\ell} \text{ for } d_2=0\\
        (A,2)^{[0,d_2,d_3,0]}_{1+\ell+\frac{d_2+d_3}{2},\ell} \text{ for } d_2>0
    \end{cases}.
\end{align}
It is worth pointing out certain structural similarities that underlie the unified treatment in the following sections. First, we define the combinations
\begin{align}
    m\left(\EE\right)=&\;\frac{1}{2}\min\left(k_1+k_2,k_3+k_4\right), \label{eq:mE}\\
p\left(\EE\right)=&\;\min\left(k_1+k_2,k_3+k_4\right)-2\EE,
    \label{eq:pE}
\end{align}
which will repeatedly appear in various quantum numbers. In the kinematical configurations $\langle\OO_{k_1}\OO_{k_2}\OO_{k_3}\OO_{k_1+k_2+k_3-2\EE}\rangle$ studied in this paper, we will always have $m(\EE)=\frac{k_1+k_2}{2}$ and $p(\EE)=k_1+k_2-2\EE=-\frac{\Delta_{34}}{\varepsilon}$.

For any $\EE\geq2$, the exchanged supermultiplets come in three families (arranged in each of the three lines) characterized by their superformal primary. Consider the $\EE=2$ case that will be central in this work. In terms of our R-symmetry convention in Equation \eqref{eq:uniformquantumnumbers} and the parameter $p(2)=k_1+k_2-4\equiv p$, the superconformal primaries of exchanged supermultiplets are summarized by
\begin{enumerate}
    \item \underline{BPS scalars:} 
    \begin{itemize}
        \item $[p,0]$ with scaling dimension $\Delta=\varepsilon p$
        \item $[p+2,0]$ and $[p,2]$, both with $\Delta=\varepsilon(p+2)$
        \item $[p+4,0]$, $[p+2,2]$, and $[p,4]$, each with $\Delta=\varepsilon(p+4)$
    \end{itemize}
    \item \underline{Semi-short operators:} \\
    These have spin $\ell$ and come in families labeled by twist $t=\Delta-\ell$. We have
    \begin{itemize}
        \item $[p,0]$ with $t=\varepsilon(p+2)$ 
        \item $[p+2,0]$ and $[p,2]$, both with  $t=\varepsilon\left(p+4\right)$
    \end{itemize}
    \item \underline{Long operators:} 
    \begin{itemize}
        \item $[p,0]$ with spin $\ell$ and unprotected scaling dimension $\Delta$
    \end{itemize}
\end{enumerate}
Since distinct supermultiplets will often have superconformal primaries with the same conformal quantum numbers $(\Delta,\ell)$, we will arrange superconformal block decompositions in terms of the quantum numbers $(\Delta',\ell')$ of the unique conformal primary in the top R-symmetry channel of each supermultiplet. Equivalently, such supermultiplets can be distinguished by the shortening condition of the superconformal primary and the resulting spectrum of superconformal descendant primaries. As an example, we present such spectra explicitly for 6d $\NN=(2,0)$ theories in Appendix \ref{6dsupermultipletsinE2}. 

The superselection rules in Equations \eqref{eq:6dsuperselect}-\eqref{eq:3dsuperselect} only describe what is allowed by superconformal symmetry and additional symmetries and physical principles may remove operators or even entire supermultiplets from these list. For instance, an additional selection rule in each of these SCFTs is that when $k_1=k_2$, Bose symmetry dictates that irreps $[2a,2b]$ must have even/odd $\ell$ for even/odd $b$ (corresponding to a symmetric/antisymmetric $\mathfrak{so}(d)_R$ irrep). The proliferation of superdescendant primaries allowed in supermultiplets exchanged in mixed (i.e. $k_1\neq k_2$) correlators is the main way in which our results generalize the $\langle\OO_p\OO_p\OO_q\OO_q\rangle$ superblocks presented in \cite{Dolan:2004mu,Chester:2014fya}. A physical consideration is that certain low-twist protected multiplets in \eqref{eq:6dsuperselect}-\eqref{eq:3dsuperselect} can only be constructed using the free scalar $S_1$ which the authors of \cite{Maldacena:2011jn} argued decouple from interacting SCFTs. As a result, such contributions can be set to zero.

\subsection{An alternative solution to the superconformal Ward identities}
\label{DGSdecomp}
The authors of \cite{Dolan:2004mu} proposed a different way to solve the superconformal Ward identities \eqref{eq:scwi} for four-point correlators $\langle\OO_k\OO_k\OO_k\OO_k\rangle$ in all SCFTs with R-symmetry algebra $\mathfrak{so}(\texttt{d})_R$. They demonstrate that the superconformal Ward identities are satisfied by a sum of two-variable so-called ``reduced correlator" functions $\TT^{\{k_i\}}_{I,J}(U,V)$ and single-variable functions $h^{\{k_i\}}_I(z)$, acted on by differential operators. A crucial observation is that, aside from the extremality $\EE$, the superconformal Ward identities are insensitive to the kinematical configurations of the four-point correlators $\GG^{\{k_i\}}$ they constrain. This allows us to immediately generalize their expression to write $\GG^{\{k_i\}}$ as
\begin{align}
    \GG^{\{k_i\}}&\left(U,V;\sigma,\tau\right)=\nonumber \\
    &\sum_{I=0}^{\EE-2}\sum_{J=0}^{\EE-2-I}U^{\varepsilon(I+J+2)}V^{-\varepsilon J}\sigma^I\tau^J\Delta_{\varepsilon}\left[\left(z\alpha-1\right)\left(z\bar{\alpha}-1\right)\left(\bar{z}\alpha-1\right)\left(\bar{z}\bar{\alpha}-1\right)\frac{\TT^{\{k_i\}}_{I,J}\left(U,V\right)}{U^{3\varepsilon-1}}\right]\nonumber \\
    +&\sum_{I=1}^{\EE}U^{\varepsilon I}\sigma^{I-1}F_{I}^{\{k_{i}\}}\left(z,\bar{z};\alpha,\bar{\alpha}\right),
    \label{eq:DGSdecomp}
\end{align}
where the factor $U^{-(3\varepsilon-1)}$ multiplying $\TT^{\{k_i\}}_{I,J}$ is included for later convenience and where
\begin{align}
    F_{I}^{\{k_{i}\}}\left(z,\bar{z};\alpha,\bar{\alpha}\right)=\left(D_\varepsilon\right)^{\varepsilon-1}\left[\frac{(z \alpha-1)(z \bar{\alpha}-1)h^{\{k_{i}\}}_I(z)-(\bar{z} \alpha-1)(\bar{z} \bar{\alpha}-1)h^{\{k_{i}\}}_I(\bar{z})}{z-\bar{z}}\right].
    \label{eq:Fh}
\end{align}
The operator $\Delta_{\varepsilon}$ is defined as
\begin{align}
    \Delta_{f}=\left(D_{\varepsilon}\right)^{f-1}U^{f-1},
\label{eq:Deltaepsilon}
\end{align} 
in terms of the differential operator 
\begin{align}
    D_{\varepsilon}=\partial_ z\partial_{\bar{z}}-\frac{\varepsilon}{z-\bar{z}}\left(\partial_{z}-\partial_{\bar{z}}\right).
\label{eq:Depsilon}
\end{align}
Other than when $\varepsilon=1,2,3,\dots$ (corresponding to even number of dimensions $d\geq4$), the fractional and/or negative powers of $D_\varepsilon$ render the operator $\Delta_\varepsilon$ non-local, making it \textit{a priori} unclear how to work with the functions $\TT^{\{k_i\}}_{I,J}$ and $h^{\{k_i\}}_I$ directly without more global information about the four-point function $\GG^{\{k_i\}}$. Nevertheless, we can manipulate these expressions by noting that the operator $\Delta_\varepsilon$ has as its eigenfunctions a family of two-variable symmetric functions $P^{(\varepsilon)}_{a,b}$ known as Jack polynomials \cite{Jack:1970,Dolan:2000ut,Dolan:2011dv} which satisfy
\begin{align}
    \Delta_f P^{(\varepsilon)}_{a,b}(z,\bar{z})=(a+\varepsilon+1)_{f-1}(b+1)_{f-1}P^{(\varepsilon)}_{a,b}(z,\bar{z}),
\end{align}
in terms of the Pochhammer symbol $(x)_y=\Gamma[x+y]/\Gamma[x]$. The operator $\Delta_\varepsilon$ has a non-trivial kernel which implies an ambiguity in the definition of the reduced correlator functions. As originally described in \cite{Dolan:2004mu}, we can use this ambiguity to eliminate some of the single-variable contributions $h^{\{k_i\}}_I$ in \eqref{eq:Fh} through redefinitions of $\TT^{\{k_i\}}_{I,J}$ and the other single-variable functions $h^{\{k_i\}}_{I'}$. Finally, it will be useful to note that the conformal blocks $G_{\Delta,\ell}^{\Delta_{12},\Delta_{34}}$ appearing in Equations \eqref{eq:Ablockdecomp} and \eqref{eq:superblock} can also be expanded in Jack polynomials as
\begin{align}
    G_{\Delta,\ell}^{\Delta_{12},\Delta_{34}}(z,\bar{z})=(-1)^\ell\sum_{m=0}^\infty\sum_{n=0}^\infty r_{m,n;\Delta,\ell}^{\Delta_{12},\Delta_{34}}\;P^{(\varepsilon)}_{\frac{\Delta+\ell}{2}+m,\frac{\Delta-\ell}{2}+n}(z,\bar{z}).
\end{align}
In Appendix \ref{jackpolys}, we will collect several properties of Jack polynomials and the expansion coefficients $r_{m,n;\Delta,\ell}^{\Delta_{12},\Delta_{34}}$ that will be essential for the results of this paper. 

The appearance of single-variable functions $h_I(z)$ in Equation \eqref{eq:DGSdecomp} is related to a deeper consequence of the superconformal Ward identities in \eqref{eq:scwi}. In particular, it has been observed that the four-point functions $\GG^{\{k_i\}}$ become topological along certain loci in cross-ratio space, leading to the identification of protected subsectors of these theories. This is seen by relating the cross-ratios $z, \bar{z}$ and $\alpha,\bar{\alpha}$ in $\varepsilon$-dependent ways to transform the superconformal Ward identities in Equation \ref{eq:scwi} into holomorphicity constraints on a ``twisted" $\GG^{\{k_i\}}$. Specifically, in each $\varepsilon$ we have
\begin{align}
    \varepsilon=2:&\;\hspace{2cm}\frac{\partial}{\partial\bar{z}}\GG^{\{k_i\}}\left(z,\bar{z};\frac{1}{\bar{z}},\frac{1}{\bar{z}}\right)=0, \label{eq:6dtwistedSCWI}\\
    \varepsilon=1:&\;\hspace{2cm}\frac{\partial}{\partial\bar{z}}\GG^{\{k_i\}}\left(z,\bar{z};\alpha,\frac{1}{\bar{z}}\right)=0, \label{eq:4dtwistedSCWI}\\
    \varepsilon=\frac{1}{2}:&\;\hspace{2cm}\frac{\partial}{\partial\bar{z}}\GG^{\{k_i\}}\left(\bar{z},\bar{z};\alpha,\frac{1}{\bar{z}}\right)=0. \label{eq:3dtwistedSCWI}
\end{align}
The protected subsectors encoded by these holomorphic solutions have interpretations as lower-dimensional quantum theories which for each $\varepsilon$ can be summarized as 
\begin{itemize}
    \item 6d $\NN=(2,0)$: a unitary 2d $\WW_\mathfrak{g}$ algebra 
    \item 4d $\NN=4$: a non-unitary 2d small $\NN=4$ super $\WW$-algebra 
    \item 3d $\NN=8$: a theory of 1d topological quantum mechanics 
\end{itemize}
A full treatment of each case can found in the original papers \cite{Chester:2014mea,Beem:2013sza,Beem:2014kka}. As described in these references, the subset of supermultiplets that are captured by these subsectors are protected and contain conformal primaries with quantum numbers that obey the certain linear relations. Below, we list all multiplets appearing in the superselection rules \eqref{eq:3dsuperselect}-\eqref{eq:6dsuperselect} that contain such primaries:
\begin{itemize}
    \item 6d $\NN=(2,0)$:\hspace{1cm}$\;\;\mathcal{D}\left[d_1,0\right]_{\;2d_1,0},\;\;\;\;\mathcal{D}\left[d_1,2\right]_{\;2d_1+4,0},\;\;\;\;\mathcal{B}\left[d_1,0\right]_{2d_1+4+\ell,\ell}.$
    \item 4d $\NN=4$:\hspace{2cm}$\mathcal{B}_{\left[\frac{d_2}{2},d_1,\frac{d_2}{2}\right],\;d_1+d_2,0},\;\;\;\;\;\;\mathcal{C}_{\left[\frac{d_2}{2},d_1,\frac{d_2}{2}\right],\;d_1+d_2+2+\ell,\ell}.$
    \item 3d $\NN=8$:\hspace{2cm}$(B,+)^{[0,0,d_1,0]}_{\frac{d_1}{2},0},\;\;\;\;\;\;\;\;\;\;\;(B,2)^{[0,d_2,d_1,0]}_{d_1+\frac{d_2}{2},0}.$ 
\end{itemize}
The twists in Equations \eqref{eq:6dtwistedSCWI}-\eqref{eq:3dtwistedSCWI} devolve $\GG^{\{k_i\}}$ into a meromorphic function involving only the $h^{\{k_i\}}_I$ reduced correlators. While not apparent in the initial construction in \cite{Dolan:2004mu}, the twisted constructions in each $\varepsilon$ imply that the only operators captured by $h^{\{k_i\}}_I$ are those belonging to the protected supermultiplets listed above. 

With this setup in place, the following Sections will be devoted to solving for the spectrum of superconformal descendant primaries and their block coefficients $\mathcal{C}^{\mathcal{X},\Delta_{12},\Delta_{34}}_{\mathcal{O}_{\Delta,\ell,mn}}$ using the decomposition in \eqref{eq:DGSdecomp}. We derive a mixed-correlator generalization of the differential equations that encode each $\mathfrak{so}\left(\texttt{d}\right)_R$ channel and describe how these can be solved to derive full superblocks for all such correlators in each maximally supersymmetric CFT. 

\section{Warm-up: Next-to-extremal four-point functions}
\label{E1}
We begin by studying correlators in the configuration $\langle\OO_{k_1}\OO_{k_2}\OO_{k_3}\OO_{k_1+k_2+k_3-2}\rangle$ which have extremality $\EE=1$. The superselection rules \eqref{eq:3dsuperselect}-\eqref{eq:6dsuperselect} show that these correlators only exchange BPS and semi-short operators and are strongly constrained by supersymmetry. These correlators are expected to vanish when the highest charge external operator $\OO_{k_4=k_1+k_2+k_3-2}$ is single-trace, however they are non-vanishing when we generalize to the case of external composite 1/2-BPS operators. The tensor products in \eqref{eq:so5tp} demonstrate that correlators in this configuration will exchange three  R-symmetry irreps. Using the temporary abbreviation $p(\EE=1)=k_1+k_2-2\equiv p$, these irreps can be expressed in the language of \eqref{eq:uniformquantumnumbers} as
\begin{align}
    \left([k_1,0]\otimes [k_2,0]\right)\cap\left([k_3,0]\otimes [k_4,0]\right)=
    &\;[p+2,0]\oplus[p+4,0]\oplus[p+2,2].
    \label{eq:so5tpE1}
\end{align}
These irreps are divided into symmetric and antisymmetric representations:
\begin{align}
    \left([k_1,0]\otimes [k_2,0]\right)\cap\left([k_3,0]\otimes [k_4,0]\right)\Big|_{\text{symm}}=&\;[p+2,0]\oplus[p+4,0],\\
    \left([k_1,0]\otimes [k_2,0]\right)\cap\left([k_3,0]\otimes [k_4,0]\right)\Big|_{\text{antisymm}}=&\;[p+2,2].
\end{align}
This distinction will be important for configurations where one of the OPEs involve operators with identical dimension, in which case Bose symmetry dictates that $\mathfrak{so}(\texttt{d})_R$ irreps $[2a,2b]$ must have even/odd spin $\ell$ for even/odd $b$. One instance where this comes into effect is when $k_1=k_2=1$, in which case our correlators involve the free scalar $S_1$ of dimension $\Delta=\varepsilon$. Beyond this section, we will ignore $S_1$ (and composite operators made thereof) and only consider external operators $\OO_{k_i}$ with $k_i\geq2$.

\subsection{$\frak{so}(\texttt{d})_R$ channel equations}

The contribution of superconformal descendant blocks in each $\mathfrak{so}(\texttt{d})_R$ channel is captured by the channel functions $A_{mn}^{k_{12},k_{34}}$ appearing in Equations \eqref{eq:Adecomp} and \eqref{eq:Ablockdecomp}, where the labels $(m,n)$ translate to irreps with uniformized Dynkin labels $[2n,2(m-n)]$ defined in \eqref{eq:uniformquantumnumbers}. The decomposition of $\GG^{\{k_i\}}$ in Equation \eqref{eq:DGSdecomp} demonstrates that $\EE=1$ correlators can be expressed in terms of just one single-variable reduced correlator $h^{\{k_i\}}_1$. The three channels appearing in \eqref{eq:so5tpE1} are encoded in $A_{mn}^{k_{12},k_{34}}$ and it will prove instructive to write $\Delta_{12}=\varepsilon\left(k_1-k_2\right)$ and $\Delta_{34}=\varepsilon\left(2\EE-k_1-k_2\right)$ throughout the rest of this work.
In terms of $m(\EE=1)=\frac{k_1+k_2}{2}\equiv m$, the three $\mathfrak{so}(\texttt{d})_R$ channel equations take the form
\begin{align}
    A^{\{k_{i}\}}_{m\;m}=&\;\frac{1}{\mathcal{Y}_{m}^{k_{12},k_{34}}}U^{\varepsilon}\left(D_\varepsilon\right)^{\varepsilon-1}\left[\frac{z^2h^{\{k_{i}\}}_1(z)-\bar{z}^2h^{\{k_{i}\}}_1(\bar{z})}{z-\bar{z}}\right], 
    \label{eq:AE1h1}\\
    A^{\{k_{i}\}}_{m\;m-1}=&\;U^{\varepsilon}\left(D_\varepsilon\right)^{\varepsilon-1}\Bigg[\frac{\Delta _{12}}{2 \left(\Delta _{34}-2 \varepsilon \right)}\frac{z^2h^{\{k_{i}\}}_1(z)-\bar{z}^2h^{\{k_{i}\}}_1(\bar{z})}{z-\bar{z}}\nonumber \\
    &\;\hspace{5.75cm}
    +\frac{z(z-2)h^{\{k_{i}\}}_1(z)-\bar{z}(\bar{z}-2)h^{\{k_{i}\}}_1(\bar{z})}{2(z-\bar{z})}\Bigg], 
    \label{eq:AE1h2}\\
    A^{\{k_{i}\}}_{m-1\;m-1}=&\;U^{\varepsilon}\left(D_\varepsilon\right)^{\varepsilon-1}\Bigg[\frac{1}{4} \left(\frac{\Delta _{12}^2}{\Delta _{34}-2\varepsilon -2}-\frac{\Delta _{12}^2-1}{\Delta _{34}-2 \varepsilon -1}+1\right)\frac{z^2h^{\{k_{i}\}}_1(z)-\bar{z}^2h^{\{k_{i}\}}_1(\bar{z})}{z-\bar{z}} \nonumber \\
    &\;\hspace{3.1cm}+\frac{\Delta _{12}}{\Delta _{34}-2 (\varepsilon +1)}\frac{z(z-2)h^{\{k_{i}\}}_1(z)-\bar{z}(\bar{z}-2)h^{\{k_{i}\}}_1(\bar{z})}{2(z-\bar{z})}\nonumber \\
    &\;\hspace{6.2cm}+\frac{(1-z)h^{\{k_{i}\}}_1(z)-(1-\bar{z})h^{\{k_{i}\}}_1(\bar{z})}{z-\bar{z}}\Bigg],
    \label{eq:AE1h3}
    \end{align}
where
\begin{align}
    \mathcal{Y}_{\frac{p}{2}}^{k_{12},k_{34}}=\frac{p!\left(\frac{k_{12}-k_{34}}{2}\right)!}{\left(\frac{p+k_{12}}{2}\right)!\left(\frac{p-k_{34}}{2}\right)!},
\label{eq:Rgluing}
\end{align}
is a kinematical factor.
With these ingredients, we can construct the superconformal blocks $\mathfrak{G}_\mathcal{X}^{\Delta_{12},\Delta_{34}}$ as follows. First, we decompose each channel function $A_{mn}^{k_{12},k_{34}}$ into a sum over contributions from each exchanged supermultiplet $\XX$ by writing 
\begin{align}
    A_{mn}^{k_{12},k_{34}}\left(U,V\right)=\sum_{\mathcal{X}\in (\OO_1\times \OO_2)\cap(\OO_3\times \OO_4)}\lambda_{k_1 k_2 \mathcal{X}}\lambda_{k_3 k_4 \mathcal{X}}\;A_{mn;\Delta_\XX',\ell_\XX'}^{k_{12},k_{34}}\left(U,V\right).
\end{align}
The channel equations \eqref{eq:AE1h1}-\eqref{eq:AE1h3} hold for each $A_{mn;\Delta_\XX,\ell_\XX}^{k_{12},k_{34}}$ individually and yield a sum of superconformal descendant primary blocks in the $\mathfrak{so}(\texttt{d})_R$ channel $[2n,2(m-n)]$, weighted by the coefficients $\mathcal{C}^{\mathcal{X},\Delta_{12},\Delta_{34}}_{\mathcal{O}_{\Delta,\ell,mn}}$. We similarly partition the reduced correlator $h^{\{k_{i}\}}_1$ into ``reduced blocks" $h^{\{k_{i}\}}_{1;\ell_\XX}$ that generate each supermultiplet $\XX$:
\begin{align}
    h^{\{k_{i}\}}_1(z)=\sum_{\mathcal{X}\in (\OO_1\times \OO_2)\cap(\OO_3\times \OO_4)}\lambda_{k_1 k_2 \mathcal{X}}\lambda_{k_3 k_4 \mathcal{X}}\;\NN_\XX\;h^{\{k_{i}\}}_{1;\ell_\XX'}\left(z\right).
\end{align}
We organized Equations \eqref{eq:AE1h1}-\eqref{eq:AE1h3} in terms of the ingredients 
\begin{align}
    \left\{\frac{z^2h^{\{k_{i}\}}_1(z)-\bar{z}^2h^{\{k_{i}\}}_1(\bar{z})}{z-\bar{z}},\frac{z(z-2)h^{\{k_{i}\}}_1(z)-\bar{z}(\bar{z}-2)h^{\{k_{i}\}}_1(\bar{z})}{2(z-\bar{z})},\frac{(1-z)h^{\{k_{i}\}}_1(z)-(1-\bar{z})h^{\{k_{i}\}}_1(\bar{z})}{z-\bar{z}}\right\},
\label{eq:hrecuringredients}
\end{align}
in anticipation of the fact that for an appropriate function $h^{\{k_{i}\}}_{1;\ell_\XX}$ defined in \eqref{eq:h1at}, each of these expressions generates a linear combination of $2(\varepsilon+1)$-dimensional conformal blocks. The determination of these functions is found in Appendix \ref{hrecurrencerelations} and using these results, we find the conformal blocks contributing to each channel. For example, $\EE=1$ correlators will universally exchange a 1/2-BPS multiplet defined by a superconformal primary $(\Delta,\ell)=(2\varepsilon-\Delta_{34},0)$ in the $[2m,0]$ channel. Using our method with $h^{\{k_i\}}_{1,0}$, we find
\begin{align}
    &\;\;\;\;\;\;A^{\{k_{i}\}}_{m\;m;2\varepsilon-\Delta_{34},0}=U^{\frac{\Delta_{34}}{2}}G^{\Delta_{12},\Delta_{34}}_{2\varepsilon-\Delta_{34},0},
    \label{eq:AE1h1BPS}\\
    &\;\;\;A^{\{k_{i}\}}_{m\;m-1;2\varepsilon-\Delta_{34},0}=\frac{\varepsilon  \left(2 \varepsilon -\Delta _{12}-\Delta _{34}\right)}{2 \left(2 \varepsilon -\Delta _{34}\right) \left(2 \varepsilon +1-\Delta _{34}\right)}U^{\frac{\Delta_{34}}{2}}G^{\Delta_{12},\Delta_{34}}_{2 \varepsilon -\Delta _{34}+1,1}, 
    \label{eq:AE1h2BPS}\\
    &A^{\{k_{i}\}}_{m-1\;m-1;2\varepsilon-\Delta_{34},0}=\nonumber \\&\;\frac{\varepsilon  (\varepsilon +1) \left(2 \varepsilon -\Delta _{12}-\Delta _{34}\right) \left(2 \varepsilon +2-\Delta _{12}-\Delta _{34}\right) \left(\Delta _{12}-\Delta _{34}+2 \varepsilon +2\right)}{8 \left(-\Delta _{34}+2 \varepsilon +3\right) \left(\Delta _{34}-2 \varepsilon -1\right){}^2 \left(\Delta _{34}-2 (\varepsilon +1)\right)^2}U^{\frac{\Delta_{34}}{2}}G^{\Delta_{12},\Delta_{34}}_{2 \varepsilon -\Delta _{34}+2,2}. 
    \label{eq:AE1h3BPS}
    \end{align}
Superconformal blocks are specified up to an overall factor and we will fix our normalization by demanding that the coefficient $\mathcal{C}^{\mathcal{X},\Delta_{12},\Delta_{34}}_{\mathcal{O}_{\Delta_\XX,\ell_\XX,mn}}$ associated with the superconformal primary equals 1. In the above example, we achieved this by multiplying each channel by 
\begin{align}
    \NN_{\XX}=\mathcal{Y}_{\frac{2-k_{34}}{2}}^{k_{12},k_{34}}=\frac{2 \left(2 \varepsilon -\Delta _{34}\right)}{\Delta _{12}-\Delta _{34}+2 \varepsilon}.
\label{eq:AE1BPSnorm}
\end{align}
Correlators with any $\EE\geq1$ will universally exchange this BPS multiplet and its normalized superblock will continue to take the form in Equations \eqref{eq:AE1h1BPS}-\eqref{eq:AE1h3BPS}. An example of this is the stress tensor multiplet exchanged in any $\langle\OO_k\OO_k\OO_k\OO_k\rangle$.

\section{Next-to-next-to-extremal four-point functions}
\label{E2}
We move our attention to the more complicated class of four-point correlators in the configuration $\langle\OO_{k_1}\OO_{k_2}\OO_{k_3}\OO_{k_1+k_2+k_3-4}\rangle$ with extremality $\EE=2$. In the following Subsections we will outline the representation theory of these correlators and specialize the decomposition in \eqref{eq:DGSdecomp} to $\EE=2$. We will redefine the reduced correlators $\TT^{\{k_i\}}_{0,0}$ and $h^{\{k_i\}}_1$ to eliminate $h^{\{k_i\}}_2$ (except when $\varepsilon=1$) and derive the contribution of these functions to each $\frak{so}(\texttt{d})_R$ channel equation. Finally, we will describe how these results can be combined with the recursion relations in Appendices \ref{hrecurrencerelations} and \ref{Trecurrencerelations} to construct superconformal blocks.

\subsection{$\frak{so}(\texttt{d})_R$ channel equations}
\label{E2Rchannels}

The tensor products in \eqref{eq:so5tp} demonstrate that correlators in this configuration will exchange six  R-symmetry irreps. Using the abbreviation $p(\EE=2)=k_1+k_2-4\equiv p$, the exchanged irreps are listed as
\begin{align}
    \left([k_1,0]\otimes [k_2,0]\right)\cap\left([k_3,0]\otimes [k_4,0]\right)=
    &\;[p,0]\oplus[p+2,0]\oplus[p,2]\oplus[p+4,0]\oplus[p+2,2]\oplus[p,4].
    \label{eq:sodtpE2}
\end{align}
These irreps are divided into symmetric and antisymmetric representations:
\begin{align}
    \left([k_1,0]\otimes [k_2,0]\right)\cap\left([k_3,0]\otimes [k_4,0]\right)\Big|_{\text{symm}}=&\;[p,0]\oplus[p+2,0]\oplus[p+4,0]\oplus[p,4],\\
    \left([k_1,0]\otimes [k_2,0]\right)\cap\left([k_3,0]\otimes [k_4,0]\right)\Big|_{\text{antisymm}}=&\;[p,2]\oplus[p+2,2].
\end{align}
Again, this distinction will be important for configurations where one or more of the OPEs involve operators with identical dimension, in which case Bose symmetry restricts the parity of exchanged Lorentz irreps. In the configuration $\langle\OO_{k_1}\OO_{k_2}\OO_{k_3}\OO_{k_1+k_2+k_3-4}\rangle$, this only comes into effect when $k_1=k_2=2$. The relaxation of this selection rule for mixed correlators and the resulting proliferation of superdescendant primary operators contributing to each supermultiplet poses a significant complication in deriving the superblocks in this paper as compared to those previously derived in \cite{Dolan:2004mu,Chester:2014fya}. 

When $\EE=2$, the decomposition in Equation \eqref{eq:DGSdecomp} involves one two-variable reduced correlator $\TT^{\{k_i\}}_{0,0}\equiv\TT^{\{k_i\}}$ and two single-variable reduced correlators $h^{\{k_i\}}_1,h^{\{k_i\}}_{2}$:
\begin{align}
    \GG^{\{k_i\}}\left(U,V;\sigma,\tau\right)=&\;U^{2\varepsilon}\Delta_{\varepsilon}\left[\left(z\alpha-1\right)\left(z\alpha-1\right)\left(z\alpha-1\right)\left(z\alpha-1\right)\frac{\TT^{\{k_i\}}\left(U,V\right)}{U^{3\varepsilon-1}}\right] \nonumber \\
     +&\;U^{\varepsilon}\left(D_\varepsilon\right)^{\varepsilon-1}\left[\frac{(z \alpha-1)(z \bar{\alpha}-1)h^{\{k_{i}\}}_1(z)-(\bar{z} \alpha-1)(\bar{z} \bar{\alpha}-1)h^{\{k_{i}\}}_1(\bar{z})}{z-\bar{z}}\right] \nonumber \\
     +&\;U^{2\varepsilon}\sigma\left(D_\varepsilon\right)^{\varepsilon-1}\left[\frac{(z \alpha-1)(z \bar{\alpha}-1)h^{\{k_{i}\}}_2(z)-(\bar{z} \alpha-1)(\bar{z} \bar{\alpha}-1)h^{\{k_{i}\}}_2(\bar{z})}{z-\bar{z}}\right]
    \label{eq:DGSdecompE2}
\end{align}

We can again determine the way in which the reduced correlator functions $\TT^{\{k_i\}}_{I,J}$ and $h^{\{k_i\}}_I$ contribute to each channel by comparing Equations \eqref{eq:Adecomp} and \eqref{eq:DGSdecomp}. Equation \eqref{eq:DGSdecomp} shows that $\EE=2$ correlators are determined in terms of one two-variable function $\TT^{\{k_i\}}_{0,0}\equiv\TT^{\{k_i\}}$ and two one-variable functions $h_i$. To avoid clutter, we will partition the channel functions into contributions from each type of reduced correlator by writing
\begin{align}
    A_{mn}^{k_{12},k_{34}}\left(U,V\right)=A_{mn}^{k_{12},k_{34};\TT}\left(U,V\right)+A_{mn}^{k_{12},k_{34};h_1}\left(U,V\right)+A_{mn}^{k_{12},k_{34};h_2}\left(U,V\right).
\label{eq:Amnsplit}
\end{align}
The six channels appearing in \eqref{eq:sodtpE2} are encoded in $A_{mn}^{k_{12},k_{34}}$ in terms of   $m=\frac{k_1+k_2}{2}$. For the reader's convenience, we include an implementation of the following results in the \texttt{Mathematica} notebook accompanying our arXiv submission.

\subsubsection{Two-variable contribution}

Suppressing the dependence on $U,V$ and again choosing the abbreviation $\Delta_{12}=\varepsilon\left(k_1-k_2\right)$ and $\Delta_{34}=\varepsilon\left(2\EE-k_1-k_2\right)$ for $\EE=2$, the $\TT^{\{k_i\}}$-dependent contributions to the top three $\mathfrak{so}(\texttt{d})_R$ channel equations take the form
\begin{align}
    A^{\{k_{i}\};\TT}_{m\;m}=&\;\frac{1}{\mathcal{Y}_{m}^{k_{12},k_{34}}}U^{2\varepsilon}\Delta_\varepsilon\left[\frac{\mathcal{T}^{\{k_i\}}}{U^{3(\varepsilon-1)}}\right], 
    \label{eq:AT1}\\
    A^{\{k_{i}\};\TT}_{m\;m-1}=&\;\frac{1}{\mathcal{Y}_{m-1}^{k_{12},k_{34}}}U^{2\varepsilon}\Delta_\varepsilon\left[\left(
    \frac{\Delta _{12}}{\Delta _{34}-4 \varepsilon }
    +\frac{V-1}{U}\right)\frac{\mathcal{T}^{\{k_i\}}}{U^{3(\varepsilon-1)}}\right], 
    \label{eq:AT2}\\
    A^{\{k_{i}\};\TT}_{m\;m-2}=&\;U^{2\varepsilon}\Delta_\varepsilon\left[\left(\frac{\left(\Delta _{12}-\Delta _{34}+2\varepsilon\right) \left(\Delta _{12}+\Delta _{34}-2\varepsilon\right)}{4 \left(\Delta _{34}-2 \varepsilon\right) \left(\Delta_{34}-3 \varepsilon\right)}
    +\frac{\Delta _{12}}{2 \left(\Delta _{34}-2 \varepsilon\right)}\frac{V-1}{U}+\frac{V+1}{2U}
    \right)\frac{\mathcal{T}^{\{k_i\}}}{U^{3(\varepsilon-1)}}\right].
    \label{eq:AT3}
    \end{align}
The remaining three $\mathfrak{so}(\texttt{d})_R$ channel equations are more complicated and depend on $\texttt{d}$. The expressions simplify when we project onto the three physical SCFTs with sixteen supercharges by specifying $\texttt{d}=\frac{2(2\varepsilon+1)}{\varepsilon}$ for $\varepsilon=\frac{1}{2},1,2$. We then have
    \begin{align}
    &\;A^{\{k_{i}\};\TT}_{m-1\;m-1}=\nonumber \\
    &\;\frac{1}{\mathcal{Y}_{m-1}^{k_{12},k_{34}}}U^{2\varepsilon}\Delta_\varepsilon\Bigg[\Bigg(
    \frac{\left(\Delta _{12}^2+4 \varepsilon \right)(2 \varepsilon +1)+\Delta _{34}^2-2 \Delta _{34} (3 \varepsilon +1)}{2 \left(4 \varepsilon +1-\Delta _{34}\right) \left(4 \varepsilon +2-\Delta _{34}\right) (\varepsilon +1)}
     \nonumber \\
    &\;\hspace{4.5cm}
    +\frac{\Delta _{12} (2 \varepsilon +1)}{\left(\Delta _{34}-4 \varepsilon -2\right)(\varepsilon+1)}\frac{V-1}{U}
    +\frac{-2}{\varepsilon +1}\frac{V+1}{2U} +\frac{\left(V-1\right)^2}{U^2}\Bigg)\frac{\mathcal{T}^{\{k_i\}}}{U^{3(\varepsilon-1)}}\Bigg],
    \label{eq:AT4}\\
    &\;A^{\{k_{i}\};\TT}_{m-1\;m-2}=\nonumber\\
  &\;U^{2\varepsilon}\Delta_\varepsilon\Bigg[\Bigg(
    \frac{\Delta _{12} \left(\left(\Delta _{34}-2 \varepsilon \right)^2-\Delta _{12}^2\right)}{4 \left(2 \varepsilon -\Delta _{34}\right) \left(3 \varepsilon +1-\Delta _{34}\right) \left(4 \varepsilon +2-\Delta _{34}\right)}
    \nonumber \\
    &\;\hspace{1.5cm}+\left(\frac{\Delta _{12}^2 \left(10 \varepsilon +2-3 \Delta _{34}\right)}{4 \left(2 \varepsilon -\Delta _{34}\right) \left(3 \varepsilon +1-\Delta _{34}\right) \left(4 \varepsilon +2-\Delta _{34}\right)}+\frac{\Delta _{34}-2 \varepsilon -2}{4 \left(3 \varepsilon +1-\Delta _{34}\right)}\right)\frac{V-1}{U}\nonumber \\
    &\;\hspace{2.5cm}+\frac{\Delta _{12} \left(2 \varepsilon -2-\Delta _{34}\right)}{\left(\Delta _{34} -2\varepsilon\right) \left(4 \varepsilon +2-\Delta _{34}\right)}\frac{V+1}{2U}
    +\frac{\Delta _{12}}{2 \left(\Delta _{34}-2 \varepsilon\right)}\frac{\left(V-1\right)^2}{U^2}+\frac{V^2-1}{2U^2}\Bigg)\frac{\mathcal{T}^{\{k_i\}}}{U^{3(\varepsilon-1)}}\Bigg],
    \label{eq:AT5}
\end{align}
\newpage
\begin{align}
    &\;A^{\{k_{i}\};\TT}_{m-2\;m-2}= \nonumber \\
    &\;U^{2\varepsilon}\Delta_\varepsilon\Bigg[\Bigg(
    \frac{\left(\Delta _{12}+\Delta _{34}-2 \varepsilon \right) \left(\Delta _{12}+\Delta _{34}-2 (\varepsilon +1)\right) \left(\Delta _{12}-\Delta _{34}+2 \varepsilon \right) \left(\Delta _{12}-\Delta _{34}+2(\varepsilon +1)\right)}{16 \left(2 \varepsilon +1-\Delta _{34}\right) \left(2 \varepsilon +2-\Delta _{34}\right) \left(3 \varepsilon +1-\Delta _{34}\right) \left(3 \varepsilon +2-\Delta_{34}\right)}\nonumber \\
    &\;\hspace{1.5cm}
    +\frac{-\Delta _{12} \left(\Delta _{12}-\Delta _{34}+2 \varepsilon +2\right) \left(\Delta _{12}+\Delta _{34}-2 (\varepsilon +1)\right)}{4 \left(2 \varepsilon +1-\Delta _{34}\right) \left(2 \varepsilon +2-\Delta _{34}\right) \left(3 \varepsilon +2-\Delta _{34}\right)}\frac{V-1}{U}\nonumber \\
    &\;\hspace{1.5cm}
    +\frac{\left(2 \varepsilon -\Delta _{34}\right) \left(\Delta _{12}+\Delta _{34}-2 (\varepsilon +1)\right) \left(\Delta _{12}-\Delta _{34}+2(\varepsilon +1)\right)}{2 \left(2 \varepsilon +1-\Delta_{34}\right) \left(2 \varepsilon +2-\Delta_{34}\right) \left(3 \varepsilon +2-\Delta_{34}\right)}\frac{V+1}{2U}
     \nonumber \\
    &\;\hspace{1.5cm}
    +\frac{\Delta _{12}^2+\Delta _{34}-2 (\varepsilon +1)}{4 \left(2 \varepsilon +1-\Delta _{34}\right) \left(2 \varepsilon +2-\Delta _{34}\right)}\frac{\left(V-1\right)^2}{U^2}
    \nonumber \\
    &\;\hspace{8.25cm}
    +\frac{\Delta _{12}}{\Delta _{34}-2 (\varepsilon +1)}\frac{V^2-1}{2U^2}+\frac{\left(V+1\right)^2}{4U^2}
    \Bigg)\frac{\mathcal{T}^{\{k_i\}}}{U^{3(\varepsilon-1)}}\Bigg]. 
    \label{eq:AT6}
\end{align}
We organized each $A^{\{k_{i}\};\TT}_{mn}$ into the operator $U^{2\varepsilon}\Delta_{\varepsilon}$ acting on $U^{-3(\varepsilon-1)}\TT^{\{k_i\}}_{\Delta,\ell}$ multiplied by one of the following rational functions 
\begin{align}
    \left\{1,\frac{V-1}{U},\frac{V+1}{2U},\frac{(V-1)^2}{U^2},\frac{(V+1)^2}{4U^2},\frac{V^2-1}{2U^2}\right\},
\label{eq:recuringredients}
\end{align}
to facilitate the construction of superconformal blocks as described in Subsection \ref{superblockhowto}. Appendix \ref{Trecurrencerelations} will demonstrate that each ingredient in \eqref{eq:recuringredients}, combined with the action of $\Delta_\varepsilon$ and its inverse, will generate a linear combination of conformal blocks that will ultimately contribute to the superconformal descendant blocks in each channel. In both the coefficients in these equations and the coefficients in front of the blocks that the ingredients \eqref{eq:recuringredients} give rise to, an overall factor of $\Delta_{12}$ will switch off odd-parity blocks which are forbidden in operator product expansions of operators with equal dimension. In the configuration $\langle\OO_{k_1}\OO_{k_2}\OO_{k_3}\OO_{k_1+k_2+k_3-4}\rangle$, the only instance where $\Delta_{34}=0$ is when $k_1=k_2=2$ such that we also have $\Delta_{12}=0$. Our configuration means that $\Delta_{34}\leq0$ such that denominator factors involving $\Delta_{34}$ are never singular. Finally, we note that these channel equations also determine the contributions of $\TT_{0,0}$ to all $\EE>2$ four-point correlators.

\subsubsection{Single-variable contribution}
As alluded to in Section \ref{DGSdecomp}, the non-trivial kernel of $\Delta_\varepsilon$ will allow us to simplify the decomposition \eqref{eq:DGSdecompE2} through redefinitions of other reduced correlators $\TT^{\{k_i\}}$ and $h^{\{k_i\}}_i$.  The first redefinition we will make is 
\begin{align}
    \frac{\TT^{\{k_i\}}}{U^{3(\varepsilon-1)}}\rightarrow\frac{\TT^{\{k_i\}}}{U^{3(\varepsilon-1)}}-\left(\Delta_\varepsilon\right)^{-1}\left(D_{\varepsilon}\right)^{\varepsilon-1}\left[\frac{z^2h^{\{k_i\}}_2(z)-\bar{z}^2h^{\{k_i\}}_2(\bar{z})}{z-\bar{z}}\right].
\label{eq:Tredef}
\end{align}
The effect of this is to remove the $h^{\{k_i\}}_2$-dependent contribution in the top R-symmetry channel function $A^{\{k_{i}\}}_{mm;\Delta,\ell}$ and to modify the dependence on $h^{\{k_i\}}_2$ in other channels. While \eqref{eq:Tredef} eliminates $h_2^{\{k_i\}}$ entirely in 3d, a redefinition of $h_1^{\{k_i\}}$ is needed to eliminate it in 6d. The technicalities of these $h_1^{\{k_i\}}$ redefinitions were described in Sections 6.2 and 7.3 of \cite{Dolan:2004mu} and we will briefly review their outcome case-by-case in Section \ref{reducedblocks}. 

The complete elimination of $h_2^{\{k_i\}}$is not possible in 4d. The remaining single-variable contributions are essential for canceling unphysical blocks in $A^{\{k_{i}\};\TT}_{mn}$, which is consistent with the widely-used organization of $\GG^{\{k_i\}}$ in \cite{Nirschl:2004pa}. Since $h_2^{\{k_i\}}$ can be removed in 3d and 6d and to avoid a lengthy summary of \cite{Nirschl:2004pa}, we will only consider the channel function $A^{\{k_{i}\};h_1}_{m\;n}$\footnote{The authors of
\cite{Dolan:2004mu} claim that the single-variable function $h^{\{k_i\}}_1$ can be completely removed in 3d, however we will choose to keep it explicit}. The $\mathfrak{so}(\texttt{d})_R$ channel equations take the form
\begin{align}
    &A^{\{k_{i}\};h_1}_{m\;m}=A^{\{k_{i}\};h}_{m\;m-1}=A^{\{k_{i}\};h}_{m\;m-2}=0,
    \label{eq:AE2h1to3}\\
    &A^{\{k_{i}\};h_1}_{m-1\;m-1}=\frac{1}{\mathcal{Y}_{m-1}^{k_{12},k_{34}}}U^{\varepsilon}\left(D_\varepsilon\right)^{\varepsilon-1}\left[\frac{z^2h^{\{k_{i}\}}_1(z)-\bar{z}^2h^{\{k_{i}\}}_1(\bar{z})}{z-\bar{z}}\right], 
    \label{eq:AE2h4}\\
    &A^{\{k_{i}\};h_1}_{m-1\;m-2}=U^{\varepsilon}\left(D_\varepsilon\right)^{\varepsilon-1}\Bigg[\frac{\Delta _{12}}{2 \left(\Delta _{34}-2 \varepsilon \right)}\frac{z^2h^{\{k_{i}\}}_1(z)-\bar{z}^2h^{\{k_{i}\}}_1(\bar{z})}{z-\bar{z}}\nonumber \\
    &\;\hspace{8.5cm}
    +\frac{z(z-2)h^{\{k_{i}\}}_1(z)-\bar{z}(\bar{z}-2)h^{\{k_{i}\}}_1(\bar{z})}{2(z-\bar{z})}\Bigg], 
    \label{eq:AE2h5} \\
    &A^{\{k_{i}\};h_1}_{m-2\;m-2}=U^{\varepsilon}\left(D_\varepsilon\right)^{\varepsilon-1}\Bigg[\frac{1}{4} \left(\frac{\Delta _{12}^2}{\Delta _{34}-2\varepsilon -2}-\frac{\Delta _{12}^2-1}{\Delta _{34}-2 \varepsilon -1}+1\right)\frac{z^2h^{\{k_{i}\}}_1(z)-\bar{z}^2h^{\{k_{i}\}}_1(\bar{z})}{z-\bar{z}} \nonumber \\
    &\;\hspace{6cm}+\frac{\Delta _{12}}{\Delta _{34}-2 (\varepsilon +1)}\frac{z(z-2)h^{\{k_{i}\}}_1(z)-\bar{z}(\bar{z}-2)h^{\{k_{i}\}}_1(\bar{z})}{2(z-\bar{z})}\nonumber \\
    &\;\hspace{9cm}+\frac{(1-z)h^{\{k_{i}\}}_1(z)-(1-\bar{z})h^{\{k_{i}\}}_1(\bar{z})}{z-\bar{z}}\Bigg]. 
    \label{eq:AE2h6}
    \end{align}
We point out that these appearances of $h_{1}^{\{k_i\}}$ are functionally the same as in the $\EE=1$ case in Equations \eqref{eq:AE1h1}-\eqref{eq:AE1h3}, noting that $m(\EE=2)=m(\EE=1)+1$. This is an example of how reduced correlator functions of a given type appear in the same way as we increase $\EE$.

The vanishing of the first three channels is indicative of the fact that, before absorption of $h^{\{k_i\}}_2$ (which contributed to all six channels), $h^{\{k_i\}}_1$ only encodes the three lowest dimension protected operators appearing in each superselection rule in \eqref{eq:3dsuperselect}-\eqref{eq:3dsuperselect}. Upon absorption of $h^{\{k_i\}}_2$, the redefined $h^{\{k_i\}}_1$ continues to describe twist $2\varepsilon-\Delta_{34}$ operators while the remaining operators previously captured by $h^{\{k_i\}}_2$ are packaged in $\TT^{\{k_i\}}$.

\subsection{Assembling superconformal blocks}
\label{superblockhowto}
With these ingredients, we can construct the superconformal blocks $\mathfrak{G}_\mathcal{X}^{\Delta_{12},\Delta_{34}}$ in Equation \eqref{eq:superblock} as follows. First, we decompose each channel function $A_{mn}^{k_{12},k_{34}}$ into a sum over contributions from each exchanged supermultiplet $\XX$ by writing 
\begin{align}
    A_{mn}^{k_{12},k_{34}}\left(U,V\right)=\sum_{\mathcal{X}\in (\OO_1\times \OO_2)\cap(\OO_3\times \OO_4)}A_{mn;\Delta_\XX',\ell_\XX'}^{k_{12},k_{34}}\left(U,V\right).
\end{align}
The labels $(\Delta_\XX',\ell_\XX')$ refer to quantum numbers in the highest R-symmetry channel of $\XX$, which will useful for our recursive generation of superblocks $\mathfrak{G}_\mathcal{X}^{\Delta_{12},\Delta_{34}}$. The separation in \eqref{eq:Amnsplit} and the channel equations \eqref{eq:AT1}-\eqref{eq:AE2h6} hold for each $A_{mn;\Delta_\XX,\ell_\XX}^{k_{12},k_{34}}$ individually and yield the sum of superconformal descendant primary blocks in the $\mathfrak{so}(\texttt{d})_R$ channel $[2n,2(m-n)]$, weighted by the coefficients $\mathcal{C}^{\mathcal{X},\Delta_{12},\Delta_{34}}_{\mathcal{O}_{\Delta,\ell,mn}}$.

The appearance of the reduced correlator functions $\TT^{\{k_i\}}$ and $h_i$ in each channel equation is suggestive of the fact that the $\mathfrak{so}(\texttt{d})_R$ channel equations are not independent and instead, interrelated by supersymmetry\footnote{Indeed, separately applying the bootstrap algorithm to the crossing equations of each channel function leads to numerical instabilities, as discussed for 3d $\NN=8$ theories in \cite{Chester:2014fya}.}. We note that for every supermultiplet $\XX$ in the superselection rules \eqref{eq:3dsuperselect}, \eqref{eq:4dsuperselect}, and \eqref{eq:6dsuperselect}, the $\mathfrak{so}(\texttt{d})_R$ channel with the highest value of $d_i$ (using the $[d_1,d_2]$ parameterization from the previous Subsection) is $[2m,0]$ and involves no more than one conformal block. Consider a supermultiplet $\XX$ with superconformal primary $\left(\Delta,\ell\right)$ and take the case where there is a block in this channel with quantum numbers $\left(\Delta',\ell'\right)$. 
\begin{align}
    A_{m\;m;\Delta',\ell'}^{k_{12},k_{34}}\left(z,\bar{z}\right)=&\;\frac{1}{\mathcal{Y}_{m}^{k_{12},k_{34}}}U^{2\varepsilon}\Delta_\varepsilon\left[\frac{\mathcal{T}^{\{k_i\}}_{\Delta',\ell'}\left(z,\bar{z}\right)}{U^{3(\varepsilon-1)}}\right]= U^{\frac{\Delta_{34}}{2}}\;G_{\Delta',\ell'}^{\Delta_{12},\Delta_{34}}\left(z,\bar{z}\right),
    \label{eq:topAT}
    \\
    \rightarrow \frac{\mathcal{T}^{\{k_i\}}_{\Delta',\ell'}\left(z,\bar{z}\right)}{U^{3(\varepsilon-1)}}=&\; \mathcal{Y}_{m}^{k_{12},k_{34}}\;\left(\Delta_{\varepsilon}\right)^{-1}\left[U^{\frac{\Delta_{34}}{2}-2\varepsilon}\;G_{\Delta',\ell'}^{\Delta_{12},\Delta_{34}}\left(z,\bar{z}\right)\right].
    \label{eq:reducedblockinversion}
\end{align}
In Section \ref{reducedblocks}, we will perform this inversion explicitly such to obtain a ``reduced block" $\TT^{\{k_i\}}_{\Delta,\ell}$. To derive full superblocks however, it will sufficient to substitute \eqref{eq:reducedblockinversion} into the channel equations and use recursion relations in Appendix \ref{Trecurrencerelations} to generate the full spectrum of superdescendant blocks in each channel. The coefficients $\mathcal{C}^{\mathcal{X},\Delta_{12},\Delta_{34}}_{\mathcal{O}_{\Delta,\ell,mn}}$ are read off as the overall factor multiplying each conformal block $G_{\Delta,\ell}^{\Delta_{12},\Delta_{34}}$. 

As an example, we will construct the $[2m-2,2]$ channel equation for a long superblock exchanged in $\langle \OO_{k_1}\OO_{k_2}\OO_{k_3}\OO_{k_1+k_2+k_3-4}\rangle$ in general dimensions. Long superblocks do not involve receive contributions from the single-variable functions $h^{\{k_i\}}_I$ so we will only need the channel expressions $A^{\{k_{i}\};\TT}_{mn;\Delta,\ell}$. Starting with the top R-symmetry channel equation in \eqref{eq:topAT} where the unique block in this channel has $(\Delta',\ell')=(\Delta+4,\ell)$, we can use Equation \eqref{eq:AT2} and the recurrence relation for $\frac{V-1}{U}\frac{\mathcal{T}^{\{k_i\}}}{U^{3(\varepsilon-1)}}$ in Equation \eqref{eq:Vm1rec} to find
\begin{align}
    &A^{\{k_{i}\}}_{m\;m-1;\Delta',\ell'}\left(z,\bar{z}\right)=\nonumber U^{\frac{\Delta_{34}}{2}}\Bigg(
    \\
    &\frac{(\Delta +3) (\Delta -2 \varepsilon +4)\left(\Delta _{12}-\Delta _{34}+2 \varepsilon \right)}{32 (\Delta -\varepsilon +3)}
    \nonumber\\
    &\times\frac{\left(\Delta -\Delta _{12}+\ell +4\right)\left(\Delta +\Delta _{12}+\ell +4\right) \left(\Delta +\Delta _{34}+\ell +4\right) \left(\Delta -\Delta _{34}+2 \varepsilon +\ell +2\right)}{ (\Delta -\varepsilon +4) \left(2 \varepsilon -\Delta _{34}\right) (\Delta +\ell +3) (\Delta +\ell +4)^2 (\Delta +\ell +5)}G^{\Delta_{12},\Delta_{34}}_{\Delta +5,\ell +1}\nonumber \\
    +&\frac{\left(\Delta _{12}-\Delta _{34}+2 \varepsilon \right) \left(-\Delta -\Delta _{34}+4 \varepsilon +\ell -4\right)}{2 \left(2 \varepsilon -\Delta _{34}\right) \left(-\Delta -\Delta _{34}+2 \varepsilon +\ell -2\right)}G^{\Delta_{12},\Delta_{34}}_{\Delta +3,\ell +1}\nonumber \\
    +&\frac{\Delta _{12} \left(\Delta _{12}-\Delta _{34}+2 \varepsilon \right)}{2 \left(2 \varepsilon -\Delta _{34}\right) \left(4 \varepsilon -\Delta _{34}\right)}\nonumber \\
    &\times\Bigg(\frac{\Delta _{34} \left(\Delta _{34}-6 \varepsilon +2\right) \left((\Delta +2) (\Delta -2 \varepsilon +4)+\ell ^2+2 \varepsilon  \ell \right)}{(\Delta +\ell +2) (\Delta +\ell +4) (-\Delta +2 \varepsilon +\ell -4) (-\Delta +2 \varepsilon +\ell -2)}
    \nonumber \\
    &\;\;\;\;\;\;-\frac{(\Delta -\ell+2) (\Delta -2 \varepsilon +\ell +4) (\Delta +2 \varepsilon +\ell +2) (\Delta -4 \varepsilon -\ell +4)}{(\Delta +\ell +2) (\Delta +\ell +4) (-\Delta +2 \varepsilon +\ell -4) (-\Delta +2 \varepsilon +\ell -2)}\Bigg)G^{\Delta_{12},\Delta_{34}}_{\Delta +4,\ell }\nonumber \\
    +&\;\frac{\ell  \left(\Delta _{12}-\Delta _{34}+2 \varepsilon \right) (2 \varepsilon +\ell -1) \left(\Delta +\Delta _{34}-2 \varepsilon +\ell +4\right)}{2 \left(2 \varepsilon -\Delta _{34}\right) \left(\Delta +\Delta _{34}+\ell +2\right) (\varepsilon +\ell -1) (\varepsilon +\ell )} G^{\Delta_{12},\Delta_{34}}_{\Delta +3,\ell -1}
    \nonumber \\
    +&\frac{(\Delta +3) \ell (2 \varepsilon +\ell -1) (\Delta -2 \varepsilon +4)\left(\Delta _{12}-\Delta _{34}+2 \varepsilon \right)\left(\Delta -2 \varepsilon -\ell +4+\Delta _{12}\right)}{32 \left(2 \varepsilon -\Delta _{34}\right)(\Delta -\varepsilon +3) (\Delta -\varepsilon +4) }
    \nonumber \\
    &\times\frac{\left(\Delta -2 \varepsilon -\ell +4-\Delta _{12}\right)\left(\Delta -\ell+2-\Delta _{34}\right) \left(\Delta-2 \varepsilon -\ell +4 +\Delta _{34}\right)}{ (\varepsilon +\ell -1) (\varepsilon +\ell ) (\Delta -2 \varepsilon -\ell +3) (\Delta -2 \varepsilon -\ell +4)^2 (\Delta -2 \varepsilon -\ell +5)}G^{\Delta_{12},\Delta_{34}}_{\Delta +5,\ell -1}\Bigg).
\end{align}
This second-highest channel expression already takes a complicated form and we will relegate the remaining channels to the ancilliary \texttt{Mathematica} file, which contains channel calculations for $\langle\OO_2\OO_2\OO_k\OO_k\rangle$ and $\langle\OO_2\OO_3\OO_k\OO_{k+1}\rangle$ can be easily modified to other $\EE=2$ configurations. One noteworthy feature is the overall factor of $\Delta_{12}$ in front of the $(\Delta+4,\ell)$ block, which is responsible for the absence of this operator in correlators $\langle\OO_k\OO_k\OO_l\OO_l\rangle$ with the stricter Bose selection rules discussed earlier in this Section.

As described in Section \ref{E1}, superconformal blocks are defined only up to a multiplicative constant and that as a convention, we choose the constant such that the superconformal primary block coefficient is equal to one. While we did not include this factor above for brevity, to implement this convention for long superblocks we would multiply all $A^{\{k_{i}\}}_{mn;\Delta',\ell'}$ with the factor 
\begin{align}
    &\NN_{\text{long}}=\nonumber\\
    &\frac{\left(\Delta+\ell +\Delta _{34}\right) \left(\Delta+\ell +2+\Delta _{34}\right) \left(\Delta-\ell-2 \varepsilon +2+\Delta _{34}\right) \left(\Delta-\ell-2 \varepsilon  +\Delta _{34}\right)}
    {\left(\Delta-\ell -4 \varepsilon  +4+\Delta _{34}\right) \left(\Delta -4 \varepsilon -\ell +2+\Delta _{34}\right) \left(\Delta-2 \varepsilon +\ell +2+\Delta _{34}\right) \left(\Delta -2 \varepsilon +\ell +4+\Delta _{34}\right)}.
\end{align}

Care needs to be taken when applying this procedure to the protected multiplets in each SCFT. Naively carrying out our procedure using the quantum numbers appearing in Equations \eqref{eq:3dsuperselect},\eqref{eq:4dsuperselect},\eqref{eq:6dsuperselect} can lead to divergent coefficients $\mathcal{C}^{\mathcal{X},\Delta_{12},\Delta_{34}}_{\mathcal{O}_{\Delta,l,mn}}$, operators with unphysical twists, and non-unitary operators with negative block coefficients. Moreover, our ability to absorb single-variable contributions $h^{\{k_i\}}_I$ via redefinitions of $\TT^{\{k_i\}}$ varies with $\varepsilon$ and in the following Subsections, we will need to specify these contributions case-by-case. 

All of these issues are canceled by $\varepsilon$-dependent properties of conformal blocks and the correct organization of the single-variable functions $h^{\{k_i\}}_I$. For brevity, we will only briefly describe the subtleties and single-variable contributions in each case and instead, explicitly implement these procedures for all superblocks exchanged in four-point functions of the form $\langle\OO_2\OO_2\OO_k\OO_{k}\rangle$ and $\langle\OO_2\OO_3\OO_k\OO_{k+1}\rangle$ in the ancilliary \texttt{Mathematica} notebook. In each case we will omit the twist $2\varepsilon-\Delta_{34}$ semi-short multiplets which lie at the unitarity bound of a long multiplet and are understood to contain conserved higher-spin currents which are only present in non-interacting theories \cite{Maldacena:2011jn}.

Correlators with $\EE>1$ will generically exchange protected low-twist supermultiplets $\XX$ which do not have a block in the $[2m,0]$ channel, such that the RHS of Equation \eqref{eq:topAT} is zero. Using properties of the Jack polynomials described in Appendix \ref{jackpolys}, one finds that the space of solutions to the equation
\begin{align}
    \Delta_\varepsilon\left[\frac{\mathcal{T}^{\{k_i\}}_{\Delta,\ell}\left(z,\bar{z}\right)}{U^{3(\varepsilon-1)}}\right]=0,
\end{align}
is spanned by restricted expansions of Jack polynomials which can only realize twist $-\Delta_{34}$ and $2\varepsilon-\Delta_{34}$. Since such operators never appear in the second and third highest channels $[2m-2,2],$ and $[2m-4,4]$, we conclude that $\mathcal{T}^{\{k_i\}}_{\Delta,\ell}$ does not encode the supermultiplet $\XX$. Instead, such multiplets are captured entirely by $h^{\{k_i\}}_1$, in direct analogy to the superblocks appearing in $\EE=1$ correlators.

\subsubsection{6d $\NN=(2,0)$}

For 6d $\NN=(2,0)$, we will follow the example of \cite{Beem:2015aoa} by absorbing $h^{\{k_{i}\}}_2$ into $\TT^{\{k_{i}\}}$ and $h^{\{k_{i}\}}_1$. We summarize the contributions $A^{\{k_{i}\};f}_{mn;\Delta,\ell}$ that contribute to each $\mathfrak{G}_\mathcal{X}^{\Delta_{12},\Delta_{34}}$ in Table \ref{6dshortAtable}, again in terms of the parameter $p(\EE=2)=k_1+k_2-4\equiv p$. As a check, we independently derived the 6d $\NN=(2,0)$ superblocks for all multiplets in $\langle\OO_{2}\OO_{2}\OO_{k}\OO_{k}\rangle$ and $\langle\OO_{2}\OO_{3}\OO_{k}\OO_{k+1}\rangle$ using the small $z,\bar{z}$ expansion strategy introduced in \cite{Chester:2014fya} and found perfect agreement with the blocks derived from the reduced correlator organization described in this section.
\begin{table}[H]
\begin{tabular}{|l||l|l|l|l|}
\hline $\mathfrak{osp}(8^*|4)$ multiplet $\XX$& $A^{\{k_{i}\};\TT}_{mn;\;\Delta',\ell'}$ & $A^{\{k_{i}\};h_1}_{mn;\;\ell'}$  \\
\hline \hline 
$\mathcal{L}[p,0]_{\Delta\geq 2p+6+\ell,\ell}$ & $A^{\{k_{i}\};\TT}_{mn;\;\Delta+4,\ell}$ & --- \\
\hline $\mathcal{B}[p,2]_{2p+8+\ell,\ell}$ & $A^{\{k_{i}\};\TT}_{mn;\;2p+11+\ell,\ell+1}$ & --- \\
\hline $\mathcal{B}[p+2,0]_{2p+8+\ell,\ell}$ & $A^{\{k_{i}\};\TT}_{mn;\;2p+10+\ell,\ell+2}$ & $A^{\{k_{i}\};h_1}_{mn;\;2p+8+\ell,\ell+4}$  \\
\hline $\mathcal{D}[p,4]_{2p+8,0}$ & $A^{\{k_{i}\};\TT}_{mn;\;2p+10,0}$ & --- \\
\hline $\mathcal{D}[p+2,2]_{2p+8,0}$ & $A^{\{k_{i}\};\TT}_{mn;\;2p+9,1}$ & $A^{\{k_{i}\};h_1}_{mn;\;2p+7,3}$  \\
\hline $\mathcal{D}[p+4,0]_{2p+8,0}$ & $A^{\{k_{i}\};\TT}_{mn;\;2p+8,0}$ & $A^{\{k_{i}\};h_1}_{mn;\;2p+6,2}$  \\
\hline $\mathcal{D}[p,2]_{2p+4,0}$ & --- & $A^{\{k_{i}\};h_1}_{mn;\;2p+5,1}$ \\
\hline $\mathcal{D}[p+2,0]_{2p+4,0}$ & --- & $A^{\{k_{i}\};h_1}_{mn;\;2p+4,0}$  \\
\hline
$\mathcal{D}[p,0]_{2p,0}$ & --- & $\delta_{m,p}\delta_{n,p}$ \\
\hline
\end{tabular}
\caption{The contribution of different kinds of channel functions $A^{\{k_{i}\}}_{mn;\Delta',\ell'}$ to superblocks $\mathfrak{G}_\mathcal{X}^{\Delta_{12},\Delta_{34}}$ in 6d $\NN=(2,0)$ correlators $\langle\OO_{k_1}\OO_{k_2}\OO_{k_3}\OO_{k_1+k_2+k_3-4}\rangle$.}
\label{6dshortAtable}
\end{table}

\subsubsection{4d $\NN=4$}
The $\varepsilon=1$ case is the most involved, in that plugging in the semi-short quantum numbers appearing in Equation \eqref{eq:4dsuperselect} into $A^{\{k_{i}\};\TT}_{mn;\Delta,\ell}$ will give rise to various spurious superdescendants with negative block coefficients and non-unitary values of the twist $t=\Delta-\ell$. Neither of the single-variable functions $h^{\{k_i\}}_I$ can be completely absorbed into $\TT^{\{k_i\}}$ through redefinitions like \eqref{eq:Tredef} and their contributions must be intricately tuned to cancel unphysical contributions from $A^{\{k_{i}\};\TT}_{mn}$. This case is also the most well-documented, starting with \cite{Dolan:2001tt,Dolan:2004mu,Nirschl:2004pa} and implemented in several bootstrap studies, e.g. in \cite{Beem:2013qxa,Alday:2014qfa,Beem:2016wfs,Bissi:2020jve}. For this reason and for brevity, we will not translate the formalism of \cite{Nirschl:2004pa} into our generalization of that in \cite{Dolan:2004mu} and merely state that both languages are equivalent\footnote{We refer the reader to Appendix 7.3 of \cite{Dolan:2004mu} which demonstrates this more explicitly for $\langle\OO_{2}\OO_{2}\OO_{2}\OO_{2}\rangle$.}. We will briefly outline their analogue of the reduced correlators $h^{\{k_i\}}_I$ in Subsection \ref{4dreducedblocks}. For the reader's convenience, we also include an implementation of our formulae for an arbitrary long block in $\langle\OO_{k_1}\OO_{k_2}\OO_{k_3}\OO_{k_1+k_2+k_3-4}\rangle$ in the ancilliary \texttt{Mathematica} file. This will be useful for accessing superconformal descendant data which is harder to read off in reduced block decompositions.

\subsubsection{3d $\NN=8$}
\label{3dsuperblocks}
When $\varepsilon=\frac{1}{2}$, the redefinition of \eqref{eq:Tredef} completely eliminates the single-variable function $h^{\{k_i\}}_2$, as described in Section 6.2.1 of \cite{Dolan:2004mu}. This transfers the contributions of certain protected multiplets to $\TT^{\{k_i\}}$. With these ingredients, we summarize the contributions $A^{\{k_{i}\};f}_{mn;\Delta,\ell}$ that contribute to each $\mathfrak{G}_\mathcal{X}^{\Delta_{12},\Delta_{34}}$ in Table \ref{3dAtable} and implement this in the ancillary \texttt{Mathematica} file. Upon converting between conventions, we found that all of our results exactly matched those used in the multi-correlator bootstrap study in \cite{Agmon:2019imm}, which the authors derived using the method in \cite{Chester:2014fya}\footnote{We thank Shai Chester for sharing a \texttt{Mathematica} notebook with these results.}.

\begin{table}[h]
\begin{tabular}{|l||l|l|l|l|}
\hline $\mathfrak{osp}(8|4)$ multiplet multiplet $\XX$& $A^{\{k_{i}\};\TT}_{mn;\;\Delta',\ell'}$ & $A^{\{k_{i}\};h_1}_{mn;\;\ell'}$  \\
\hline \hline 
$\left(A,0\right)^{[0,0,p,0]}_{\Delta\geq p/2+1+\ell,\ell}$ & $A^{\{k_{i}\};\TT}_{mn;\;\Delta+4,\ell}$& ---   \\
\hline $\left(A,2\right)^{[0,1,p,0]}_{p/2+2+\ell,\ell}$ & $A^{\{k_{i}\};\TT}_{mn;\;p/2+5+\ell,\ell+1}$& ---  \\
\hline
$\left(A,+\right)^{[0,0,p+2,0]}_{p/2+2+\ell,\ell}$ & $A^{\{k_{i}\};\TT}_{mn;\;p/2+4+\ell,\ell+2}$& ---  \\
\hline $\left(B,2\right)^{[0,2,p,0]}_{p/2+2,0}$ & $A^{\{k_{i}\};\TT}_{mn;\;p/2+4,0}$ & ---  \\
\hline $\left(B,2\right)^{[0,1,p+2,0]}_{p/2+2,0}$ & $A^{\{k_{i}\};\TT}_{mn;\;p/2+3,1}$ & ---   \\
\hline $\left(B,+\right)^{[0,0,p+4,0]}_{p/2+2,0}$ & $A^{\{k_{i}\};\TT}_{mn;\;p/2+2,0}$ & ---  \\
\hline $\left(B,2\right)^{[0,1,p-2,0]}_{p/2+1,0}$ & --- & $A^{\{k_{i}\};h_1}_{mn;\;p/2+2,1}$  \\
\hline $\left(B,+\right)^{[0,0,p+2,0]}_{p/2+1,0}$ & --- & $A^{\{k_{i}\};h_1}_{mn;\;p/2+1,0}$ \\
\hline 
$\left(B,+\right)^{[0,0,p,0]}_{p/2,0}$ & --- & $\delta_{m,p}\delta_{n,p}$  \\
\hline
\end{tabular}
\caption{The contribution of different kinds of channel functions $A^{\{k_{i}\}}_{mn;\Delta',\ell'}$ to superblocks $\mathfrak{G}_\mathcal{X}^{\Delta_{12},\Delta_{34}}$ in 3d $\NN=8$ correlators $\langle\OO_{k_1}\OO_{k_2}\OO_{k_3}\OO_{k_1+k_2+k_3-4}\rangle$.}
\label{3dAtable}
\end{table}

\section{Reduced superblocks}
\label{reducedblocks}
In this section, we demonstrate that for each $\varepsilon$, one can explicit evaluate \eqref{eq:reducedblockinversion} to express the reduced block $\mathcal{T}^{\{k_i\}}_{\Delta,\ell}$ in terms of an ordinary $2\left(\varepsilon+1\right)$-dimensional bosonic block. If one can correctly assign $\mathcal{T}^{\{k_i\}}_{\Delta,\ell}$ and the single-variable reduced blocks $h^{\{k_i\}}_{1,\ell}$ defined in Equation \eqref{eq:h1at} with the multiplets appearing in $\langle \OO_{k_1}\OO_{k_2}\OO_{k_3}\OO_{k_1+k_2+k_3-4}\rangle$, one can potentially study four-point correlators $\langle \OO_{k_1}\OO_{k_2}\OO_{k_3}\OO_{k_1+k_2+k_3-4}\rangle$ in terms of a substantially simpler block expansion. After deriving the reduced block $\mathcal{T}^{\{k_i\}}_{\Delta,\ell}$, we will comment on $\varepsilon$-dependent subtleties of reduced blocks in each maximally supersymmetric CFT.

\subsection{Two-variable reduced blocks}
To derive a block expansion of the two-variable reduced correlator $\TT^{\{k_i\}}$, we will return to Equation \eqref{eq:reducedblockinversion} which arose from the universal highest R-symmetry channel $[2m,0]$ containing no more than one conformal block. We eschew difficulties interpreting the operator $\Delta_\varepsilon$ by expanding the conformal block $G_{\Delta,\ell}^{\Delta_{12},\Delta_{34}}$ in Jack polynomials $P^{(\varepsilon)}_{a,b}$, which are eigenfunctions of $\Delta_\varepsilon$. In particular, after absorbing the factor of $\mathcal{Y}_{m}^{k_{12},k_{34}}$ for later convenience, we have that 
\begin{align}
    \mathcal{T}^{\{k_i\}}_{\Delta,\ell}\left(U,V\right)=&\;U^{3(\varepsilon-1)}\left(\Delta_{\varepsilon}\right)^{-1}\left[U^{\frac{\Delta_{34}}{2}-2\varepsilon}\;G_{\Delta,\ell}^{\Delta_{12},\Delta_{34}}\left(U,V\right)\right]\nonumber \\
    =&\:U^{3(\varepsilon-1)}\left(\Delta_\varepsilon\right)^{-1}\left[U^{\frac{\Delta_{34}}{2}-2\varepsilon}(-1)^{\ell}\sum_{m=0}^\infty\sum_{n=0}^\infty \;r_{m,n;\Delta,\ell}^{\Delta_{12},\Delta_{34}}\;P^{(\varepsilon)}_{\frac{\Delta+\ell}{2}+m,\frac{\Delta-\ell}{2}+n}\left(z,\bar{z}\right)\right].
\end{align}
Using properties of Jack polynomials that we summarize in Appendix \ref{jackpolys} and in particular, the non-trivial relation 
\begin{align}
    &\frac{r_{m,n;\Delta,\ell}^{\Delta_{12},\Delta_{34}}}{\left(\frac{\Delta+\ell}{2}+m+\frac{\Delta_{34}}{2}-\varepsilon+1\right)_{\varepsilon-1}\left(\frac{\Delta-\ell}{2}+n+\frac{\Delta_{34}}{2}-2\varepsilon+1\right)_{\varepsilon-1}}=\nonumber \\
    &\hspace{3cm}\frac{\Gamma \left(\frac{1}{2} \left( \Delta-\ell -4 \varepsilon +\Delta _{34}+2\right)\right) \Gamma \left(\frac{1}{2} \left(\Delta+\ell -2 \varepsilon +\Delta _{34}+2\right)\right)}{\Gamma \left(\frac{1}{2} \left(\Delta+\ell +\Delta _{34}\right)\right) \Gamma \left(\frac{1}{2} \left(\Delta-\ell -2 \varepsilon +\Delta _{34}\right)\right)}\;r_{m,n;\Delta,\ell}^{\Delta_{12},\Delta_{34}-2\left(\varepsilon-1\right)},
\end{align}
obeyed by the expansion coefficients $r_{m,n;\Delta,\ell}^{\Delta_{12},\Delta_{34}}$ written explicitly in Equation \eqref{eq:rcoeff}, we find that we can express the two-variable reduced block as
\begin{align}
    &\mathcal{T}^{\{k_i\}}_{\Delta,\ell}\left(U,V\right)=\nonumber \\
    &\frac{\Gamma \left[\frac{1}{2} \left(\Delta+\ell -2 \varepsilon +\Delta _{34}+2\right)\right]\Gamma \left[\frac{1}{2} \left( \Delta-\ell -4 \varepsilon +\Delta _{34}+2\right)\right]}{\Gamma \left[\frac{1}{2} \left(\Delta+\ell +\Delta _{34}\right)\right] \Gamma \left[\frac{1}{2} \left(\Delta-\ell-2 \varepsilon +\Delta _{34}\right)\right]}\;U^{\frac{\Delta_{34}}{2}+\varepsilon-3}\;G_{\Delta,\ell}^{\Delta_{12},\Delta_{34}-2(\varepsilon-1)}\left(U,V\right).
\label{eq:Tat}
\end{align}
This resembles the conformal block one would expect in a correlator in a kinematical configuration with $(\tilde{\Delta}_{12},\tilde{\Delta}_{34})=(\Delta_{12},\Delta_{34}-2\left(\varepsilon-1\right))$. This mirrors our result for the single-variable reduced blocks $h^{\{k_i\}}_{1,\ell}$ in Equation \eqref{eq:h1at}.

The non-trivial kernel of $\Delta_\varepsilon$ makes one question the uniqueness of the reduced block $\mathcal{T}^{\{k_i\}}_{\Delta,\ell}$. We constructed $\mathcal{T}^{\{k_i\}}_{\Delta,\ell}$ as the inhomogeneous solution to the operator Equation \eqref{eq:topAT} and any additional reduced block $\tilde{\mathcal{T}}^{\{k_i\}}_{\Delta,\ell}$ would be a homogeneous solution satisfying
\begin{align}
    \Delta_\varepsilon\left[\frac{\tilde{\mathcal{T}}^{\{k_i\}}_{\Delta,\ell}\left(z,\bar{z}\right)}{U^{3(\varepsilon-1)}}\right]=0.
\end{align}
Solutions to this equation involve the restricted Jack polynomial expansions that gave rise to $h^{\{k_i\}}_{1,\ell}(z)$ in Appendix \ref{hrecurrencerelations} and inserting these into the other channels generates unphysical twist $2\varepsilon-\Delta_{34}$ and twist $-\Delta_{34}$ blocks. Phrased differently, Section \ref{E2} and Appendices \ref{hrecurrencerelations} and \ref{Trecurrencerelations} demonstrated through recursion relations for the five remaining channel equations that the original contributions $\mathcal{T}^{\{k_i\}}_{\Delta,\ell}$ and $h^{\{k_i\}}_{1,\ell}(z)$ gives rise to the correct spectrum of conformal primaries in each channel (for each supermultiplet), so we conclude $\tilde{\mathcal{T}}^{\{k_i\}}_{\Delta,\ell}=0$ for physical four-point correlators. The idea is then that one can expand the two-variable reduced correlator $\TT^{\{k_i\}}$ in a block decomposition
\begin{align}
    \TT^{\{k_i\}}(U,V)=\sum_{\mathcal{X}\in (\OO_1\times \OO_2)\cap(\OO_3\times \OO_4)}\lambda_{k_1 k_2 \mathcal{X}}\lambda_{k_3 k_4 \mathcal{X}}\;\NN_{\XX}\TT^{\{k_i\}}_{\Delta_\XX,\ell_\XX}\left(z\right),
\end{align}
where $\NN_\XX$ is the superblock normalization constant described previously. The R-symmetry channel equations in Subsection \ref{E2Rchannels} and the recursion relations in Appendices \ref{Trecurrencerelations} and \ref{hrecurrencerelations} justify that these reduced blocks encode full superblocks and that in principle, no constraints should be lost by bootstrapping the functions $\TT^{\{k_i\}}$ and $h^{\{k_i\}}_1$. 

\subsection{6d $\NN=(2,0)$}
\label{6dreducedblocks}
When $\varepsilon=2$, the decomposition of solutions to the superconformal Ward identities for $\EE=2$ in Section \ref{DGSdecomp} takes the form
\begin{align}
    \GG^{\{k_i\}}\left(U,V;\sigma,\tau\right)=&\;U^4\Delta_{2}\left[\left(z\alpha-1\right)\left(z\bar{\alpha}-1\right)\left(\bar{z}\alpha-1\right)\left(\bar{z}\alpha-1\right)\frac{\TT^{\{k_i\}}\left(U,V\right)}{U^5}\right]\nonumber \\
    +&\;\sum_{I=1}^{2}U^{2I}\sigma^{I-1}F_{I}^{\{k_{i}\}}\left(z,\bar{z};\alpha,\bar{\alpha}\right).
    \label{eq:6dDGSdecomp}
\end{align}
According to Equation \eqref{eq:Tat}, the reduced correlator $\TT^{\{k_i\}}$ can be expanded in the reduced blocks
\begin{align}
    &\mathcal{T}^{\{k_i\}}_{\Delta,\ell}\left(U,V\right)=\frac{4}{\left(\Delta+\ell+\Delta _{34}-2\right)\left(\Delta-\ell+\Delta _{34} -6\right)}\;U^{\frac{\Delta_{34}}{2}-1}\;G_{\Delta,\ell}^{\Delta_{12},\Delta_{34}-2}\left(U,V\right).
\label{eq:Tat6d}
\end{align}
This expression generalizes the $\langle \OO_2\OO_2\OO_p\OO_p\rangle$ result in \cite{Beem:2015aoa,Alday:2020tgi} derived using a quadratic Casimir commutation relation in \cite{Dolan:2011dv}. In writing Equation \eqref{eq:6dDGSdecomp}, we eliminated $h^{\{k_i\}}_{2}$ using the redefinitions \eqref{eq:Tredef} and 
\begin{align}
    h^{\{k_i\}}_{1}(z)\rightarrow h^{\{k_i\}}_{1}(z)+f(z),
\end{align}
where $f(z)$ is chosen such that $\partial_zf(z)=-z^2\partial_zh^{\{k_i\}}_2(z)$. The remaining function $h^{\{k_i\}}_{1}$ naturally encodes the operators listed in Section \ref{DGSdecomp} that contribute to the $\WW_\mathfrak{g}$ algebra under the twist. Indeed, one could have derived the $\varepsilon=2$ single-variable reduced block
\begin{align}
    h^{\{k_i\}}_{1,\ell}(z)
    =&\;\frac{-1}{\ell +1}z^{\frac{\Delta _{34}}{2}-1}g^{\Delta_{12},\Delta_{34}-2}_{\ell+4-\Delta_{34},\ell}(z),
\label{eq:h1at6d}
\end{align}
by directly integrating the relation
\begin{align}
    \GG^{\{k_i\}}\left(z,\bar{z};\bar{z}^{-1},\bar{z}^{-1}\right)=-z^2\partial_zh^{\{k_i\}}_1(z),
\label{eq:6dtwistedG}
\end{align}
and relating it to the twisted block decomposition of $\GG^{\{k_i\}}\left(z,\bar{z};\bar{z}^{-1},\bar{z}^{-1}\right)$ shown in \cite{Woolley:2025qbd}. With these ingredients in place, we summarize the assignment of reduced blocks for each supermultiplet in Table \ref{6dreducedblocktable}.

\begin{table}[h]
\begin{tabular}{|l||l|l|l|l|}
\hline $\mathfrak{osp}(8^*|4)$ multiplet $\XX$ & $\TT^{\{k_i\}}_{\Delta',\ell'}$ & $h^{\{k_i\}}_{1,\ell'}$ & $\NN_\XX$\\
\hline \hline 
$\mathcal{L}[p,0]_{\Delta,\ell}$ & $\TT^{\{k_i\}}_{\Delta+4,\ell}$ & --- & $\frac{\left(\Delta+\ell +\Delta _{34} +2\right)\left(\Delta-\ell+\Delta _{34}-2\right)}{\left(\Delta+\ell +\Delta _{34} -2\right)\left(\Delta -\ell+\Delta _{34} -6\right)}$ \\
\hline $\mathcal{B}[p,2]_{2p+8+\ell,\ell}$ & $\TT^{\{k_i\}}_{2p+11+\ell,\ell+1}$ & --- & $\frac{2 (\ell +2) (\ell +5)}{(\ell +1) (\ell +4)}$ \\
\hline $\mathcal{B}[p+2,0]_{2p+8+\ell,\ell}$ & $\TT^{\{k_i\}}_{2p+10+\ell,\ell+2}$ & $h^{\{k_i\}}_{1,\ell+4}$ & $\frac{2 \left(4-\Delta _{34}\right) (\ell +3)}{\left(\Delta _{12}-\Delta _{34}+4\right) (\ell +1)}$ \\
\hline $\mathcal{D}[p,4]_{2p+8,0}$ & $\TT^{\{k_i\}}_{2p+10,0}$ & --- & $\frac{8}{3}$\\
\hline $\mathcal{D}[p+2,2]_{2p+8,0}$ & $\TT^{\{k_i\}}_{2p+9,1}$ & $h^{\{k_i\}}_{1,3}$ & $\frac{4 \left(4-\Delta _{34}\right)}{\Delta _{12}-\Delta _{34}+4}$ \\
\hline $\mathcal{D}[p+4,0]_{2p+8,0}$ & $\TT^{\{k_i\}}_{2p+8,0}$ & $h^{\{k_i\}}_{1,2}$ & $\frac{4 \left(8-\Delta _{34}\right) \left(6-\Delta _{34}\right)}{\left(\Delta _{12}-\Delta _{34}+4\right) \left(\Delta _{12}-\Delta _{34}+8\right)}$ \\
\hline $\mathcal{B}[p,0]_{2p+4+\ell,\ell}$ & --- & $h^{\{k_i\}}_{1,\ell+2}$ & $\frac{\ell +3}{\ell +1}$\\
\hline $\mathcal{D}[p,2]_{2p+4,0}$ & --- & $h^{\{k_i\}}_{1,1}$ & 2 \\
\hline $\mathcal{D}[p+2,0]_{2p+4,0}$ & --- & $h^{\{k_i\}}_{1,0}$ & $\frac{2 \left(4 -\Delta _{34}\right)}{\Delta _{12}-\Delta _{34}+4 }$ \\
\hline
$\mathcal{D}[p,0]_{2p,0}$ & --- & $h^{\{k_i\}}_{1,-2}$ & 1 \\
\hline
\end{tabular}
\caption{The normalized contributions of two-variable and single-variable reduced blocks to $\TT^{\{k_i\}}$ and $h^{\{k_i\}}_{1}$ in 6d $\NN=(2,0)$ extremality $\EE=2$ correlators.}
\label{6dreducedblocktable}
\end{table}

\subsubsection{Alternative formulation}
The decomposition of $\GG^{\{k_i\}}$ in \eqref{eq:6dDGSdecomp} is not unique, in that one can clearly add and subtract terms to $\TT^{\{k_i\}}$ and $h^{\{k_i\}}_1$ which vanish upon applying the chiral algebra twist in \eqref{eq:6dtwistedG}. While the form \eqref{eq:6dDGSdecomp} is convenient for conformal block decompositions of the reduced correlators $\TT^{\{k_i\}}$ and $h^{\{k_i\}}_1$, we can make contact with holographic studies in \cite{Rastelli:2017ymc,Zhou:2017zaw,Alday:2020lbp,Alday:2020tgi,Behan:2021pzk} by writing 
\begin{align}
\mathcal{G}^{\{k_i\}}\left(U,V;\sigma,\tau\right)=\mathcal{F}^{\{k_i\}}\left(U,V;\sigma,\tau\right)+\Upsilon\circ\mathcal{H}^{\{k_i\}}\left(U,V;\sigma,\tau\right).
\label{eq:RZdecomp}
\end{align}
 The $\mathcal{F}^{\{k_i\}}$ contribution is a 6d uplift of the $\WW$-algebra correlator arising from the construction in \cite{Beem:2014kka} and can be fixed exactly, as elaborated upon in \cite{Woolley:2025qbd}. The $\Upsilon\circ\mathcal{H}^{\{k_i\}}$ contribution encodes both $\WW$-algebra and non-$\WW$-algebra data and vanishes exactly upon twisting. The differential operator $\Upsilon$ is defined in \cite{Dolan:2004mu} (with typos corrected in \cite{Rastelli:2017ymc}) to take the form\footnote{The authors of \cite{Rastelli:2017ymc} not only corrected typos in \cite{Dolan:2004mu}, but also modified the definition of $\Upsilon$ such that the reduced correlator $\TT$ differs from that in \cite{Dolan:2004mu,Beem:2015aoa} by a factor $U^5$.}
\begin{align}
\Upsilon & =\sigma^2 \mathcal{D}^{\prime} U V+\tau^2 \mathcal{D}^{\prime} U+\mathcal{D}^{\prime} V-\sigma \mathcal{D}^{\prime} V(U+1-V)-\tau \mathcal{D}^{\prime}(U+V-1)-\sigma \tau \mathcal{D}^{\prime} U(V+1-U), \nonumber \\
\mathcal{D}^{\prime} & =D_2-\frac{2}{V}\left(D_0^{+}-D_1^{+}+2 \partial_\sigma \sigma\right) \tau \partial_\tau+\frac{2}{U V}\left(-V D_1^{+}+2\left(V \partial_\sigma \sigma+\partial_\tau \tau-1\right)\right)\left(\partial_\sigma \sigma+\partial_\tau \tau\right), \nonumber \\
D_2 & =\partial_z \partial_{\bar{z}}-\frac{2}{z-\bar{z}}\left(\partial_z-\partial_{\bar{z}}\right), \quad D_0^{+}=\partial_z+\partial_{\bar{z}}, \quad D_1^{+}=z \partial_z+\bar{z} \partial_{\bar{z}}.
\label{eq:diffops}
\end{align}
The form of $\Upsilon$ implies that the function $\mathcal{H}^{\{k_i\}}$ is a polynomial in $\sigma$ and $\tau$ of degree 2 less than $\mathcal{G}^{\{k_i\}}$ and we can expand it in $U,V$-dependent reduced correlators as
\begin{align}
    \HH^{\{k_i\}}(U,V;\sigma,\tau)=\sum_{I=0}^{\EE-2}\sum_{J=0}^{\EE-2-I}U^{2I}\left(\frac{U}{V}\right)^{2J}\sigma^I\tau^J\HH^{\{k_i\}}_{I,J}\left(U,V\right).
\end{align}
These two arrangements of $\GG^{k_i}$ have different properties that are technically practical for different applications. While the reduced correlators $\TT^{k_i}$ and $h^{k_i}_1$ in \eqref{eq:6dDGSdecomp} admit reduced block decompositions, they do not individually satisfy crossing symmetry. By explicitly acting with the differential operator $\Delta_2$ in Equation \eqref{eq:6dDGSdecomp}, we use the linearly independent monomials in $\alpha,\bar{\alpha}$ to algebraically eliminate all derivatives of $\TT^{\{k_i\}}$. Then, imposing the chiral algebra crossing equation on Equation \eqref{eq:6dtwistedG} eliminates the remaining derivatives of $h^{\{k_i\}}_1$, yielding the reduced $1\leftrightarrow3$ crossing equation
\begin{align}
    \TT^{k_1k_2k_3k_4}\left(U,V\right)-&\frac{U^4}{V^4}\TT^{k_3k_2k_1k_4}\left(V,U\right)=\nonumber\\
    &-U^3\left(\frac{h^{k_1k_2k_3k_4}_1(z)-h^{k_1k_2k_3k_4}_1(\bar{z})}{\left(z-\bar{z}\right)^3}-\frac{U}{V}\frac{h^{k_3k_2k_1k_4}_1(1-z)-h^{k_3k_2k_1k_4}_1(1-\bar{z})}{\left(z-\bar{z}\right)^3}\right).
\label{eq:6dreducedcrossing}
\end{align}
By contrast, the terms $\FF^{\{k_i\}}$
and $\HH^{\{k_i\}}$ individually satisfy their own crossing equations. For $\FF^{\{k_i\}}$ this follows from the crossing symmetry of the underlying $\WW$-algebra correlator, while we can use the crossing properties of the differential operator $\Upsilon$ derived in \cite{Rastelli:2017ymc} to deduce that for general $\EE$,
\begin{align}
\HH^{k_1k_2k_3k_4}_{I,J}\left(U,V\right)=\HH^{k_2k_1k_3k_4}_{J,I}\left(\frac{U}{V},\frac{1}{V}\right)=\left(\frac{U}{V}\right)^4\HH^{k_3k_2k_1k_4}_{I,\EE-2-I-J}\left(V,U\right).
\label{eq:6dHcross}
\end{align}
These relations may help in determining reduced blocks for $\EE>2$. 

Another benefit of the decomposition in \eqref{eq:RZdecomp} is its simplification of holographic calculations. It was observed in \cite{Rastelli:2017ymc} that with this organization, the Mellin transformation of $\FF^{\{k_i\}}$ can be discarded while reduced correlator functions $\HH^{\{k_i\}}_{I,J}$ can be transformed into a relatively simple reduced Mellin amplitude in the examples $\langle\OO_k\OO_k\OO_k\OO_k\rangle$ for $k=2,3,4$ at $O\left(c_\mathfrak{g}^{-1}\right)$. In a similar vein, the author of \cite{Zhou:2017zaw} derived a formula for the reduced Mellin amplitude at $O\left(c_\mathfrak{g}^{-1}\right)$ for the $\EE=2$ correlators $\langle\OO_{k_1}\OO_{k_2}\OO_{k_3}\OO_{k_1+k_2+k_3-4}\rangle$ studied in this work.

It can prove useful to study a combination of these decompositions, as was done for example in an analytic bootstrap computation in \cite{Alday:2020tgi}. By noting that the action of $\Delta_2$ and $\Upsilon$ on $\sigma,\tau$-independent functions is related by
\begin{align}
    U^{4}\Delta_2\left[\left(z\alpha-1\right)\left(z\bar{\alpha}-1\right)\left(\bar{z}\alpha-1\right)\left(\bar{z}\alpha-1\right)\frac{\TT^{\{k_i\}}(U,V)}{U^5}\right]=\Upsilon\circ \TT^{\{k_i\}}(U,V),
\end{align} 
we can bridge these formulations by solving the equation
\begin{align}
    \Upsilon\circ\HH^{\{k_i\}}(z,\bar z;\alpha,\bar\alpha)=&\;F^{\{k_i\}}(z,\bar z;\alpha,\bar\alpha)+\Upsilon\circ\mathcal{T}^{\{k_i\}}(z,\bar z)-\mathcal{F}^{\{k_i\}}(z,\bar z;\alpha,\bar\alpha)\nonumber \\
    \equiv&\;\Upsilon\circ\left[U^3\frac{h^{\{k_i\}}(z)-h^{\{k_i\}}(\bar{z})}{\left(z-\bar{z}\right)^3}+\mathcal{T}^{\{k_i\}}(z,\bar z)+f^{\{k_i\}}(z,\bar z)\right],
\label{eq:6ddecompconvert}
\end{align}
for difference function $f^{\{k_i\}}$. The functions $\mathcal{F}^{\{k_i\}}$ (and consequently $f^{\{k_i\}}$) are determined by a $\WW$-algebra calculation, as described in \cite{Woolley:2025qbd}. In $\langle\OO_2\OO_k\OO_2\OO_k\rangle$ for example, we find
\begin{align}
    \FF^{2k2k}\left(U,V;\sigma,\tau\right)=&\;U^4\sigma^2 +\delta _{2k} \left(1+\frac{U^4}{V^4}\tau ^2 \right)+\frac{2 k}{c_\mathfrak{g}} \left(\frac{(k-1)U^2}{V^2}\tau+\frac{U^4}{V^2}\sigma\tau+U^2\sigma \right),\\f^{2k2k}\left(U,V\right)=&\;\delta _{2,k}\frac{U^2 \left(U^2+V^2\right) }{V^2 \left(U^2-2 U (V+1)+(V-1)^2\right)}\nonumber \\
+&\;\frac{U^4}{U^2-2 U (V+1)+(V-1)^2}+\frac{1}{c_\mathfrak{g}}\frac{2 k U^3 (k+U+V-1)}{V \left(U^2-2 U (V+1)+(V-1)^2\right)},
\end{align}
where $c_\mathfrak{g}$ is the central charge of a 6d (2,0) theory of type $\mathfrak{g}$.

\subsection{4d $\NN=4$}
\label{4dreducedblocks}
When $\varepsilon=1$, the operators $\Delta_{\varepsilon}$ and $D_\varepsilon$ become factors of 1 and the recurrence relations in Appendix \ref{Trecurrencerelations} reduce to multiplicative recurrence relations obeyed by 4d conformal blocks. The decomposition of four-point functions $\GG^{\{k_i\}}$ in \eqref{eq:DGSdecompE2} has the form
\begin{align}
    &\GG^{\{k_i\}}\left(U,V;\sigma,\tau\right)=\left(z\alpha-1\right)\left(z\bar{\alpha}-1\right)\left(\bar{z}\alpha-1\right)\left(\bar{z}\bar{\alpha}-1\right)\TT^{\{k_i\}}\left(U,V\right)\nonumber \\
    &\hspace{8cm}+\sum_{I=1}^{2}U^{I}\sigma^{I-1}F_{I}^{\{k_{i}\}}\left(z,\bar{z};\alpha,\bar{\alpha}\right),
    \label{eq:4dDGSdecomp}
\end{align}
where $\TT^{\{k_i\}}$ can be expanded in the simplest example of  Equation \eqref{eq:Tat}, namely
\begin{align}
    &\mathcal{T}^{\{k_i\}}_{\Delta,\ell}\left(U,V\right)=U^{\frac{\Delta_{34}}{2}-2}\;G_{\Delta,\ell}^{\Delta_{12},\Delta_{34}}\left(U,V\right).
\label{eq:Tat4d}
\end{align}
This is the same reduced block as that originally derived in \cite{Dolan:2001tt}. As described in \cite{Dolan:2004mu}, we make contact with the results in \cite{Dolan:2001tt} using the redefinition of $\TT^{\{k_i\}}$ in \eqref{eq:Tredef}, the choice 
\begin{align}
    h^{\{k_i\}}_{2,\ell}(z)=&\; \frac{1}{z} h^{\{k_i\}}_{1,\ell-1}(z),
\end{align}
and the redefinition
\begin{align}
    h^{\{k_i\}}_1(z)\rightarrow&\; h^{\{k_i\}}_1(z)-zh^{\{k_i\}}_2(z).
\end{align}
As reviewed in \cite{Bissi:2020jve}, these allow us to re-express the sum over $h^{\{k_i\}}_I$ contributions into 
\begin{align}
    &\FF^{\{k_i\}}_{\hat{f}}(z,\bar{z};\alpha,\bar{\alpha})\equiv\sum_{I=1}^{2}U^{I}\sigma^{I-1}F_{I}^{\{k_{i}\}}\left(z,\bar{z};\alpha,\bar{\alpha}\right)=-k\nonumber \\
    &+\frac{\left(\bar{\alpha}z-1\right)\left(\alpha\bar{z}-1\right)\left(f^{\{k_i\}}\left(z,\alpha\right)+f^{\{k_i\}}\left(\bar{z},\bar{\alpha}\right)\right)-\left(\bar{\alpha}z-1\right)\left(\alpha\bar{z}-1\right)\left(f^{\{k_i\}}\left(z,\alpha\right)+f^{\{k_i\}}\left(\bar{z},\bar{\alpha}\right)\right)}{\left(z-\bar{z}\right)\left(\alpha-\bar{\alpha}\right)},
\end{align}
where the constant $k=f^{\{k_i\}}\left(z,\frac{1}{z}\right)$. The functions $f^{\{k_i\}}\left(z,\alpha\right)$ have the form\footnote{The functions $f^{\{k_i\}}\left(z,\alpha\right)$ have nothing to do with $f^{\{k_i\}}(z,\bar{z})$ appearing in the 6d context of \eqref{eq:6ddecompconvert}.}
\begin{align}
    f^{\{k_i\}}\left(z,\alpha\right)=k+\left(\alpha-\frac{1}{z}\right)\hat{f}^{\{k_i\}}\left(z,\alpha\right),
\end{align}
in terms of $\hat{f}\left(z,\alpha\right)$ which has the decomposition
\begin{align}
    &\hat{f}^{\{k_i\}}\left(z,\alpha\right)
    =\sum_{n=0}^{1}\sum_{\ell\geq-n}^\infty b_{n,\ell}\;P_n^{\left(a,b\right)}\left(2\alpha-1\right)z^{2}h^{\{k_i\}}_{1,\ell}(z).
\end{align}
The $P_n^{\left(a,b\right)}$ are Jacobi polynomials with $\left(a,b\right)=\left(\frac{|k_{12}-k_{34}|}{2},\frac{|k_{23}-k_{14}|}{2}\right)$ and $b_{n,\ell}$ are related to the OPE coefficients $\lambda_{k_1 k_2 \mathcal{X}}\lambda_{k_3 k_4 \mathcal{X}}$. The appearance of these functions is related to the chiral algebra subsector alluded to at the end of Subsection \ref{DGSdecomp}. In particular, the survival of an R-symmetry cross ratio $\alpha$ reflects the fact that 4d $\NN=4$ theories have small $\NN=4$ super-$\WW$-algebras as their chiral algebras. Finally, the reduced $1\leftrightarrow3$ crossing equation is
\begin{align}
&\TT^{k_1k_2k_3k_4}\left(U,V\right)-\frac{U^2}{V^2}\TT^{k_3k_2k_1k_4}\left(V,U\right)=-\frac{1}{\left(z\alpha-1\right)\left(z\bar{\alpha}-1\right)\left(\bar{z}\alpha-1\right)\left(\bar{z}\bar{\alpha}-1\right)}\nonumber\\
&\times\left(\FF^{k_1k_2k_3k_4}_{\hat{f}}\left(U,V;\sigma,\tau\right)-\frac{U^2}{V^2}\tau^2\;\FF^{k_3k_2k_1k_4}_{\hat{f}}\left(V,U;\frac{\sigma}{\tau},\frac{1}{\tau}\right)\right).
\label{eq:4dreducedcrossing}
\end{align}
\subsection{3d $\NN=8$}
\label{3dreducedblocks}
Finally, when $\varepsilon=\frac{1}{2}$ we have 
\begin{align}
    \GG^{\{k_i\}}\left(U,V;\sigma,\tau\right)=&\;U\Delta_{\frac{1}{2}}\left[\left(z\alpha-1\right)\left(z\bar{\alpha}-1\right)\left(\bar{z}\alpha-1\right)\left(\bar{z}\bar{\alpha}-1\right)\frac{\TT^{\{k_i\}}\left(U,V\right)}{U^{\frac{1}{2}}}\right]\nonumber \\
    +&\;U^{\frac{1}{2}}\left(D_\varepsilon\right)^{\varepsilon-1}\left[\frac{(z \alpha-1)(z \bar{\alpha}-1)h^{\{k_{i}\}}_1(z)-(\bar{z} \alpha-1)(\bar{z} \bar{\alpha}-1)h^{\{k_{i}\}}_1(\bar{z})}{z-\bar{z}}\right].
    \label{eq:DGSdecomp3d}
\end{align}
The most surprising instance of Equation \eqref{eq:Tat} is when $\varepsilon=\frac{1}{2}$ where despite $\Delta_{\frac{1}{2}}$ being a non-local operator, we derived that $\TT^{\{k_i\}}$ can be decomposed in reduced blocks of the explicit form
\begin{align}
    &\mathcal{T}^{\{k_i\}}_{\Delta,\ell}\left(U,V\right)=\nonumber \\
    &\frac{\Gamma \left[\frac{1}{2} \left( \Delta-\ell -2 +\Delta _{34}+2\right)\right] \Gamma \left[\frac{1}{2} \left(\Delta+\ell -1 +\Delta _{34}+2\right)\right]}
    {\Gamma \left[\frac{1}{2} \left(\Delta+\ell +\Delta _{34}\right)\right] \Gamma \left[\frac{1}{2} \left(\Delta-\ell-1 +\Delta _{34}\right)\right]}\;U^{\frac{\Delta_{34}-5}{2}}\;G_{\Delta,\ell}^{\Delta_{12},\Delta_{34}+1}\left(U,V\right).
\label{eq:Tat3d}
\end{align}
According to Appendix \ref{hrecurrencerelations}, the single-variable reduced correlator has a decomposition in terms of 
\begin{align}
    h^{\{k_i\}}_{1,\ell}(z)
    =&\;(-1)^{\frac{3}{2}}\frac{ \Gamma \left[\ell+1\right]}{\Gamma\left[\frac{1}{2}\right] \Gamma\left[\ell+\frac{1}{2}\right]}z^{\frac{\Delta _{34}-5}{2}}g^{\Delta_{12},\Delta_{34}+1}_{\ell+1-\Delta_{34},\ell}(z).
\end{align}
When studying this single-variable reduced block decomposition, we run into an apparent contradiction. Recall that by the construction described near the end of Subsection \ref{DGSdecomp}, the twist $(z,\bar{z};\alpha,\bar{\alpha})\rightarrow(\bar{z},\bar{z};\alpha,\bar{z}^{-1})$ causes the $\TT^{\{k_i\}}$ contribution to vanish identically while the remaining $h^{\{k_i\}}_I$ functions encode a correlation function in the resulting 1d topological sector derived in \cite{Chester:2014mea}. If one performs the redefinition of $\TT^{\{k_i\}}$ in \eqref{eq:Tat}, one finds that when $\varepsilon=\frac{1}{2}$, the function $h^{\{k_i\}}_2$ is completely eliminated. This is unlike in 4d and 6d, where one additionally needed to redefine $h^{\{k_i\}}_1$.
The paradox is that there will generically be BPS multiplets $\left(B,+\right)^{[0,0,p+4,0]}_{p/2+2,0}$, $\left(B,2\right)^{[0,1,p+2,0]}_{p/2+2,0}$, and $\left(B,2\right)^{[0,2,p,0]}_{p/2+2,0}$ that should be captured by $h^{\{k_i\}}_2$, but now seem to be completely encoded by $\TT^{k_i}$. Unlike in 4d and 6d, the generation of superblocks using $A^{\{k_{i}\};\TT}_{mn;\;\Delta',\ell'}$ in 3d does not result in $2\varepsilon-\Delta_{34}$ operators that require cancellation by $h^{\{k_i\}}_{1,\ell}$, so it would appear that these multiplets are no longer encoded in the single-variable sector. This puzzle exacerbated by the claim in \cite{Dolan:2004mu} that one can also absorb $h^{\{k_i\}}_1$ into $\TT^{k_i}$, leaving no single-variable reduced correlators to produce the 1d sector upon twisting. For example, this makes it unclear how to re-derive the sum rules for 1d OPE coefficients as was done in \cite{Chester:2014mea}, but now using the formalism of \cite{Dolan:2004mu} in the twisted limit.

Postponing a resolution of this contradiction to further work, we tabulate our organization of 3d $\NN=8$ reduced blocks using the same organization as Section \ref{3dsuperblocks} in Table \ref{3dreducedblocktable}. In particular, we use $\TT^{\{k_i\}}_{\Delta,\ell}$ to describe long and semi-short multiplets and by keeping $h^{\{k_i\}}_1$ explicit, we obtain reduced superblocks for the three lowest-twist BPS operators. We emphasize that this organization gives the correct results for full superblocks. The aforementioned puzzle stems from the lack of $h^{\{k_i\}}_{1,\ell}$ contributions to $\left(B,+\right)^{[0,0,p+4,0]}_{p/2+2,0}$, $\left(B,2\right)^{[0,1,p+2,0]}_{p/2+2,0}$, and $\left(B,2\right)^{[0,2,p,0]}_{p/2+2,0}$.

\begin{table}[h]
\begin{tabular}{|l||l|l|l|l|}
\hline $\mathfrak{osp}(8|4)$ multiplet $\XX$ & $\TT^{\{k_i\}}_{\Delta',\ell'}$ & $h^{\{k_i\}}_{1,\ell'}$ & $\NN_\XX$\\
\hline \hline 
$\left(A,0\right)^{[0,0,p,0]}_{\Delta\geq p/2+1+\ell,\ell}$ & $\TT^{\{k_i\}}_{\Delta+4,\ell}$ & ---   & $\frac{\left(\Delta+\ell+\Delta_{34}\right) \left(\Delta+\ell +\Delta _{34} +2\right)\left(\Delta-\ell +\Delta _{34} +1\right) \left(\Delta-\ell +\Delta _{34} -1\right)}
{\left(\Delta+\ell+\Delta _{34} +1\right) \left(\Delta+\ell +\Delta _{34} +3\right)\left(\Delta -\ell+\Delta _{34} +2\right) \left(\Delta-\ell  +\Delta _{34}\right)}$ \\
\hline $\left(A,2\right)^{[0,1,p,0]}_{p/2+2+\ell,\ell}$ & $\TT^{\{k_i\}}_{p/2+5+\ell,\ell+1}$& ---  & $\frac{(\ell +2) (2 \ell +1)}{2 (\ell +1) (2 \ell +5)}$\\
\hline
$\left(A,+\right)^{[0,0,p+2,0]}_{p/2+2+\ell,\ell}$ & $\TT^{\{k_i\}}_{p/2+4+\ell,\ell+2}$& ---  & $\frac{\left(1-\Delta _{34}\right) (2 \ell +1) (2 \ell +3)}{2 \left(\Delta _{12}-\Delta _{34}+1\right) (\ell +1) (\ell +2)}$\\
\hline $\left(B,2\right)^{[0,2,p,0]}_{p/2+2,0}$ & $\TT^{\{k_{i}\}}_{p/2+4,0}$ & ---  & $\frac{1}{3}$\\
\hline $\left(B,2\right)^{[0,1,p+2,0]}_{p/2+2,0}$ & $\TT^{\{k_{i}\}}_{p/2+3,1}$ & ---   & $\frac{1-\Delta _{34}}{\Delta _{12}-\Delta _{34}+1}$\\
\hline $\left(B,+\right)^{[0,0,p+4,0]}_{p/2+2,0}$ & $\TT^{\{k_{i}\}}_{p/2+2,0}$ & ---  & $\frac{2 \left(\Delta _{34}-2\right) \left(2 \Delta _{34}-3\right)}{\left(\Delta _{12}-\Delta _{34}+1\right) \left(\Delta _{12}-\Delta _{34}+2\right)}$\\
\hline
$\left(A,+\right)^{[0,0,p,0]}_{p/2+\ell,\ell}$ & --- & $h^{\{k_{i}\}}_{1,\ell+2}$  & $\frac{(2 \ell +1) (2 \ell +3)}{4 (\ell +1) (\ell +2)}$\\
\hline $\left(B,2\right)^{[0,1,p-2,0]}_{p/2+1,0}$ & --- & $h^{\{k_{i}\}}_{1,1}$ & $\frac{1}{2}$\\
\hline $\left(B,+\right)^{[0,0,p+2,0]}_{p/2+1,0}$ & --- & $h^{\{k_{i}\}}_{1,0}$ & $\frac{2(1- \Delta _{34})}{\Delta _{12}-\Delta _{34}+1}$\\
\hline 
$\left(B,+\right)^{[0,0,p,0]}_{p/2,0}$ & --- & $h^{\{k_i\}}_{1,-2}$  & 1\\
\hline
\end{tabular}
\caption{The normalized contributions of two-variable and single-variable reduced blocks to $\TT^{\{k_i\}}$ and $h^{\{k_i\}}_{1}$ in 3d $\NN=8$ extremality $\EE=2$ correlators.}
\label{3dreducedblocktable}
\end{table}

Another important conceptual concern is that even with explicit reduced blocks $\mathcal{T}^{\{k_i\}}_{\Delta,\ell}$ and $h^{\{k_i\}}_{1;\ell}$, actually implementing these blocks in a crossing problem in analogy to Equations \eqref{eq:6dreducedcrossing} and \eqref{eq:4dreducedcrossing} will lead to ambiguities stemming from the operator $\Delta_{\frac{1}{2}}$. In particular, starting from the full crossing equations in \eqref{eq:12cross}-\eqref{eq:23cross}, one can see that when expressed in terms of \eqref{eq:DGSdecompE2}, one relates a $D_{\frac{1}{2}}$-exact contribution to another that is exact with respect to the crossed form of $D_{\frac{1}{2}}$. The non-trivial kernel of this operator makes such an expression ambiguous. The analogous issue was address in a 6d $\NN=(1,0)$ crossing equation in Appendix C of \cite{Chang:2017xmr}, where the R-symmetry in half-maximal supersymmetry is not constraining enough to be able to algebraically eliminate the derivatives in $D_2$. The authors found that the freedom required one to include an additional $z,\bar{z}$-dependent term in the reduced crossing equation\footnote{We note that this constant was forced to vanish in an analytically-known solution to crossing, namely 6d Generalized Free Field Theory (GFF).}. What makes the situation worse in 3d is that our formal definition of $\Delta_{\frac{1}{2}}$ is in terms of the functional space spanned by Jack polynomials, which do not obey simple relations under crossing. Unlike for integer $\varepsilon\geq1$ where $D_{\varepsilon}$ is a well-defined operator which is invariant under $1\leftrightarrow3$ crossing, it is unclear how $D_{\frac{1}{2}}$ and its crossed form may be related.

\section{Future directions}
The result of this paper was to explicitly write down the superblocks for four-point functions $\langle\OO_{k_1}\OO_{k_2}\OO_{k_3}\OO_{k_1+k_2+k_3-4}\rangle$ of 1/2-BPS operators $\OO_{k_i}$ in the maximally supersymmetric CFTs in $d=3,4,6$. We observed that these superblocks, which encode all operators in a supermultiplet, can be repackaged into reduced superblocks involving (at most) a single $2(\varepsilon+1)$-dimensional conformal block and a global $SL(2,\mathbb{R})$ block. 

The applications of our full superblocks in studying next-to-next-to-extremal correlators in the numerical and analytic superconformal bootstrap are straightforward and manifold. The availability of reduced superblocks presents an obvious simplification of such studies. For instance, the relaxation of the Bose selection rule in $\langle\OO_{k_1}\OO_{k_2}\OO_{k_3}\OO_{k_1+k_2+k_3-4}\rangle$ for $k_1\neq k_2$ implies that a long superblock will generically contain 74 bosonic blocks. Reduced superblocks repackage all of this information in a single block defined in Equation \eqref{eq:Tat}. This is especially useful when dealing with the convoluted form of bosonic blocks in 6d as well as in 3d, where blocks are not known in closed-form and must be approximated. With this, reduced blocks streamline numerical and analytic bootstrap studies in 6d and assuming one can unambiguously formulate a reduced crossing equation for $\varepsilon=\frac{1}{2}$, in 3d.

    One immediate future direction is to resolve the apparent contradiction posed by the absorption of $h^{\{k_i\}}_2$ and the absence of reduced blocks for certain BPS multiplets in 3d $\NN=8$ theories. One possibility would be to avoid the redefinition \eqref{eq:Tredef} and keep $h^{\{k_i\}}_2$ explicit as in 4d. This would still not explain why it can ever be an option to eliminate all $h^{\{k_i\}}_I$ functions, as claimed in \cite{Dolan:2004mu}. In the following subsections, we outline two broader avenues for generalization.

\subsection{Higher extremality}
The derivation of reduced blocks for $\EE>2$ is considerably more challenging, assuming $\varepsilon\neq1$. This is a consequence of the proliferation of two-variable reduced correlator functions $\TT^{\{k_i\}}_{I,J}$ and the way they appear in the $\frac{1}{2}(\EE+1)(\EE+2)$ channel equations $A_{mn}^{k_{12},k_{34}}$. Take for example, the $\EE=3$ correlator $\langle\OO_{k_1}\OO_{k_2}\OO_{k_3}\OO_{k_1+k_2+k_3-6}\rangle$ where each $k_i\geq3$. According to the organization in \eqref{eq:DGSdecomp}, these correlators will have three undetermined functions $\TT^{\{k_i\}}_{0,0}$, $\TT^{\{k_i\}}_{1,0}$, and $\TT^{\{k_i\}}_{0,1}$ which appear in the ten $\mathfrak{so}(\texttt{d})_R$ equations. In terms of $m(\EE=3)=\frac{k_1+k_2}{2}\equiv m$, the four new channel equations take the form\footnote{We organized these equations in terms of the ingredients in \eqref{eq:recuringredients}. While this basis was useful in $\EE=2$, the following equations do not make it clear when terms vanish due to $\Delta_{12}=0$, which implies the parity of the blocks they encode. There is presumably a basis that makes such features manifest.}
\begin{align}
    A^{\{k_{i}\};\TT}_{m\;m}=&\;\frac{1}{\mathcal{Y}_{m}^{k_{12},k_{34}}}U^{3\varepsilon}\Delta_\varepsilon\left[\frac{\mathcal{T}_{1,0}^{\{k_i\}}}{U^{3(\varepsilon-1)}}\right]+\frac{1}{\mathcal{Y}_{m}^{k_{12},k_{34}}}\frac{U^{3\varepsilon}}{V^\varepsilon}\Delta_\varepsilon\left[\frac{\mathcal{T}_{0,1}^{\{k_i\}}}{U^{3(\varepsilon-1)}}\right], 
    \label{eq:AE3T1}\\
    A^{\{k_{i}\};\TT}_{m\;m-1}=&\;\frac{1}{\mathcal{Y}_{m-1}^{k_{12},k_{34}}}U^{3\varepsilon}\Delta_\varepsilon\left[\left(
    \frac{3 \Delta _{12}+\Delta _{34}-6 \varepsilon }{2 \left(\Delta _{34}-6 \varepsilon\right)}+\frac{V-1}{U}\right)\frac{\mathcal{T}_{1,0}^{\{k_i\}}}{U^{3(\varepsilon-1)}}\right]\nonumber\\
    +&\;\frac{1}{\mathcal{Y}_{m-1}^{k_{12},k_{34}}}\frac{U^{3\varepsilon}}{V^\varepsilon}\Delta_\varepsilon\left[\left(
    \frac{3 \Delta _{12}-\Delta _{34}+6 \varepsilon }{2 \left(\Delta _{34}-6 \varepsilon\right)}
    +\frac{V-1}{U}\right)\frac{\mathcal{T}_{0,1}^{\{k_i\}}}{U^{3(\varepsilon-1)}}\right], 
    \label{eq:AE3T2}\\
    A^{\{k_{i}\};\TT}_{m\;m-2}=&\;\frac{1}{\mathcal{Y}_{m-2}^{k_{12},k_{34}}}U^{3\varepsilon}\Delta_\varepsilon\Bigg[\Bigg(\frac{\left(\Delta _{12}+\Delta _{34}-4 \varepsilon \right) \left(3 \Delta _{12}-\Delta _{34}+2 \varepsilon \right)}{4 \left(\Delta _{34} -4 \varepsilon\right) \left(\Delta _{34}-5 \varepsilon\right)}\nonumber\\
    &\;\hspace{4cm}+\frac{2 \Delta _{12}+\Delta _{34}-4 \varepsilon }{2 \left(\Delta _{34}-4 \varepsilon\right)}\frac{V-1}{U}+\frac{V+1}{2U}
    \Bigg)\frac{\mathcal{T}_{1,0}^{\{k_i\}}}{U^{3(\varepsilon-1)}}\Bigg]\nonumber \\
    +&\;\frac{1}{\mathcal{Y}_{m-2}^{k_{12},k_{34}}}\frac{U^{3\varepsilon}}{V^\varepsilon}\Delta_\varepsilon\Bigg[\Bigg(\frac{\left(\Delta _{12}-\Delta _{34}+4 \varepsilon \right)\left(3 \Delta _{12}+\Delta _{34}-2 \varepsilon \right) }{4 \left(\Delta _{34} -4 \varepsilon\right) \left(\Delta _{34}-5 \varepsilon\right)}\nonumber\\
    &\;\hspace{4cm}+\frac{2 \Delta _{12}-\Delta _{34}+4 \varepsilon }{2 \left(\Delta _{34}-4 \varepsilon\right)}\frac{V-1}{U}+\frac{V+1}{2U}
    \Bigg)\frac{\mathcal{T}_{0,1}^{\{k_i\}}}{U^{3(\varepsilon-1)}}\Bigg],
    \label{eq:AE3T3}\\
    A^{\{k_{i}\};\TT}_{m\;m-3}=&\;U^{3\varepsilon}\Delta_\varepsilon\Bigg[\Bigg(\frac{\left(2 \varepsilon-\Delta _{12}-\Delta _{34}\right)\left(\Delta _{12}-\Delta _{34}+2 \varepsilon \right)\left(\Delta _{12}+\Delta _{34}-4 \varepsilon \right)}{8 \left(2 \varepsilon -\Delta _{34}\right) \left(3 \varepsilon -\Delta _{34}\right) \left(4 \varepsilon -\Delta _{34}\right)}\nonumber\\
    &\;\hspace{2cm}+\frac{\left(\Delta _{12}-\varepsilon \right) \left(\Delta _{12}+\Delta _{34}-2 \varepsilon \right)}{4 \left(2 \varepsilon -\Delta _{34}\right) \left(3 \varepsilon -\Delta _{34}\right)}\frac{V-1}{U}+\frac{\Delta _{12}+\Delta _{34}-2 \varepsilon }{2 \left(\Delta _{34}-2 \varepsilon\right)}\frac{V+1}{2U}
    \Bigg)\frac{\mathcal{T}_{1,0}^{\{k_i\}}}{U^{3(\varepsilon-1)}}\Bigg]\nonumber \\
    +&\;\frac{U^{3\varepsilon}}{V^\varepsilon}\Delta_\varepsilon\Bigg[\Bigg(\frac{\left(2 \varepsilon-\Delta _{12}-\Delta _{34}\right) \left(\Delta _{12}-\Delta _{34}+2 \varepsilon \right) \left(\Delta _{12}-\Delta _{34}+4 \varepsilon \right)}{8 \left(2 \varepsilon -\Delta _{34}\right) \left(3 \varepsilon -\Delta _{34}\right) \left(4 \varepsilon -\Delta _{34}\right)}\nonumber\\
    &\;\hspace{2cm}+\frac{\left(\Delta _{12}+\varepsilon \right) \left(\Delta _{12}-\Delta _{34}+2 \varepsilon \right)}{4 \left(2 \varepsilon -\Delta _{34}\right) \left(3 \varepsilon -\Delta _{34}\right)}\frac{V-1}{U}+\frac{\Delta _{12}-\Delta _{34}+2 \varepsilon }{2 \left(\Delta _{34}-2 \varepsilon\right)}\frac{V+1}{2U}
    \Bigg)\frac{\mathcal{T}_{0,1}^{\{k_i\}}}{U^{3(\varepsilon-1)}}\Bigg].
    \label{eq:AE3T4}
\end{align}
While $\TT^{\{k_i\}}_{1,0}$, and $\TT^{\{k_i\}}_{0,1}$ also contribute to the lowest six channels which we have omitted here, the $\TT^{\{k_i\}}_{0,0}$ contributions will remain the same as in the $\EE=2$ cases \eqref{eq:AT1}-\eqref{eq:AT6}. To derived reduced block decompositions of $\TT^{\{k_i\}}_{1,0}$, and $\TT^{\{k_i\}}_{0,1}$ in analogy to Section \ref{reducedblocks}, one would need to devise a way to invert $A^{\{k_{i}\};\TT}_{mm}$ in terms of the (at most) single conformal primary it encodes. If such reduced blocks exist, their form will be constrained by the following observations:
\begin{itemize}
    \item The reduced correlators are interrelated under crossing symmetry. Consider $1\leftrightarrow2$ crossing using the relation 
    \begin{align}
        \Delta_\varepsilon\big|_{(U,V)\rightarrow\left(\frac{U}{V},\frac{1}{V}\right)}=V^{2\varepsilon}\Delta_\varepsilon V^{-\varepsilon-1},
    \end{align}
    in \cite{Dolan:2004mu}. While $\TT^{k_1k_2k_3k_4}_{0,0}$ maps to its crossed form as in $\EE=2$, we have that
    \begin{align}
        \mathcal{T}_{1,0;\Delta,\ell}^{k_1k_2k_3k_4}(U,V)=\mathcal{T}_{0,1;\Delta,\ell}^{k_2k_1k_3k_4}\left(\frac{U}{V},\frac{1}{V}\right).
    \label{eq:T10TO1cross}
    \end{align}
    Assuming reduced blocks for $\TT^{\{k_i\}}_{1,0}$, and $\TT^{\{k_i\}}_{0,1}$ involve a combination of ordinary conformal blocks (which obey a simple relation under $1\leftrightarrow2$ crossing), this relation halves the number of reduced blocks one needs to determine, especially when $k_i=3$. We saw similar identities for the $\HH^{\{k_i\}}_{I,J}$ reduced correlators in 6d in Subsection \ref{6dreducedblocks}.
    \item Superconformal representation theory dictates that among the new multiplets appearing at $\EE=3$, one long multiplet and one semi-short multiplet will not have a block in the $[2m,0]$ channel, so the reduced blocks for these multiplets must satisfy
    \begin{align}
\Delta_\varepsilon\left[\frac{\mathcal{T}_{1,0;\Delta,\ell}^{\{k_i\}}(U,V)}{U^{3(\varepsilon-1)}}\right]=-\frac{1}{V^\varepsilon}\Delta_\varepsilon\left[\frac{\mathcal{T}_{0,1;\Delta,\ell}^{\{k_i\}}(U,V)}{U^{3(\varepsilon-1)}}\right].
\label{eq:E3constraint}
    \end{align}
    Such multiplets instead have a single block in the second highest $[2(m-1),2]$ channel, which simplifies due to Equation \eqref{eq:E3constraint}.
    \item Finally, we notice that 
    \begin{align}
        \sum_{n=m}^{m-3}A^{\{k_{i}\};\TT}_{mn;\Delta,\ell}=U^{3\varepsilon}\Delta_\varepsilon\left[\frac{V}{U}\frac{\mathcal{T}_{1,0;\Delta,\ell}^{\{k_i\}}}{U^{3(\varepsilon-1)}}\right],
        \label{eq:channelsum}
    \end{align}
    which is equal to a complicated combination of blocks. When $\Delta_{12}=0$, one can similarly isolate $\mathcal{T}_{0,1;\Delta,\ell}^{\{k_i\}}$ through a sum over $(-1)^{m-n}A^{\{k_{i}\};\TT}_{mn;\Delta,\ell}$.  
\end{itemize}
Using these observations for the configuration $\langle\OO_{3}\OO_{3}\OO_{3}\OO_{3}\rangle$, one could in principle extract $\mathcal{T}_{1,0;\Delta,\ell}^{3333}$ by inverting \eqref{eq:channelsum} as we did in \eqref{eq:Tat} and then deduce $\mathcal{T}_{0,1;\Delta,\ell}^{3333}$ using \eqref{eq:T10TO1cross}. We hope to find a more transparent presentation of these reduced blocks however.

In closing, we point out that $\EE=3$ correlators will also contain a single-variable reduced correlator $h^{\{k_i\}}_3$ which contributes to all 10 channels and other than in 4d (and possibly 3d), can be absorbed into $\TT^{\{k_i\}}_{1,0}$, and $\TT^{\{k_i\}}_{0,1}$, and $h^{\{k_i\}}_1$.

\subsection{Half-maximally supersymmetric CFTs}
In \cite{Dolan:2004mu}, the authors wrote down an analogous decomposition for solutions to the superconformal Ward identities associated with SCFTs with eight real Poincaré supercharges, i.e. the 6d $\NN=(1,0)$, 5d $\NN=1$, 4d $\NN=2$, and 3d $\NN=4$ theories. In our conventions, this decomposition took the form 
\begin{align}
    \GG^{\{k_i\}}\left(U,V;\alpha\right)=&\;
    \sum_{I=0}^{\EE-2}U^{\varepsilon(I+2)}\alpha^I\Delta_{\varepsilon}\left[\left(z\alpha-1\right)\left(\bar{z}\alpha-1\right)\frac{b^{\{k_i\}}_{I}\left(U,V\right)}{U^{3\varepsilon-1}}\right]\nonumber\\
    +&\;U^\varepsilon\left(D_\varepsilon\right)^{\varepsilon-1}\left[\frac{(z \alpha-1)h^{\{k_{i}\}}(z)-(\bar{z} \alpha-1)h^{\{k_{i}\}}(\bar{z})}{z-\bar{z}}\right],
    \label{eq:DGSdecomphalfsusy}
\end{align}
where the $\mathfrak{so}(3)_R$ R-symmetry algebra factor results in one only having a single R-symmetry cross-ratio $\alpha$. In this expression, \cite{Dolan:2004mu} claim that the single-variable function $h^{\{k_i\}}$ can be absorbed into the two-variable reduced correlator $b^{\{k_i\}}_0$ in all $\varepsilon\neq1$. This function only has an interpretation in 4d and 3d, where it encodes the non-unitary chiral algebra and 1d topological subsectors of these theories, respectively.
The only point we wish to make in this subsection is that the operator $\Delta_\varepsilon$ again appears in these decompositions and the $\mathfrak{so}(3)_R$ channel equations one derives from them. This means that the manipulations leading to Equations \eqref{eq:Tat} and \eqref{eq:h1at} should be equally applicable in finding a reduced block decomposition of $b^{\{k_i\}}_0$, including the new $\varepsilon=\frac{3}{2}$ case of 5d $\NN=1$ SCFTs.

The superblocks of half-maximally supersymmetric CFTs have been written down, e.g. in \cite{Bobev:2017jhk,Chang:2017xmr,Baume:2019aid}. Reduced superblocks in position space have been written down for $\langle\OO_2\OO_2\OO_2\OO_2\rangle$\footnote{In the half-maximally supersymmetric setting, by $\OO_k$ we still refer to a 1/2-BPS operator with $\Delta_k=\varepsilon k$.} in 4d $\NN=2$ theories in \cite{Dolan:2001tt} and for 6d $\NN=(1,0)$ theories in \cite{Chang:2017xmr}. More recently, the authors of \cite{Chester:2025jxg} derived reduced correlators in Mellin space for half-maximally supersymmetric CFTs in $3\leq d\leq 6$ using a strategy introduced in \cite{Virally:2025nnl} that handles the Mellin space analogue of the operator $\Delta_\varepsilon$. The only shortcoming in this work was the lack of a reduced exchange diagram (which in corresponds to a conformal block in position space) for a certain 1/2-BPS multiplet in 3d. This was related to the definition of such a multiplet as a negative $\ell$ limit of a long multiplet, which is hard to interpret in Mellin space. Negative $\ell$ limits are commonplace in position space and it would be interesting to see if this issue can be resolved using the methods in this work\footnote{We are grateful to Shai Chester for discussions on these points.}.

We note however, that the ambiguities in trying to formulate a reduced crossing equation are more severe with half-maximal supersymmetry, where the single R-symmetry cross-ratio $\alpha$ is insufficient to algebraically eliminate the action of $\Delta_\varepsilon$, even when it is a well-defined differential operator in integer $\varepsilon$. This issue was addressed and overcome in the 6d $\NN=(1,0)$ $\langle\OO_2\OO_2\OO_2\OO_2\rangle$ setting in Appendix C of \cite{Chang:2017xmr}. 

\section*{Acknowledgments}

We are very grateful to Costis Papageorgakis, Shai Chester, Franceso Aprile, Himanshu Raj, Paul Richmond, and Vasilis Niarchos for helpful discussions. We also acknowledge the warm hospitality and stimulating environments at both the ``$20^\text{th}$ Asian Winter School
on Strings, Particles, and Cosmology" at IISER Bhopal and ICTS-TIFR Bangalore, where this work was finalized. This project was funded by a Science and Technology Facilities Council (STFC) studentship. 

\begin{appendix}

\section{Superconformal block conventions}
\label{superblockconventions}
In this Appendix, we define our conventions for the $\mathfrak{so}(d+1,1)$ conformal blocks $G^{\Delta_{12},\Delta_{34}}_{\Delta,\ell}$ and $\mathfrak{so}(\texttt{d})_R$ R-symmetry polynomials $Y_{mn}^{a,b}$ that appear throughout the main text. Conformal blocks encode the contribution of a conformal primary and all conformal descendants and can be defined as solutions to the $\mathfrak{so}(d+1,1)$ conformal quadratic Casimir equation \cite{Dolan:2003hv}. Suppressing the dependence on $\varepsilon$, the conformal Casimir equation takes the form
\begin{align}
    \hat{\CC}_{\;\Delta,\ell}^{\Delta_{12},\Delta_{34}}\;G^{\Delta_{12},\Delta_{34}}_{\Delta,\ell}(z,\bar{z})=c_{\Delta,\ell}\;G^{\Delta_{12},\Delta_{34}}_{\Delta,\ell}(z,\bar{z}),
\label{eq:Casimireq}
\end{align}
in terms of the quadratic conformal Casimir operator
\begin{align}
    \hat{\CC}_{\;\Delta,\ell}^{\Delta_{12},\Delta_{34}}=D_z+D_{\bar{z}}+2\varepsilon\frac{z \bar{z}}{z-\bar{z}}\left(\left(1-z\right)\partial_z-\left(1-\bar{z}\right)\partial_{\bar{z}}\right),
\end{align}
where we define\footnote{Note that the operator $D_z$ is not to be confused with the operator $D_\varepsilon$ defined in \eqref{eq:Depsilon}.}
\begin{align}
    D_z=z^2\partial_{z}\left(1-z\right)\partial_z+\frac{\Delta_{12}-\Delta_{34}}{2}z^2\partial_z+\frac{\Delta_{12}\Delta_{34}}{4}z.
\end{align}
The Casimir operator has eigenvalues
\begin{align}
    c_{\Delta,\ell}=\frac{1}{2}\left(\ell\left(\ell+2\varepsilon\right)+\Delta\left(\Delta-2\varepsilon-2\right)\right),
\end{align}
which are independent of $\Delta_{ij}$. Analytic solutions to the conformal Casimir equation can be found in even dimensions, such as for when $d=4,6$ in \cite{Dolan:2000ut,Dolan:2003hv}. When $d=4$, we have that 
\begin{align}
&\;G_{\Delta, \ell}^{\Delta_{12}, \Delta_{34}}(z, \bar{z})= \nonumber \\
&\;\frac{(z \bar{z})^{\frac{\Delta-\ell}{2}}}{z-\bar{z}}\left((-z)^{\ell} z\;{ }_2 F_1\left(\frac{\Delta+\ell-\Delta_{12}}{2}, \frac{\Delta+\ell+\Delta_{34}}{2}, \Delta+\ell; z\right)\right. \nonumber \\
&\;\hspace{2cm}\times\left.{}_2 F_1\left(\frac{\Delta-\ell-2-\Delta_{12}}{2}, \frac{\Delta-\ell-2+\Delta_{34}}{2}, \Delta-\ell-2; \bar{z}\right)-(z \leftrightarrow \bar{z})\right).
\end{align}
In $d=6$, we have that 
\begin{align}
&G_{\Delta, \ell}^{\Delta_{12}, \Delta_{34}}(z, \bar{z})= \nonumber \\
&\mathcal{F}_{0,0}-\frac{(\ell+3)}{\ell+1} \mathcal{F}_{-1,1}+\frac{2(\Delta-4) \Delta_{12} \Delta_{34}(\ell+3)}{(\Delta+\ell)(\Delta+\ell-2)(\Delta-\ell-4)(\Delta-\ell-6)} \mathcal{F}_{0,1} \nonumber \\
&+\frac{(\Delta-4)(\ell+3)\left(\Delta-\Delta_{12}-\ell-4\right)\left(\Delta+\Delta_{12}-\ell-4\right)\left(\Delta+\Delta_{34}-\ell-4\right)\left(\Delta-\Delta_{34}-\ell-4\right)}{16(\Delta-2)(\ell+1)(\Delta-\ell-5)(\Delta-\ell-4)^2(\Delta-\ell-3)} \mathcal{F}_{0,2} \nonumber \\
&-\frac{(\Delta-4)\left(\Delta-\Delta_{12}+\ell\right)\left(\Delta+\Delta_{12}+\ell\right)\left(\Delta+\Delta_{34}+\ell\right)\left(\Delta-\Delta_{34}+\ell\right)}{16(\Delta-2)(\Delta+\ell-1)(\Delta+\ell)^2(\Delta+\ell+1)} \mathcal{F}_{1,1},
\end{align}
\newpage
\noindent
in terms of 
\begin{align}
&\mathcal{F}_{m,n}(z, \bar{z}) = \nonumber \\
&\frac{(z \bar{z})^{\frac{\Delta-\ell}{2}}}{(z-\bar{z})^3}\left((-z)^{\ell} z^{m+3} \bar{z}^n{ }_2 F_1\left(\frac{\Delta+\ell-\Delta_{12}}{2}+m, \frac{\Delta+\ell+\Delta_{34}}{2}+m, \Delta+\ell+2 m ; z\right)\right. \nonumber \\
&\hspace{1.5cm}\times\left.{}_2 F_1\left(\frac{\Delta-\ell-\Delta_{12}}{2}-3+n, \frac{\Delta-\ell+\Delta_{34}}{2}-3+n, \Delta-\ell-6+2 n ; \bar{z}\right)-(z \leftrightarrow \bar{z})\right).
\end{align}
These conventions were chosen such that for any $\varepsilon$, the leading small $z,\bar{z}$ behavior is
\begin{align}
    G_{\Delta, \ell}^{\Delta_{12}, \Delta_{34}}(z, \bar{z})\sim(-1)^\ell z^{\frac{\Delta-\ell}{2}}\bar{z}^{\frac{\Delta+\ell}{2}}\hspace{1cm}\text{as }z,\bar{z}\rightarrow0.
\end{align}
Analytic formulae for conformal blocks in odd dimensions are not known, although in Appendix \ref{jackpolys} we will express conformal blocks in general dimensions $d=2(\varepsilon+1)$ in infinite expansions of two-variable symmetric functions known as Jack polynomials\footnote{We note that the author of \cite{Song:2025avg} presented a 3d conformal block in terms of fractional derivatives acting on a product of $_4F_3$ hypergeometric functions, though we will not use this result here.}. It is also useful to point out that the Casimir equation \eqref{eq:Casimireq} implies relations between conformal blocks $G_{\Delta, \ell}^{\Delta_{12}, \Delta_{34}}$ and $G_{\Delta, 2\varepsilon-\ell}^{\Delta_{12}, \Delta_{34}}$. This is useful for our construction of superconformal blocks in Subsection \ref{superblockhowto}, where we will run into conformal blocks with negative spin $\ell$. We use that in our conventions,
\begin{align}
    &G^{\Delta_{12}, \Delta_{34}}_{\Delta,-\varepsilon}(U,V)=0,\hspace{1cm}G^{\Delta_{12}, \Delta_{34}}_{\Delta,-\ell}(U,V)=(-1)^{2 \varepsilon }
    \frac{\Gamma (\ell -2 \varepsilon +1) \Gamma (\varepsilon-\ell+1)}{\Gamma (1-\ell ) \Gamma (\ell -\varepsilon +1)}G^{\Delta_{12}, \Delta_{34}}_{\Delta,\ell-2\varepsilon}(U,V).
\end{align}
\par
The harmonic functions $Y_{mn}^{a,b}$ are the analogue of conformal blocks for the compact R-symmetry algebra $\mathfrak{so}(\texttt{d})_R$ and can be computed using the scheme described in \cite{Nirschl:2004pa,Alday:2020dtb}. Explicitly, $Y_{mn}^{a,b}(\sigma,\tau)$ are polynomials 
\begin{align}
Y_{mn}^{a,b}(\sigma,\tau)=\sum_{i=0}^{m}\sum_{j=0}^{m-i}c_{i,j}\sigma^i\tau^j,
\end{align}
where we solve for the coefficients $c_{i,j}$ using the $SO(\text{\texttt{d}})$ Casimir eigenfunction equation
\begin{align}
    L^2Y_{mn}^{a,b}(\sigma,\tau)=-2C^{a,b}_{mn}Y_{mn}^{a,b}(\sigma,\tau),
\end{align}
with quadratic Casimir operator\footnote{We correct a typo in the definition of $L^2$ in \cite{Alday:2020dtb} as compared with \cite{Nirschl:2004pa}.}
\begin{align}
    L^2=2\mathcal{D}_\text{\texttt{d}}^{a,b}-(a+b)(a+b+\text{\texttt{d}}-2),
\end{align}
in terms of operators
\begin{align}
    \mathcal{D}_\text{\texttt{d}}^{a,b}=&\;\mathcal{D}_\text{\texttt{d}}+(1-\sigma-\tau)(a\partial_\tau+b\partial_\sigma)-2a\sigma\partial_\sigma-2b\tau\partial_\tau,  \\
    \mathcal{D}_\text{\texttt{d}}=&\;(1-\sigma-\tau)(\partial_\sigma\sigma\partial_\sigma+\partial_\tau\tau\partial_\tau)-4\sigma\tau\partial_\sigma\partial_\tau-(\text{\texttt{d}}-2)(\sigma\partial_\sigma+\tau\partial_\tau),
\end{align}
and eigenvalues
\begin{align}
    C_{mn}^{a,b}=\left(m+\frac{a+b}{2}\right)\left(m+\frac{a+b}{2}+\text{\texttt{d}}-3\right)+\left(n+\frac{a+b}{2}\right)\left(n+\frac{a+b}{2}+1\right).
\end{align}
We normalize $Y_{mn}^{a,b}(\sigma,\tau)$ such that the $\sigma^m$ term has coefficient $c_{m,0}=1$. For consistency with crossing symmetry, we must normalize solutions corresponding to $1\leftrightarrow2$ crossed channels such that the $\tau^m$ term has coefficient $c_{0,m}=1$.

\section{Jack Polynomials}
\label{jackpolys}
Jack polynomials $P^{(\varepsilon)}_{a,b}$ form a useful basis for expanding symmetric 2-variable functions, such as conformal blocks and the reduced correlators in the main text. They can be defined in terms of Gegenbauer polynomials as
\begin{align}
    P^{(\varepsilon)}_{a,b}(z,\bar{z})=\frac{(a-b)!}{\left(2\varepsilon\right)_{a-b}}\left(z \bar{z}\right)^{\frac{1}{2}\left(a+b\right)}C^{(\varepsilon)}_{a-b}\left(\frac{z+\bar{z}}{2\left(z \bar{z}\right)^{1/2}}\right),
\end{align}
and satisfy an orthogonality condition given by
\begin{align}
    \int^{1}_{-1}C_m^{(\varepsilon)}(x)C_n^{(\varepsilon)}(x)\left(1-x^2\right)^{\varepsilon-\frac{1}{2}}dx=\delta_{m,n}\frac{2^{1-2\varepsilon}\pi\Gamma\left[m+2\varepsilon\right]}{m!(m+\varepsilon)\Gamma\left[\varepsilon\right]^2}.
\end{align}
For $d=4,6$, Jack polynomials can be written explicitly as
\begin{align}
    P^{\left(1\right)}_{a,b}(z,\bar{z})=&\; \frac{1}{\left(a-b+1\right)\left(z-\bar{z}\right)}\left(z^{a+1}\bar{z}^{b}-z^{b}\bar{z}^{a+1}\right),\\
    P^{(2)}_{a,b}(z,\bar{z})=&\;\frac{6}{\left(a-b+2\right)\left(z-\bar{z}\right)^3}\left(\frac{z^{a+3}\bar{z}^{b}-z^b\bar{z}^{a+3}}{a-b+3}-\frac{z^{a+2}\bar{z}^{b+1}-z^{b+1}\bar{z}^{a+2}}{a-b+1}\right).
\end{align}
It will be useful to note several identities for Jack polynomials with any $\varepsilon$ when multiplied by powers of $U=z \bar{z}$ and $V=(1-z)(1-\bar{z})$. Using that
\begin{align}
    U^fP^{(\varepsilon)}_{a,b}=&\;P^{(\varepsilon)}_{a+f,b+f},\\
    (z+\bar z)P^{(\varepsilon)}_{a,b}=&\;\frac{a-b+2\varepsilon}{a-b+\varepsilon}P^{(\varepsilon)}_{a+1,b}+\frac{a-b}{a-b+\varepsilon}P^{(\varepsilon)}_{a,b+1},
\end{align}
we find that
\begin{align}
    V P^{(\varepsilon)}_{a,b}=&\;P^{(\varepsilon)}_{a+1,b+1}-\frac{a-b+2\varepsilon}{a-b+\varepsilon}P^{(\varepsilon)}_{a+1,b}-\frac{a-b}{a-b+\varepsilon}P^{(\varepsilon)}_{a,b+1}+P_{a,b}, \\
    V^2P^{(\varepsilon)}_{a,b}=&\;P^{(\varepsilon)}_{a+2,b+2}+\left(\frac{2 (a-b)}{a-b+\varepsilon }-4\right) P^{(\varepsilon)}_{a+2,b+1}+\frac{2 (b-a) }{a-b+\varepsilon }P^{(\varepsilon)}_{a+1,b+2}\nonumber \\
    +&\;\frac{(a-b+2 \varepsilon ) (a-b+2 \varepsilon +1) }{(a-b+\varepsilon ) (a-b+\varepsilon +1)}P^{(\varepsilon)}_{a+2,b}+\frac{(a-b) (a-b-1) }{(a-b+\varepsilon -1) (a-b+\varepsilon )}P^{(\varepsilon)}_{a,b+2}\nonumber \\
    +&\;\frac{2 \left(4 \varepsilon  (a-b)+2 (a-b)^2 +\varepsilon ^2+\varepsilon -2\right) }{(a-b+\varepsilon -1) (a-b+\varepsilon +1)}P^{(\varepsilon)}_{a+1,b+1}\nonumber \\
    +&\;\left(\frac{2 (a-b)}{a-b+\varepsilon }-4\right) P^{(\varepsilon)}_{a+1,b}+\frac{2 (b-a) }{a-b+\varepsilon }P^{(\varepsilon)}_{a,b+1}+P^{(\varepsilon)}_{a,b}.
\end{align}
We can write more generally that 
\begin{align}
    V^{f}P^{(\varepsilon)}_{a,b}=&\sum_{i=0}^f\sum_{j=0}^f \VV^{f,i,j}_{a-b}P^{(\varepsilon)}_{a+i,b+j},
\end{align}
where in 4d we found that for $f\geq0$,
\begin{align}
    \VV^{f,i,j}_{\beta}=(-1)^{i+j}\binom{f}{i}\binom{f}{j}\frac{\beta+1+i-j}{\beta+1},
\end{align}
while in 6d,
\begin{align}
    \VV^{f,i,j}_{\beta}=(-1)^{i+j}\binom{f}{i}\binom{f}{j}\frac{\left((\beta+4+i-j)\beta+3+i-3j+\frac{2ij}{f}\right)\left(\beta+2+i-j\right)}{\left(\beta+1\right)\left(\beta+2\right)\left(\beta+3\right)}.
\end{align}
As discussed in the main text, conformal blocks in $d=2(1+\varepsilon)$ dimensions can be expressed in the expansion
\begin{align}
    G^{\Delta_{12},\Delta_{34}}_{\Delta,\ell}(z,\bar{z})=(-1)^{\ell}\sum_{m=0}^\infty\sum_{n=0}^\infty \;r_{m,n;\Delta,\ell}^{\Delta_{12},\Delta_{34}}\;P^{(\varepsilon)}_{\frac{\Delta+\ell}{2}+m,\frac{\Delta-\ell}{2}+n}(z,\bar{z}).
\label{eq:Gjackpoly}
\end{align}
The action of the conformal quadratic Casimir operator $\hat{\CC}_{\;\Delta,\ell}^{\Delta_{12},\Delta_{34}}$ on Jack polynomials
\begin{align}
    \hat{\CC}_{\Delta,\ell}^{\Delta_{12},\Delta_{34}}\;P^{(\varepsilon)}_{a,b}=&\;\left(a(a-1)+b(b-1-2\varepsilon)\right)P^{(\varepsilon)}_{a,b}\nonumber \\
    -&\;\frac{a-b+2\varepsilon}{a-b+\varepsilon}\left(a-\frac{\Delta_{12}}{2}\right)\left(a+\frac{\Delta_{34}}{2}\right)P^{(\varepsilon)}_{a+1,b}\nonumber \\
    -&\;\frac{a-b}{a-b+\varepsilon}\left(b-\frac{\Delta_{12}}{2}-\varepsilon\right)\left(b+\frac{\Delta_{34}}{2}-\varepsilon\right)P^{(\varepsilon)}_{a,b+1}, 
\end{align}
\noindent
implies that the expansion coefficients $r_{m,n;\Delta,\ell}^{\Delta_{12},\Delta_{34}}$ satisfy
\begin{align}
    &\left(m\left(\Delta+\ell+m-1\right)+n\left(\Delta-\ell+n-1-2\varepsilon\right)\right)r^{\Delta_{12},\Delta_{34}}_{m,n;\Delta,\ell}=\nonumber \\
    &\;\;\;\;\;\;\;\;\frac{\ell+m-n-1+2\varepsilon}{\ell+m-n-1+\varepsilon}\left(\frac{\Delta+\ell-\Delta_{12}}{2}+m-1\right)\left(\frac{\Delta+\ell+\Delta_{34}}{2}+m-1\right)r^{\Delta_{12},\Delta_{34}}_{m-1,n;\Delta,\ell}\nonumber \\
    &\;\;\;\;+\frac{\ell+m-n+1}{\ell+m-n+1+\varepsilon}\left(\frac{\Delta-\ell-\Delta_{12}}{2}+n-1-\varepsilon\right)\left(\frac{\Delta-\ell+\Delta_{34}}{2}+n-1-\varepsilon\right)r^{\Delta_{12},\Delta_{34}}_{m,n-1;\Delta,\ell}.
\end{align}
This recursion relation was solved in \cite{Dolan:2003hv} so that the coefficients take the form\footnote{In this expression, we corrected some apparent typos in \cite{Dolan:2003hv} and checked using \texttt{Mathematica} that this form reproduces the other results appearing in that work. In practice, one needs to regularize the ${}_4F_3$ to obtain sensible answers for specific $\varepsilon$ and integer $\ell$.}
\begin{align}
    r^{\Delta_{12},\Delta_{34}}_{m,n;\Delta,\ell}=&\;\left(\frac{\Delta+\ell-\Delta_{12}}{2} \right)_m \left(\frac{\Delta+\ell +\Delta_{34}}{2}\right)_m \left(\frac{\Delta-\ell -\Delta_{12}-2 \varepsilon}{2}\right)_n \left(\frac{\Delta-\ell +\Delta_{34}-2 \varepsilon}{2}\right)_n\nonumber \\
    \times&\;\frac{(2 \varepsilon )_\ell}{(\varepsilon )_\ell}\frac{(\varepsilon +\ell+m-n) }{(\varepsilon +\ell+m)}
    \frac{ (\varepsilon +1-\Delta )_\ell (\ell+m)! (2 \varepsilon )_{\ell+m-n}}{ m! n! (2 \varepsilon +1-\Delta )_\ell  (\ell+m-n)! (\Delta+\ell )_m (\varepsilon )_{\ell+m} (\Delta-\ell -\varepsilon )_n}\nonumber \\
    \times&\;\, _4F_3\left(
    \begin{matrix}
        \varepsilon, -\ell-m+n,-\ell, \Delta -1\\
        2 \varepsilon, -\ell-m, -\ell+n+\Delta -\varepsilon
    \end{matrix}
;1\right).
\label{eq:rcoeff}
\end{align}
Jack polynomials also have the useful property that they are eigenfunctions of the operator $\Delta_\varepsilon$, satisfying 
\begin{align}
    \Delta_f P^{(\varepsilon)}_{a,b}=(a+\varepsilon+1)_{f-1}(b+1)_{f-1}P^{(\varepsilon)}_{a,b}.
\label{eq:eigenjackpoly}
\end{align}
We used these properties to re-derive and generalize the recurrence relations for conformal blocks in Appendix \ref{Trecurrencerelations}, as well as to compute the reduced block $\mathcal{T}^{\{k_i\}}_{\Delta,\ell}$ in Equation \eqref{eq:Tat}.

\section{Recurrence relations for $h^{\{k_i\}}_{1,\ell}(z)$, conformal blocks, and $\left(D_{\varepsilon}\right)^{\varepsilon-1}$}
\label{hrecurrencerelations}
In this Appendix, we describe the way in which the contributions of certain protected supermultiplets to correlators $\langle\OO_{k_1}\OO_{k_2}\OO_{k_3}\OO_{k_1+k_2+k_3-2\EE}\rangle$ are encoded by the single-variable reduced correlators $h^{\{k_i\}}_1$. We will demonstrate how we can use the particularly simple form of the highest $\mathfrak{so}(\texttt{d})_R$ channel equation to generate (a subset of) the superdescendant primary content of the rest of the supermultiplet. Some of the formulae in this section are long-winded and we include all of them in the ancillary \texttt{Mathematica} notebook attached to our arXiv submission.

An important simplification in the Jack polynomial expansions of conformal blocks occurs for certain low integer twists, such as those appearing in the superblocks associated with certain protected supermultiplets. For twists $t=\Delta-\ell=2\varepsilon-\Delta_{34}$, the Jack polynomial expansion coefficients $r^{\Delta_{12},\Delta_{34}}_{m,n;\Delta,\ell}$ develop a factor of $\delta_{n,0}$, which collapses the expansion in \eqref{eq:Gjackpoly} to the single sum
\begin{align}
    G^{\Delta_{12},\Delta_{34}}_{\ell+2\varepsilon-\Delta_{34}+\alpha,\ell+\alpha}(z,\bar{z})=U^{\varepsilon-\frac{\Delta_{34}}{2}}(-1)^{\ell+\alpha}\sum_{m=0}^\infty \;\tilde{r}_{m;\ell+2\varepsilon-\Delta_{34}+\alpha,\ell+\alpha}^{\Delta_{12},\Delta_{34}}\;P^{(\varepsilon)}_{\ell+\alpha+m,0}(z,\bar{z}),
\label{eq:restrictedGjackpolyalpha}
\end{align}
with the restricted expansion coefficients
\begin{align}
    \tilde{r}_{m;\ell+2\varepsilon-\Delta_{34}+\alpha,\ell+\alpha}^{\Delta_{12},\Delta_{34}}=&\;
    \frac{(2 \varepsilon)_{\ell+\alpha}}{(\varepsilon )_{\ell+\alpha}}
    \frac{\left(\Delta _{34}-\ell -\alpha -\varepsilon+1\right)_{\ell+\alpha}}{\left(\Delta _{34}-\ell -\alpha+1\right)_{\ell +\alpha }}\nonumber \\
    \times&\;
    \frac{(2 \varepsilon )_{m+\ell+\alpha}}{(\varepsilon )_{m+\ell+\alpha}}
    \frac{(\ell +\alpha +\varepsilon )_m  \left(\ell +\alpha +\varepsilon -\frac{\Delta _{12}}{2}-\frac{\Delta _{34}}{2}\right)_m}{m!\left(2 \left(\ell +\alpha +\varepsilon-\frac{\Delta _{34}}{2}\right)\right)_m}\nonumber\\
    \times&\;\, _3F_2\left(
    \begin{matrix}
    \varepsilon, -\ell-\alpha, \ell +\alpha +2 \varepsilon -\Delta _{34}-1\\
    2 \varepsilon ,\varepsilon -\Delta _{34}
    \end{matrix};1\right).
\label{eq:restrictedcoeff}
\end{align}
This simplification is useful for manipulating the $h^{\{k_{i}\}}_{1}$ functions appearing throughout the main text. As described in Section \ref{E1}, we can decompose $h^{\{k_{i}\}}_{1}$ into ``blocks" that account for its contribution to each supermultiplet $\XX$ by writing
\begin{align}
    h^{\{k_{i}\}}_1(z)=\sum_{\mathcal{X}\in (\OO_1\times \OO_2)\cap(\OO_3\times \OO_4)}\lambda_{k_1 k_2 \mathcal{X}}\lambda_{k_3 k_4 \mathcal{X}}\;\NN_\XX\;h^{\{k_{i}\}}_{1,\ell_\XX}\left(z\right).
\end{align}
where the label $\ell_\XX$ is related to the quantum numbers of the superconformal primary of $\XX$ and $\NN_\XX$ is a superblock normalization constant. 

The selection rules in Equations \eqref{eq:3dsuperselect}-\eqref{eq:6dsuperselect} demonstrate that extremality-$\EE$ correlators of the form $\langle\OO_{k_1}\OO_{k_2}\OO_{k_3}\OO_{k_1+k_2+k_3-2\EE}\rangle$ exchange protected semi-short multiplets defined by superconformal primaries with quantum numbers $\left(\Delta,\ell\right)=\left(\ell+2\varepsilon-\Delta_{34},\ell\right)$, where we abbreviated $\Delta_{12}=\varepsilon k_{12}$ and $\Delta_{34}=\varepsilon\left(2\EE-k_{1}-k_2\right)$. These supermultiplets contain a single conformal block in the highest $\mathfrak{so}(\texttt{d})_R$ representation $\left[k_1+k_2-2\EE,0\right]$ with quantum numbers $\left(\Delta',\ell'\right)=\left(\Delta+2,\ell+2\right)$. This is seen in the $\mathfrak{so}(\texttt{d})_R$ channel equation $A^{\{k_{i}\};\XX;h_1}_{\tilde{m}\;\tilde{m};\Delta',\ell'}$ where we abbreviated the quantity $\tilde{m}\left(\EE\right)=\frac{k_1+k_2-2\EE}{2}$, which has the universal single-variable contribution 
\begin{align}
    A^{\{k_{i}\};\XX;h_1}_{\tilde{m}\;\tilde{m};\Delta',\ell'}=&\;\frac{1}{\mathcal{Y}_{\tilde{m}\;\tilde{m}}^{k_{12},k_{34}}}U^{\varepsilon}\left(D_\varepsilon\right)^{\varepsilon-1}\left[\frac{z^2h^{\{k_{i}\}}_{1,\ell'}(z)-\bar{z}^2h^{\{k_{i}\}}_{1,\ell'}(\bar{z})}{z-\bar{z}}\right]=\mathcal{C}^{\mathcal{X},\Delta_{12},\Delta_{34}}_{\mathcal{O}_{\Delta',\ell',\tilde{m}\;\tilde{m}}}\;U^{\frac{\Delta_{34}}{2}}\;G^{\Delta_{12},\Delta_{34}}_{\Delta',\ell'}.
\label{eq:topAh1}
\end{align}
With the observation in \cite{Dolan:2004mu} that
\begin{align}
    P^{(\varepsilon)}_{m,0}(z,\bar{z})=\frac{(-1)^{\varepsilon+1}\Gamma\left[m+1\right]}{\left(2\varepsilon\right)_{m}\Gamma\left[\varepsilon\right]^2}\left(D_\varepsilon\right)^{\varepsilon-1}\left[\frac{z^{m+2\varepsilon-1}-\bar{z}^{m+2\varepsilon-1}}{z-\bar{z}}\right],
\end{align}
we can use Equations \eqref{eq:restrictedcoeff} and \eqref{eq:topAh1} to determine that the contribution of protected multipets with superconformal primary $\left(\Delta,\ell\right)=\left(\ell+2\varepsilon-\Delta_{34},\ell\right)$ is encoded by the single-variable reduced block $h^{\{k_i\}}_{1,\ell+2}$ where we have defined
\begin{align}
    &\;h^{\{k_i\}}_{1,\ell}(z)
    =(-1)^{\varepsilon+1}\frac{ \Gamma \left[\ell+1\right]}{\Gamma [\varepsilon] \Gamma\left[\ell+\varepsilon\right]}z^{\frac{\Delta _{34}}{2}+\varepsilon-3}g^{\Delta_{12},\Delta_{34}-2(\varepsilon-1)}_{\ell+2\varepsilon-\Delta_{34},\ell}(z),
\label{eq:h1at}
\end{align}
with $g^{\Delta_{12},\Delta_{34}}_{\Delta,\ell}$ being the global $SL(2,\mathbb{R})$ block 
\begin{align}
    g_{\Delta,\ell}^{\Delta_{12},\Delta_{34}}(z)=(-1)^\ell z^{\frac{\Delta+\ell}{2}}{ }_2 F_1\left(\frac{\Delta+\ell}{2}-\frac{\Delta_{12}}{2},\frac{\Delta+\ell}{2}+\frac{\Delta_{34}}{2},\Delta+\ell,z\right).
\label{eq:sl2Rblock}
\end{align}
As a convention, we have left out the factors of $\mathcal{Y}_{\tilde{m}}^{k_{12},k_{34}}$ and $\mathcal{C}^{\mathcal{X},\Delta_{12},\Delta_{34}}_{\mathcal{O}_{\Delta',\ell',\tilde{m}\;\tilde{m}}}$ appearing in Equation \eqref{eq:topAh1} for later convenience. The reduced block $h^{\{k_i\}}_{1,\ell}$ resembles the global $SL(2,\mathbb{R})$ block found in a chiral algebra correlator with $\left(\tilde{\Delta}_{12},\tilde{\Delta}_{34}\right)=\left(\Delta_{12},\Delta_{34}-2\left(\varepsilon-1\right)\right)$. While this is natural given the chiral algebras obtainable by twists of 4d $\NN=4$ and 6d $\NN=(2,0)$ SCFTs, the analogous twist in 3d $\NN=8$ SCFTs produces a 1d topological quantum mechanics theory with no chiral algebra interpretation.

The interdependence of R-symmetry channels is made apparent by the appearance of $h^{\{k_i\}}_{1,\ell}$ in the other channels \eqref{eq:AE1h2} and \eqref{eq:AE1h3}. To determine the spectrum of conformal primaries that $h^{\{k_i\}}_{1,\ell}$ generates in these equations, we need to compute the other two combinations in \eqref{eq:hrecuringredients}. Using the definition
\begin{align}
    U^{\varepsilon-\frac{\Delta_{34}}{2}}\left(D_\varepsilon\right)^{\varepsilon-1}\left[\frac{z^2h^{\{k_{i}\}}_{1,\ell'}(z)-\bar{z}^2h^{\{k_{i}\}}_{1,\ell'}(\bar{z})}{z-\bar{z}}\right]=G^{\Delta_{12},\Delta_{34}}_{\Delta',\ell'},
\end{align}
we can examine $\frac{z(z-2)}{2}h^{\{k_i\}}_{1,\ell'}(z)$ and $(1-z)h^{\{k_i\}}_{1,\ell'}(z)$ to find that
\begin{align}
    &U^{\varepsilon-\frac{\Delta_{34}}{2}}\left(D_\varepsilon\right)^{\varepsilon-1}\left[\frac{z(z-2)h^{\{k_{i}\}}_{1,\ell'}(z)-\bar{z}(\bar{z}-2)h^{\{k_{i}\}}_{1,\ell'}(\bar{z})}{2(z-\bar{z})}\right]=\nonumber\\
   &\frac{\ell +2}{\varepsilon +\ell +1}G^{\Delta_{12},\Delta_{34}}_{\Delta'-1,\ell'-1}
   +\frac{\Delta _{12} \left(2 \varepsilon -2-\Delta _{34}\right)}{2 \left(2 (\varepsilon +\ell +1)-\Delta _{34}\right) \left(2 (\varepsilon +\ell +2)-\Delta _{34}\right)}G^{\Delta_{12},\Delta_{34}}_{\Delta',\ell'}\nonumber\\
    +&
    \frac{(\varepsilon +\ell +2) \left(2 \varepsilon +\ell +1-\Delta_{34}\right) \left(2 (\varepsilon +\ell +2)-\Delta _{12}-\Delta _{34}\right) \left(\Delta _{12}-\Delta _{34}+2 (\varepsilon +\ell +2)\right)}{4 \left(2 \varepsilon +2 \ell +3-\Delta _{34}\right) \left(2 \varepsilon +2 \ell +5-\Delta _{34}\right) \left(2 (\varepsilon +\ell +2)-\Delta _{34}\right){}^2}G^{\Delta_{12},\Delta_{34}}_{\Delta'+1,\ell'+1},\\
    &U^{\varepsilon-\frac{\Delta_{34}}{2}}\left(D_\varepsilon\right)^{\varepsilon-1}\left[\frac{(1-z)h^{\{k_{i}\}}_{1,\ell'}(z)-(1-\bar{z})h^{\{k_{i}\}}_{1,\ell'}(\bar{z})}{z-\bar{z}}\right]=\nonumber\\
    &\frac{(\ell +1) (\ell +2)}{(\varepsilon +\ell ) (\varepsilon +\ell +1)}G^{\Delta_{12},\Delta_{34}}_{\Delta'-2,\ell'-2}
    +\frac{\Delta _{12}\left(2 \varepsilon -2-\Delta _{34}\right)(\ell +2)}{(\varepsilon +\ell +1) \left(2 (\varepsilon +\ell )-\Delta _{34}\right) \left(2 (\varepsilon +\ell +2)-\Delta _{34}\right)}G^{\Delta_{12},\Delta_{34}}_{\Delta'-1,\ell'-1}\nonumber \\
    -&\;\frac{1}{8} \Bigg(\frac{2 \Delta _{12}^2 (\ell +2)^2}{2 (\varepsilon +\ell +1)-\Delta _{34}}-\frac{2 \Delta _{12}^2 (\ell +3)^2}{2 (\varepsilon +\ell +2)-\Delta _{34}}+\frac{\left(\Delta _{12}^2-1\right) (\ell +3) (\ell +4)}{2 \varepsilon +2 \ell +5-\Delta _{34}}+2\nonumber \\
    &\hspace{9cm}+\frac{\left(\Delta _{12}^2-1\right) (\ell +1) (\ell +2)}{\left(\Delta _{34}-2 \varepsilon -2 \ell -1\right)}\Bigg)G^{\Delta_{12},\Delta_{34}}_{\Delta',\ell'}\nonumber \\
    +&\;\frac{\Delta _{12} \left(2 \varepsilon -2-\Delta _{34}\right)\left(2 \varepsilon +\ell +1-\Delta _{34}\right)(\varepsilon +\ell +2)}{4 \left(2 \varepsilon +2 \ell +3-\Delta _{34}\right) \left(2 \varepsilon +2 \ell +5-\Delta _{34}\right)}\nonumber \\
    &\times\frac{\left(2 (\varepsilon +\ell +2)-\Delta _{12}-\Delta _{34}\right) \left(2 (\varepsilon +\ell +2)+\Delta _{12}-\Delta _{34}\right)}{ \left(2 (\varepsilon +\ell +1)-\Delta _{34}\right) \left(2 (\varepsilon +\ell +3)-\Delta _{34}\right) \left(\Delta _{34}-2 (\varepsilon +\ell +2)\right){}^2}G^{\Delta_{12},\Delta_{34}}_{\Delta'+1,\ell'+1}\nonumber \\
    +&\;\frac{(\varepsilon +\ell +2) (\varepsilon +\ell +3) \left(2 \varepsilon +\ell +1-\Delta _{34}\right) \left(2 \varepsilon +\ell +2-\Delta _{34}\right)\left(2 (\varepsilon +\ell +2)-\Delta _{12}-\Delta _{34}\right) }{16 \left(2 \varepsilon +2 \ell +3-\Delta _{34}\right) \left(2 \varepsilon +2 \ell +5-\Delta _{34}\right)^2}\nonumber \\
    &\times
    \frac{\left(2 (\varepsilon +\ell +3)-\Delta _{12}-\Delta _{34}\right)\left(2 (\varepsilon +\ell +2)+\Delta _{12}-\Delta _{34}\right) \left(2 (\varepsilon +\ell +3)+\Delta _{12}-\Delta _{34}\right)}{ \left(2 \varepsilon +2 \ell +7-\Delta _{34}\right) \left(2 (\varepsilon +\ell +2)-\Delta _{34}\right)^2 \left(2 (\varepsilon +\ell +3)-\Delta _{34}\right)^2}G^{\Delta_{12},\Delta_{34}}_{\Delta'+2,\ell'+2}.
\end{align}
These expressions are used in the main text to generate (or cancel) twist $t=2\varepsilon-\Delta_{34}$ blocks in the superblocks derived in Sections \ref{E1} and \ref{E2}. We will see that $h_{1,\ell}^{\{k_i\}}$ is needed in all protected supermultiplets that give rise to twist $2\varepsilon-\Delta_{34}$ operators, i.e. beyond just the supermultipet with superconformal primary $\left(\Delta,\ell\right)=\left(\ell+2\varepsilon-\Delta_{34},\ell\right)$ discussed previously. As our terminology has suggested, the function $h_{1,\ell}^{\{k_i\}}$ will define the single-variable reduced blocks in Section \ref{reducedblocks}.

\section{Recurrence relations for $\TT^{\{k_i\}}_{\Delta,\ell}(U,V)$, conformal blocks, and $\Delta_{\varepsilon}$}
\label{Trecurrencerelations}
Here, we generalize the recurrence relations involving bosonic conformal blocks in $d=2(\varepsilon+1)$ dimensions and the operator $\Delta_{\varepsilon}$ to the $\EE=2$ kinematical configuration $\langle\OO_{k_1}\OO_{k_2}\OO_{k_3}\OO_{k_1+k_2+k_3-4}\rangle$. These relations will allow us to use the appearance of the two-variable reduced correlator $\TT^{\{k_i\}}_{0,0}$ in Equations \eqref{eq:AT1}-\eqref{eq:AT6} to generate the full spectrum of conformal primaries from a single, simple channel equation in \eqref{eq:AT1}. The formulae in this section are very long-winded and we include all of them in the ancillary \texttt{Mathematica} notebook attached to our arXiv submission.

The main generalization compared with earlier studies in \cite{Dolan:2004mu,Chester:2014fya} is the more complicated spectrum stemming from the relaxation of the Bose selection rule for the OPE of scalars with unequal dimension. The kinematical factors prefactors that contribute to the superdescendant weighting factors $\mathcal{C}^{\mathcal{X},\Delta_{12},\Delta_{34}}_{\mathcal{O}_{\Delta,\ell,mn}}$ also develop $\Delta_{ij}=\varepsilon k_{ij}$ dependences which complicate them substantially.
Along the way, we correct typos that appeared in both \cite{Dolan:2004mu} and \cite{Chester:2014fya}. In analogy to these earlier works, it will be convenient to define
\begin{align}
    F_{r,s}\left(U,V\right)=&\; D\left(\Delta,\ell,r,s\right)\;G_{\Delta,\ell}^{\Delta_{12},\Delta_{34}}\left(U,V\right),
\end{align}
where we leave the dependence on $\varepsilon$ implicit and we define
\begin{align}
    D\left(\Delta,\ell,r,s\right)=B^{\text{sgn}(r)}_{\frac{\Delta+\ell}{2},r}\;B^{\text{sgn}(s)}_{\frac{\Delta-\ell}{2},s}\;A_{\ell+1,-\frac{1}{2}\left(\text{sgn}\left(r-s\right)-1\right)\left(r-s\right)}\;A_{2-\Delta,-\frac{1}{2}\left(\text{sgn}\left(r+s\right)+1\right)\left(r+s\right)},
\end{align}
in terms of 
\begin{align}
    A_{\lambda,t}=&\;\frac{\left(\lambda+\varepsilon\right)_t\left(\lambda+\varepsilon-1\right)_t}{\left(\lambda\right)_t\left(\lambda+2\varepsilon-1\right)_t}, \\
    B^+_{\lambda,t}=&\;\frac{\left(\lambda-\frac{\Delta _{12}}{2}\right)_t \left(\lambda+\frac{\Delta _{12}}{2}\right)_t \left(\lambda+\frac{\Delta _{34}}{2}\right)_t \left(\lambda+\varepsilon -\frac{\Delta _{34}}{2}-1\right)_t}{(\lambda)_t{}^2 \left(\lambda+\frac{1}{2}\right)_t\left(\lambda-\frac{1}{2}\right)_t},\\
    B^0_{\lambda,t}=&\; 16^{-t} \frac{(\lambda)_t (\lambda+\varepsilon -1)_t}{\left(\lambda+\frac{1}{2}\right)_t\left(\lambda-\frac{1}{2}\right)_t },\\
    B^{-}_{\lambda,t}=&\;\frac{\left(\lambda+\frac{\Delta _{34}}{2}\right)_t}{\left(\lambda-\varepsilon +\frac{\Delta _{34}}{2}+1\right)_t}.
\end{align}
We will suppress dependencies on $U$ and $V$ in the following. To generate blocks in the second and third highest channels in \eqref{eq:AT2} and \eqref{eq:AT3}, we derived that
\begin{align}
    U^{2\varepsilon-\frac{\Delta_{34}}{2}}\;\Delta_\varepsilon\;\frac{V-1}{U}\;\Delta_\varepsilon^{-1}\;U^{-2\varepsilon+\frac{\Delta_{34}}{2}}G_{\Delta,\ell}^{\Delta_{12},\Delta_{34}}=&\;F_{0,-1}+F_{-1,0}+F_{1,0}+F_{0,1}\nonumber \\
    +&\;\alpha_{\Delta,\ell}^{\frac{V-1}{U}}F_{0,0}, \label{eq:Vm1rec}\\
    U^{2\varepsilon-\frac{\Delta_{34}}{2}}\;\Delta_\varepsilon\;\frac{V+1}{2U}\;\Delta_\varepsilon^{-1}\;U^{-2\varepsilon+\frac{\Delta_{34}}{2}}G_{\Delta,\ell}^{\Delta_{12},\Delta_{34}}=&\;F_{-1,-1}+F_{-1,1}+F_{1,-1}+F_{1,1}\nonumber \\
    +&\;\Delta_{12}\left(\varepsilon-1-\frac{\Delta_{34}}{2}\right)\AA_{\Delta+\ell-\varepsilon}\;\left(F_{0,-1}+F_{0,1}\right)\nonumber\\
    +&\;\Delta_{12}\left(\varepsilon-1-\frac{\Delta_{34}}{2}\right)\AA_{\Delta-\ell-3\varepsilon}\;\left(F_{-1,0}+F_{1,0}\right)\nonumber\\
    +&\;\alpha_{\Delta,\ell}^{\frac{V+1}{2U}}\;F_{0,0}.
\end{align}
\newpage
\noindent
To generate blocks in the remaining channels in \eqref{eq:AT4}-\eqref{eq:AT6}, we found more intricately that
\begin{align}
    U^{2\varepsilon-\frac{\Delta_{34}}{2}}\;\Delta_\varepsilon\;\frac{\left(V-1\right)^2}{U^2}\;\Delta_\varepsilon^{-1}\;U^{-2\varepsilon+\frac{\Delta_{34}}{2}}G_{\Delta,\ell}^{\Delta_{12},\Delta_{34}}=&\;F_{0,-2}+F_{-2,0}+F_{0,2}+F_{2,0}\nonumber \\
    +&\;2\left(1-\varepsilon\left(\varepsilon-1\right)\AA_{\ell+1}\right)\left(F_{-1,-1}+F_{1,1}\right)\nonumber\\
    +&\;2\left(1-\varepsilon\left(\varepsilon-1\right)\AA_{2-\Delta}\right)\left(F_{-1,1}+F_{1,-1}\right)\nonumber\\
    +&\;\beta^{\frac{(V-1)^2}{U^2}}_{\Delta,\ell}\;F_{0,-1}+\beta^{\frac{(V-1)^2}{U^2}}_{\Delta,-\ell-2\varepsilon}\;F_{-1,0}\nonumber\\
    +&\;\beta^{\frac{(V-1)^2}{U^2}}_{2\varepsilon+2-\Delta,\ell}\;F_{1,0}+\beta^{\frac{(V-1)^2}{U^2}}_{2\varepsilon+2-\Delta,-\ell-2\varepsilon}\;F_{0,1}\nonumber\\
    +&\;\alpha_{\Delta,\ell}^{\frac{(V-1)^2}{U^2}}F_{0,0},
\end{align}
\begin{align}
    U^{2\varepsilon-\frac{\Delta_{34}}{2}}\;\Delta_\varepsilon\;\frac{V^2-1}{2U^2}\;\Delta_\varepsilon^{-1}\;U^{-2\varepsilon+\frac{\Delta_{34}}{2}}G_{\Delta,\ell}^{\Delta_{12},\Delta_{34}}=&\;F_{-2,-1}
    +F_{-1,-2}+F_{-2,1}
    +F_{1,-2}\nonumber \\
    +&\;
    F_{2,-1}
    +F_{-1,2}
    +F_{2,1}
    +F_{1,2}\nonumber \\
    +&\;\Delta_{12}\left(\varepsilon-1-\frac{\Delta_{34}}{2}\right)\AA_{\Delta+\ell-\varepsilon}\left(F_{0,-2}+F_{0,2}\right)\nonumber\\
    +&\;\Delta_{12}\left(\varepsilon-1-\frac{\Delta_{34}}{2}\right)\AA_{\Delta-\ell-3\varepsilon}\left(F_{-2,0}+F_{2,0}\right)\nonumber\\
    +&\;\Delta_{12}\left(\varepsilon-1-\frac{\Delta_{34}}{2}\right)\left(c_{\Delta,\ell}\;F_{-1,-1}+c_{2+\Delta,\ell}\;F_{1,1}\right)\nonumber\\
    +&\;\Delta_{12}\left(\varepsilon-1-\frac{\Delta_{34}}{2}\right)\left(c_{3-\ell,1-\Delta}\;F_{-1,1}+c_{1-\ell,1-\Delta}\;F_{1,-1}\right)\nonumber\\
    +&\;\beta_{\Delta,\ell}^{\frac{V^2-1}{2U^2}}F_{0,-1}+\beta_{\Delta,-\ell-2\varepsilon}^{\frac{V^2-1}{2U^2}}F_{-1,0}\nonumber\\
    +&\;\beta_{2\varepsilon+2-\Delta,\ell}^{\frac{V^2-1}{2U^2}}F_{1,0}+\beta_{2\varepsilon+2-\Delta,-\ell-2\varepsilon}^{\frac{V^2-1}{2U^2}}F_{0,1}\nonumber\\
    +&\;\alpha_{\Delta,\ell}^{\frac{V^2-1}{2U^2}}F_{0,0}, 
\end{align}
\newpage
\begin{align}
    U^{2\varepsilon-\frac{\Delta_{34}}{2}}\;\Delta_\varepsilon\;\frac{\left(V+1\right)^2}{4U^2}\;\Delta_\varepsilon^{-1}\;U^{-2\varepsilon+\frac{\Delta_{34}}{2}}G_{\Delta,\ell}^{\Delta_{12},\Delta_{34}}=&\;
    F_{-2,-2}+F_{-2,2}+F_{2,-2}+F_{2,2}\nonumber \\
    +&\;2\Delta_{12}\left(\varepsilon-1-\frac{\Delta_{34}}{2}\right)
    \BB_{\Delta+\ell-\varepsilon-2}\left(F_{-1,-2}+F_{-1,2}\right)\nonumber\\
    +&\;2\Delta_{12}\left(\varepsilon-1-\frac{\Delta_{34}}{2}\right)
    \BB_{\Delta-\ell-3\varepsilon-2}\left(F_{-2,-1}+F_{2,-1}\right)\nonumber \\
    +&\;2\Delta_{12}\left(\varepsilon-1-\frac{\Delta_{34}}{2}\right)
    \BB_{\Delta+\ell-\varepsilon}\left(F_{1,-2}+F_{1,2}\right)\nonumber \\
    +&\;2\Delta_{12}\left(\varepsilon-1-\frac{\Delta_{34}}{2}\right)
    \BB_{\Delta-\ell-3\varepsilon}\left(F_{-2,1}+F_{2,1}\right)\nonumber \\    +&\;d_{\Delta,\ell}\left(F_{0,-2}+F_{0,2}\right)+d_{2\varepsilon+2-\Delta,\ell}\left(F_{-2,0}+F_{2,0}\right)\nonumber \\
    +&\;e_{\Delta,\ell}F_{-1,-1}+e_{\Delta+2,\ell}F_{1,1}\nonumber\\
    +&\;e_{1-\ell,1-\Delta}F_{1,-1}+e_{3-\ell,1-\Delta}F_{-1,1}\nonumber\\
    +&\;\beta^{\frac{\left(V+1\right)^2}{4U^2}}_{\Delta,\ell}F_{0,-1}+\beta^{\frac{\left(V+1\right)^2}{4U^2}}_{\Delta,-\ell-2\varepsilon}F_{-1,0}\nonumber\\
    +&\;\beta^{\frac{\left(V+1\right)^2}{4U^2}}_{2\varepsilon+2-\Delta,\ell}F_{1,0}+\beta^{\frac{\left(V+1\right)^2}{4U^2}}_{2\varepsilon+2-\Delta,-\ell-2\varepsilon}F_{0,1}\nonumber\\
    +&\;\alpha_{\Delta,\ell}^{\frac{(V+1)^2}{4U^2}}F_{0,0}.
\end{align}
The recursion relations above are written in terms of several quantities that we define in the following.
\begin{align}
    \AA_{\lambda}=&\;\frac{1}{\left(\lambda+\varepsilon\right)\left(\lambda+\varepsilon-2\right)},\hspace{1cm}\BB_{\lambda}=\frac{1}{\left(\lambda+\varepsilon+2\right)\left(\lambda+\varepsilon-2\right)}, \\
    %
    %
    %
    c_{\Delta,\ell}=&\;-\AA_{\Delta+\ell-\varepsilon}\nonumber\\
    \times&\;\left(\frac{(\ell-1) \ell (\Delta +l)}{(\Delta +\ell-4) (\ell+\varepsilon -1)}+\frac{2 (\Delta -1) \ell}{\Delta -\ell-2 \varepsilon }+\frac{2 (\Delta -4) (\ell+1) (\Delta +l)}{(\Delta +\ell-4) (-\Delta +\ell+2 \varepsilon +4)}-\frac{(\ell+1) (\ell+2)}{\ell+\varepsilon +1}\right),\\
    d_{\Delta,\ell}=&\;\frac{1}{4}
    +\frac{\Delta _{12} \left(\varepsilon -1-\frac{\Delta _{34}}{2}\right) \left(\Delta-\Delta _{12}+\ell-2\right) \left(\Delta+\Delta _{34}+\ell-2 \varepsilon \right)}{2 (\Delta+\ell-4) (\Delta+\ell-2) (\Delta+\ell)}\nonumber\\
    -&\;\frac{\left(\Delta-\Delta _{12}+\ell-4\right) \left(\Delta-\Delta _{12}+\ell-2\right) \left(\Delta+\Delta _{34}+\ell-2 \varepsilon -2\right) \left(\Delta+\Delta _{34}+\ell-2 \varepsilon \right)}{32 (\Delta+\ell-4) (\Delta+\ell-3)}\nonumber \\
    +&\;\frac{\left(\Delta-\Delta _{12}+\ell\right) \left(\Delta-\Delta _{12}+\ell+2\right) \left(\Delta+\Delta _{34}+\ell-2 \varepsilon +2\right) \left(\Delta+\ell-2 \varepsilon +4+\Delta _{34}\right)}{32 (\Delta+\ell) (\Delta+\ell+1)}\nonumber \\
    -&\;\frac{\left(\Delta+\ell-\Delta _{12}\right) \left(\Delta+\ell-2 \varepsilon +2+\Delta _{34}\right)}{4 (\Delta+\ell)},\\
    e_{\Delta,\ell}=&\;\frac{\left(\Delta -\Delta _{12}+\ell \right) \left(\Delta -\Delta _{12}-2 \varepsilon -\ell \right) \left(\Delta +\Delta _{34}-2 \varepsilon +\ell +2\right) \left(\Delta +\Delta _{34}-4 \varepsilon -\ell +2\right)}{16 (\Delta -\varepsilon ) (\Delta +\ell ) (\varepsilon +\ell -1) (\varepsilon +\ell +1) (\Delta -2 \varepsilon -\ell )}\nonumber \\
    \times&\; \left(\Delta  \left(\varepsilon +\ell ^2+2 \varepsilon  \ell -1\right)-\varepsilon  \left(\ell ^2+2 \varepsilon  (\ell +1)-2\right)\right) \nonumber\\
    -&\;\frac{\left(\Delta -\Delta _{12}+\ell -4\right)  \left(\Delta -\Delta _{12}-2 \varepsilon -\ell -4\right) \left(\Delta +\Delta _{34}-4 \varepsilon -\ell -2\right) \left(\Delta +\Delta _{34}-2 \varepsilon +\ell -2\right)}{16 (\Delta -\varepsilon -4) (\Delta +\ell -4) (\varepsilon +\ell -1) (\varepsilon +\ell +1) (\Delta -2 \varepsilon -\ell -4)}\nonumber \\
    \times&\; \left((\varepsilon -1) (\Delta -2 (\varepsilon +2))+\ell ^2 (\Delta -\varepsilon -4)+2 \varepsilon  \ell  (\Delta -\varepsilon -4)\right) \nonumber \\ 
    +&\;\frac{2 \varepsilon ^2+\varepsilon +3 \ell ^2+6 \varepsilon  \ell -3}{2 (\varepsilon +\ell -1) (\varepsilon +\ell +1)}
    +\frac{\ell  (2 \varepsilon +\ell -1) \left(-\Delta +\Delta _{12}+2 \varepsilon +\ell \right) \left(-\Delta -\Delta _{34}+4 \varepsilon +\ell -2\right)}{4 (\varepsilon +\ell -1) (\varepsilon +\ell ) (-\Delta +2 \varepsilon +\ell )}\nonumber \\
    +&\;\frac{\Delta _{12} \ell  \left(-\Delta _{34}+2 \varepsilon -2\right) \left(\Delta -\Delta _{12}+\ell -4\right) (2 \varepsilon +\ell -1) \left(\Delta +\Delta _{34}-2 \varepsilon +\ell -2\right)}{4 (\Delta +\ell -4) (\varepsilon +\ell -1) (\varepsilon +\ell ) (-\Delta +2 \varepsilon +\ell ) (-\Delta +2 \varepsilon +\ell +4)}\nonumber\\
    -&\;\frac{(\ell +1) \left(\Delta -\Delta _{12}+\ell \right) (2 \varepsilon +\ell ) \left(\Delta +\Delta _{34}-2 \varepsilon +\ell +2\right)}{4 (\Delta +\ell ) (\varepsilon +\ell ) (\varepsilon +\ell +1)}\nonumber\\
    -&\;\frac{\Delta _{12} (\ell +1) \left(-\Delta _{34}+2 \varepsilon -2\right) (2 \varepsilon +\ell ) \left(-\Delta +\Delta _{12}+2 \varepsilon +\ell +4\right) \left(-\Delta -\Delta _{34}+4 \varepsilon +\ell +2\right)}{4 (\Delta +\ell -4) (\Delta +\ell ) (\varepsilon +\ell ) (\varepsilon +\ell +1) (-\Delta +2 \varepsilon +\ell +4)}.
\end{align}
The functions $\alpha^{x}_{\Delta,\ell}$, and $\beta^{x}_{\Delta,\ell}$ depend on $\Delta, \ell, \varepsilon$, and $\Delta_{ij}$. These expressions are complicated and not illuminating so we relegate them to the ancilliary \texttt{Mathematica} notebook. Some simple examples are
\begin{align}
    \alpha^{\frac{V-1}{U}}_{\Delta,\ell}=&\; 2\AA_{\Delta+\ell-\varepsilon}\;\AA_{\Delta-\ell-3\varepsilon}\;\Delta_{12}\left(\varepsilon-1-\frac{\Delta_{34}}{2}\right)\left(\left(\Delta+\ell\right)\left(\Delta+\ell-2\varepsilon\right)-2\left(\ell+1\right)\left(\Delta-2\varepsilon\right)\right),\\
    \beta^{\frac{(V-1)^2}{U^2}}_{\Delta,\ell}=&\;4\AA_{\Delta+\ell-\varepsilon}\;\BB_{\Delta-\ell-3\varepsilon-2}\;\Delta_{12}\left(\varepsilon-1-\frac{\Delta_{34}}{2}\right)\left(\Delta\left(\Delta-3\right)+\ell\left(\ell+1\right)-2\varepsilon  \left(\Delta-\ell-3\right)\right).
\end{align}
We point out that the functions $\alpha^{\frac{V-1}{U}}_{\Delta,\ell}$, $\beta^{\frac{(V-1)^2}{U^2}}_{\Delta,\ell}$, $c_{\Delta,\ell}$, $\alpha^{\frac{V^2-1}{2U^2}}_{\Delta,\ell}$, and $\beta^{\frac{(V+1)^2}{4U^2}}_{\Delta,\ell}$ vanish when $\Delta_{ij}=0$. When $\Delta_{ij}=0$, we also note that the functions $\alpha^{\frac{\left(V-1\right)^2}{U^2}}_{\Delta,\ell},\beta^{\frac{V^2-1}{2U^2}}_{\Delta,\ell}, e_{\Delta,\ell}$ and $\alpha^{\frac{\left(V+1\right)^2}{4U^2}}_{\Delta,\ell}$ reduce to the functions $C_{\Delta,\ell}$, $a_{\Delta,\ell}$, $e_{\Delta,\ell}$, and $D^{\left(\varepsilon\right)}_{\Delta,\ell}$ defined in \cite{Chester:2014fya}, respectively, and that we have
\begin{align}
    d_{\Delta,\ell}\big|_{\Delta_{ij}=0}=\frac{1}{8}\left(1-\left(2\varepsilon-1\right)\left(2\varepsilon-3\right)\BB_{\ell+\varepsilon+1-\Delta}\right).
\end{align}

As described in Section \ref{superblockhowto}, these recurrence relations will be used to determine the spectrum of superdescendant primaries $\mathcal{O}_{\Delta,\ell,mn}$ and the associated block coefficients $\mathcal{C}^{\mathcal{X},\Delta_{12},\Delta_{34}}_{\mathcal{O}_{\Delta,\ell,mn}}$ appearing in the channel functions in \eqref{eq:superblock}. Furthermore, they prove our ability to use the reduced superblocks $\TT^{\{k_i\}}_{0,0}$ to generate full superblocks of various multiplets in \eqref{eq:6dsuperselect}-\eqref{eq:3dsuperselect}.

\section{6d $\NN=(2,0)$ superconformal multiplets in $\EE=2$ correlators}
\label{6dsupermultipletsinE2}
In this Appendix, we elaborate on the spectrum of conformal primaries in the superconformal multiplets exchanged in $\EE=2$ correlators $\langle\OO_{k_1}\OO_{k_2}\OO_{k_3}\OO_{k_4=k_1+k_2+k_3-4}\rangle$ in 6d $\NN=(2,0)$ SCFTs. As discussed in the main text, in terms of $p=\min(k_1+k_2,k_3+k_4)-4$, the $\mathfrak{so}(5)_R$ tensor product \eqref{eq:so5tp} restricts the exchanged irreps to be
\begin{align}
    \left([k_1,0]\otimes [k_2,0]\right)\cap\left([k_3,0]\otimes [k_4,0]\right)=
    &\;[p,0]\oplus[p+2,0]\oplus[p,2]\oplus[p+4,0]\oplus[p+2,2]\oplus[p,4].
    \label{eq:so5tpE2}
\end{align}
The superselection rules \eqref{eq:6dsuperselect} then dictate that such $\EE=2$ correlators exchange the supermultiplets
\begin{align}
    (\DD[k_1,0]_{2k_1,0}\;\otimes &\;\DD[k_2,0]_{2k_2,0})\cap\left(\DD[k_3,0]_{2k_3,0}\;\otimes \;\DD[k_4,0]_{2k_4,0}\right)=\nonumber\\
    &\;\DD[p,0]_{2p,0}\oplus\DD[p+2,0]_{2p+4,0}\oplus\DD[p,2]_{2p+4,0}\nonumber \\
    \oplus&\;\DD[p+4,0]_{2p+8,0}\oplus\DD[p+2,2]_{2p+8,0}\oplus\DD[p,4]_{2p+8,0}\nonumber \\
    \oplus&\;\BB[p,0]_{2p+4+\ell,\ell}\oplus\BB[p+2,0]_{2p+8+\ell,\ell}\oplus\BB[p,2]_{2p+8+\ell,\ell}\nonumber \\
    \oplus&\;\LL[p,0]_{\Delta,\ell}.
\label{eq:6dE2superselect}
\end{align}
Below, we tabulate for each of the supermultiplets in Equation \eqref{eq:6dE2superselect} the allowed superdescendant primaries with quantum numbers $(\Delta',\ell')$ in each $\mathfrak{so}(5)_R$ irrep\footnote{As noted in \cite{Buican:2016hpb}, the shortening conditions for $\BB[p,0]_{2p+4+\ell,\ell}$ are enhanced when $p=0$ and we treat this case separately in Table \ref{B00}.}. We calculated this using the \texttt{Python} implementation of the Racah-Speiser algorithm in Appendix B of \cite{Hayling:2020mbp}. Each row corresponds to an irrep in \eqref{eq:so5tpE2} and the columns are organized by dimension. 

We emphasize that Tables \ref{Dp0}-\ref{Lp0} list the spectrum of conformal primaries allowed by superconformal symmetry and that additional symmetries and physical considerations may set these contributions to zero. For instance, entries written in \textcolor{red}{red} correspond to operators that are forbidden by Bose symmetry in correlators with $\Delta_{ij}=0$. An extreme example is that since the scalar superconformal primaries of $\DD[p,2]$ supermultiplets are forbidden when $\Delta_{ij}=0$, all of their superdescendants are also excluded. Another refinement is that, despite the various possible $\BB[p,0]_{2p+4+\ell,\ell}$ superdescendants tabulated in Table \ref{Bp0}, extremality $\EE=1$ correlators $\langle\OO_{k_1}\OO_{k_2}\OO_{k_3}\OO_{k_1+k_2+k_3-2}\rangle$ only exchange operators with twist $t\leq4-\Delta_{34}$, as is seen explicitly in Section \ref{E1}.
\begin{table}[H]
\begin{tabular}{|l||c|c|c|c|c|c|c|c|c|}
\hline $\DD[p,0]_{2p,0}$: & \multicolumn{9}{|c|}{dimension $\Delta=2p$} \\
\hline $\mathfrak{s o}(5)_R$ irrep & $\Delta$ & $\Delta+1$ & $\Delta+2$ & $\Delta+3$ & $\Delta+4$ & $\Delta+5$ & $\Delta+6$ & $\Delta+7$ & $\Delta+8$ \\
\hline \hline $[p+4,0]$ & --- & --- & --- & --- & --- & --- & --- & --- & --- \\
\hline $[p+2,2]$ & --- & --- & --- & --- & --- & --- & --- & --- & --- \\
\hline $[p,4]$ & --- & --- & --- & --- & --- & --- & --- & --- & --- \\
\hline $[p+2,0]$ & --- & --- & --- & --- & --- & --- & --- & --- & --- \\
\hline $[p,2]$ & --- & --- & --- & --- & --- & --- & --- & --- & --- \\
\hline $[p,0]$ & 0 & --- & --- & --- & --- & --- & --- & --- & --- \\
\hline
\end{tabular}
\caption{All possible conformal primaries in $\DD[p,0]_{2p,0}$ multiplets exchanged in $\EE=2$ correlators $\langle\OO_{k_1}\OO_{k_2}\OO_{k_3}\OO_{k_1+k_2+k_3-4}\rangle$.}
\label{Dp0}
\end{table}
\begin{table}[H]
\begin{tabular}{|l||c|c|c|c|c|c|c|c|c|}
\hline $\DD[p+2,0]_{2p+4,0}$: & \multicolumn{9}{|c|}{dimension $\Delta=2p+4$} \\
\hline $\mathfrak{s o}(5)_R$ irrep & $\Delta$ & $\Delta+1$ & $\Delta+2$ & $\Delta+3$ & $\Delta+4$ & $\Delta+5$ & $\Delta+6$ & $\Delta+7$ & $\Delta+8$ \\
\hline \hline $[p+4,0]$ & --- & --- & --- & --- & --- & --- & --- & --- & --- \\
\hline $[p+2,2]$ & --- & --- & --- & --- & --- & --- & --- & --- & --- \\
\hline $[p,4]$ & --- & --- & --- & --- & --- & --- & --- & --- & --- \\
\hline $[p+2,0]$ & 0 & --- & --- & --- & --- & --- & --- & --- & --- \\
\hline $[p,2]$ & --- & 1 & --- & --- & --- & --- & --- & --- & --- \\
\hline $[p,0]$ & --- & --- & 2 & --- & --- & --- & --- & --- & --- \\
\hline
\end{tabular}
\caption{All possible conformal primaries in $\DD[p+2,0]_{2p+4,0}$ multiplets exchanged in $\EE=2$ correlators $\langle\OO_{k_1}\OO_{k_2}\OO_{k_3}\OO_{k_1+k_2+k_3-4}\rangle$.}
\label{Dp20}
\end{table}
\begin{table}[!h]
\begin{tabular}{|l||c|c|c|c|c|c|c|c|c|}
\hline $\DD[p,2]_{2p+4,0}$: & \multicolumn{9}{|c|}{dimension $\Delta=2p+4$} \\
\hline $\mathfrak{s o}(5)_R$ irrep & $\Delta$ & $\Delta+1$ & $\Delta+2$ & $\Delta+3$ & $\Delta+4$ & $\Delta+5$ & $\Delta+6$ & $\Delta+7$ & $\Delta+8$ \\
\hline \hline $[p+4,0]$ & --- & --- & --- & --- & --- & --- & --- & --- & --- \\
\hline $[p+2,2]$ & --- & --- & --- & --- & --- & --- & --- & --- & --- \\
\hline $[p,4]$ & --- & --- & --- & --- & --- & --- & --- & --- & --- \\
\hline $[p+2,0]$ & --- & \textcolor{red}{1} & --- & --- & --- & --- & --- & --- & --- \\
\hline $[p,2]$ & \textcolor{red}{0} & \textcolor{red}{1} & \textcolor{red}{0}, & --- & --- & --- & --- & --- & --- \\
               &  &  & \textcolor{red}{2} & &  &  &  &  &  \\
\hline $[p,0]$ & --- & \textcolor{red}{1} & \textcolor{red}{2} & \textcolor{red}{1}, & --- & --- & --- & --- & --- \\
               &  &  &  & \textcolor{red}{3} &  &  &  &  &  \\
\hline
\end{tabular}
\caption{All possible conformal primaries in $\DD[p,2]_{2p+4,0}$ multiplets exchanged in $\EE=2$ correlators $\langle\OO_{k_1}\OO_{k_2}\OO_{k_3}\OO_{k_1+k_2+k_3-4}\rangle$.}
\label{Dp2}
\end{table}
\begin{table}[H]
\begin{tabular}{|l||c|c|c|c|c|c|c|c|c|}
\hline $\DD[p+4,0]_{2p+8,0}$: & \multicolumn{9}{|c|}{dimension $\Delta=2p+8$} \\
\hline $\mathfrak{s o}(5)_R$ irrep & $\Delta$ & $\Delta+1$ & $\Delta+2$ & $\Delta+3$ & $\Delta+4$ & $\Delta+5$ & $\Delta+6$ & $\Delta+7$ & $\Delta+8$ \\
\hline 
\hline $[p+4,0]$ & 0 & --- & --- & --- & --- & --- & --- & --- & --- \\
\hline $[p+2,2]$  & ---  & 1 & --- & --- & --- & --- & --- & --- & --- \\
\hline $[p,4]$    & --- & --- &  0  & --- & --- & --- & --- & --- & --- \\
\hline $[p+2,0]$  & --- & --- &  2  & --- & --- & --- & --- & --- & --- \\
\hline $[p,2]$    & --- & --- & --- &  1  & --- & --- & --- & --- & --- \\
\hline $[p,0]$    & --- & --- & --- & --- &  0  & --- & --- & --- & --- \\
\hline
\end{tabular}
\caption{All possible conformal primaries in $\DD[p+4,0]_{2p+8,0}$ multiplets exchanged in $\EE=2$ correlators $\langle\OO_{k_1}\OO_{k_2}\OO_{k_3}\OO_{k_1+k_2+k_3-4}\rangle$.}
\label{Dp40}
\end{table}
\begin{table}[H]
\begin{tabular}{|l||c|c|c|c|c|c|c|c|c|}
\hline $\DD[p+2,2]_{2p+8,0}$: & \multicolumn{9}{|c|}{dimension $\Delta=2p+8$} \\
\hline $\mathfrak{s o}(5)_R$ irrep & $\Delta$ & $\Delta+1$ & $\Delta+2$ & $\Delta+3$ & $\Delta+4$ & $\Delta+5$ & $\Delta+6$ & $\Delta+7$ & $\Delta+8$ \\
\hline \hline $[p+4,0]$ & --- &  \textcolor{red}{1}  & --- & --- & --- & --- & --- & --- & --- \\
\hline $[p+2,2]$ &  \textcolor{red}{0}  &  \textcolor{red}{1}  &  \textcolor{red}{0}, & --- & --- & --- & --- & --- & --- \\
                 &     &     &  \textcolor{red}{2}  &     &     &     &     &     &     \\
\hline $[p,4]$   & --- &  \textcolor{red}{1}  &  \textcolor{red}{0}  &  \textcolor{red}{1}  & --- & --- & --- & --- & --- \\
\hline $[p+2,0]$ & --- &  \textcolor{red}{1}  &  \textcolor{red}{2}  &  \textcolor{red}{1}, & --- & --- & --- & --- & --- \\
                 &     &     &     &  \textcolor{red}{3}  &     &     &     &     &     \\
\hline $[p,2]$ & --- & --- &  \textcolor{red}{0}, &  \textcolor{red}{1}  &  \textcolor{red}{0}, & --- & --- & --- & --- \\
               &     &     &  \textcolor{red}{2}  &     &  \textcolor{red}{2}  &     &     &     &     \\
\hline $[p,0]$ & --- & --- & --- &  \textcolor{red}{1}  &  \textcolor{red}{0}  & \textcolor{red}{1}  & --- & --- & --- \\
\hline
\end{tabular}
\caption{All possible conformal primaries in $\DD[p+2,2]_{2p+8,0}$ multiplets exchanged in $\EE=2$ correlators $\langle\OO_{k_1}\OO_{k_2}\OO_{k_3}\OO_{k_1+k_2+k_3-4}\rangle$. Note that when $k_1=k_2$, these entire supermultiplets are absent due to the Bose selection rule discussed in the main text.}
\label{Dp22}
\end{table}
\begin{table}[H]
\begin{tabular}{|l||c|c|c|c|c|c|c|c|c|}
\hline $\DD[p,4]_{2p+8,0}$: & \multicolumn{9}{|c|}{dimension $\Delta=2p+8$} \\
\hline $\mathfrak{s o}(5)_R$ irrep & $\Delta$ & $\Delta+1$ & $\Delta+2$ & $\Delta+3$ & $\Delta+4$ & $\Delta+5$ & $\Delta+6$ & $\Delta+7$ & $\Delta+8$ \\
\hline \hline $[p+4,0]$ & --- & --- & 0 & --- & --- & --- & --- & --- & --- \\
\hline $[p+2,2]$ & --- & 1 & \textcolor{red}{0} & 1 & --- & --- & --- & --- & --- \\
\hline $[p,4]$   & 0 & \textcolor{red}{1} & 0, & \textcolor{red}{1} & 0 & --- & --- & --- & --- \\
                 &   &   & 2  &   &   &     &     &     &     \\
\hline $[p+2,0]$ & --- & --- & 0, & \textcolor{red}{1} & 2 & --- & --- & --- & --- \\
                 &     &     & 2  &   &   &     &     &     &     \\
\hline $[p,2]$ & --- &  1  & \textcolor{red}{0}, & 1, & \textcolor{red}{0}, & 1 & --- & --- & --- \\
               &     &     & \textcolor{red}{2} & 3 & \textcolor{red}{2}  &   &     &     &     \\
\hline $[p,0]$ & --- & --- &  0  & \textcolor{red}{1} & 0, & \textcolor{red}{1} & 0 & --- & --- \\
               &     &     &     &   & 2 &   &   &     &     \\
\hline
\end{tabular}
\caption{All possible conformal primaries in $\DD[p,4]_{2p+8,0}$ multiplets exchanged in $\EE=2$ correlators $\langle\OO_{k_1}\OO_{k_2}\OO_{k_3}\OO_{k_1+k_2+k_3-4}\rangle$.}
\label{Dp4}
\end{table}
\begin{table}[H]
\begin{tabular}{|l||c|c|c|c|c|c|c|c|c|}
\hline $\BB[0,0]_{4+\ell,\ell}$: & \multicolumn{9}{|c|}{dimension $\Delta=4+\ell$} \\
\hline $\mathfrak{s o}(5)_R$ irrep & $\Delta$ & $\Delta+1$ & $\Delta+2$ & $\Delta+3$ & $\Delta+4$ & $\Delta+5$ & $\Delta+6$ & $\Delta+7$ & $\Delta+8$ \\
\hline \hline $[p+4,0]$ & --- & --- & --- & --- & --- & --- & --- & --- & --- \\
\hline $[p+2,2]$ & --- & --- & --- & --- & --- & --- & --- & --- & --- \\
\hline $[p,4]$ & --- & --- & --- & --- & --- & --- & --- & --- & --- \\
\hline $[p+2,0]$ & --- & --- & $\ell+2$ & --- & --- & --- & --- & --- & --- \\
\hline $[p,2]$ & --- & $\ell+1$ & --- & $\ell+3$, & --- & --- & --- & --- & --- \\
\hline $[p,0]$ & $\ell$ & --- & $\ell+2$ & --- & $\ell+4$, & --- & --- & --- & --- \\
\hline
\end{tabular}
\caption{All possible conformal primaries in $\BB[0,0]_{4+\ell,\ell}$ multiplets. We note that these multiplets correspond to higher-spin conserved currents and, neglecting external $S_1$ operators, these multiplets are only exchanged in $\langle\OO_{k}\OO_{k}\OO_{l}\OO_{l}\rangle$. Furthermore, these multiplets can consistently be decoupled from four-point correlators of interacting theories \cite{Maldacena:2011jn}.}
\label{B00}
\end{table}
\begin{table}[H]
\begin{tabular}{|l||c|c|c|c|c|c|c|c|c|}
\hline $\BB[p,0]_{2p+4+\ell,\ell}$: & \multicolumn{9}{|c|}{dimension $\Delta=2p+4+\ell$} \\
\hline $\mathfrak{s o}(5)_R$ irrep & $\Delta$ & $\Delta+1$ & $\Delta+2$ & $\Delta+3$ & $\Delta+4$ & $\Delta+5$ & $\Delta+6$ & $\Delta+7$ & $\Delta+8$ \\
\hline \hline $[p+4,0]$ & --- & --- & --- & --- & --- & --- & --- & --- & --- \\
\hline $[p+2,2]$ & --- & --- & --- & --- & --- & --- & --- & --- & --- \\
\hline $[p,4]$ & --- & --- & --- & --- & --- & --- & --- & --- & --- \\
\hline $[p+2,0]$ & --- & --- & $\ell+2$ & --- & --- & --- & --- & --- & --- \\
\hline $[p,2]$ & --- & $\ell+1$ & \textcolor{red}{$\ell+2$} & $\ell+1$, & --- & --- & --- & --- & --- \\
               &  &  &  & $\ell+3$ &  &  &  &  &  \\
\hline $[p,0]$ & $\ell$ & \textcolor{red}{$\ell+1$} & $\ell$, & \textcolor{red}{$\ell+1$}, & $\ell$, & --- & --- & --- & --- \\
               &  &  & $\ell+2$ & \textcolor{red}{$\ell+3$} & $\ell+2$, &  &  &  &  \\
               &  &  &  &  & $\ell+4$ &  &  &  &  \\
\hline
\end{tabular}
\caption{All possible conformal primaries in $\BB[p,0]_{2p+4+\ell,\ell}$ multiplets exchanged in $\EE=2$ correlators $\langle\OO_{k_1}\OO_{k_2}\OO_{k_3}\OO_{k_1+k_2+k_3-4}\rangle$.}
\label{Bp0}
\end{table}
\begin{table}[H]
\begin{tabular}{|l||c|c|c|c|c|c|c|c|c|}
\hline $\BB[p+2,0]_{2p+8+\ell,\ell}$: & \multicolumn{9}{|c|}{dimension $\Delta=2p+8+\ell$} \\
\hline $\mathfrak{s o}(5)_R$ irrep & $\Delta$ & $\Delta+1$ & $\Delta+2$ & $\Delta+3$ & $\Delta+4$ & $\Delta+5$ & $\Delta+6$ & $\Delta+7$ & $\Delta+8$ \\
\hline \hline $[p+4,0]$ & --- & --- & $\ell+2$ & --- & --- & --- & --- & --- & --- \\
\hline $[p+2,2]$ & --- & $\ell+1$ & \textcolor{red}{$\ell+2$} & $\ell+1$, & --- & --- & --- & --- & ---\\
& & & & $\ell+3$ &  &  &  &  &   \\
\hline $[p,4]$ & --- & --- & $\ell$, & \textcolor{red}{$\ell+1$} & $\ell+2$ & --- & --- & --- & ---  \\
&  &  & $\ell+2$ &  &  &  &  &  &  \\
\hline $[p+2,0]$ & $\ell$ & \textcolor{red}{$\ell+1$}
& $\ell$, & \textcolor{red}{$\ell+1$}, & $\ell$, & --- & --- & --- & --- \\
               &  &  & $\ell+2$ & \textcolor{red}{$\ell+3$} & $\ell+2$, & & &  &  \\
               &  &  & & & $\ell+4$ & & &  &  \\
\hline $[p,2]$ & --- & $\ell\pm1$ & \textcolor{red}{$\ell$}, & $\ell\pm1$, & \textcolor{red}{$\ell$}, & $\ell+1$, & --- & --- & --- \\
               &  &  & \textcolor{red}{$\ell+2$} & $\ell+3$ & \textcolor{red}{$\ell+2$} & $\ell+3$ &  &  &  \\
\hline $[p,0]$ & --- & --- & $\ell$ & \textcolor{red}{$\ell\pm1$} & $\ell$, & \textcolor{red}{$\ell+1$} & $\ell+2$ & --- & --- \\
               &  &  & $\ell\pm2$ &  & $\ell+2$ &  &  &  &  \\
\hline
\end{tabular}
\caption{All possible conformal primaries in $\BB[p+2,0]_{2p+8+\ell,\ell}$ multiplets exchanged in $\EE=2$ correlators $\langle\OO_{k_1}\OO_{k_2}\OO_{k_3}\OO_{k_1+k_2+k_3-4}\rangle$.}
\label{Bp20}
\end{table}
\begin{table}[H]
\begin{tabular}{|l||c|c|c|c|c|c|c|c|c|}
\hline $\BB[p,2]_{2p+8+\ell,\ell}$: & \multicolumn{9}{|c|}{dimension $\Delta=2p+8+\ell$} \\
\hline $\mathfrak{s o}(5)_R$ irrep & $\Delta$ & $\Delta+1$ & $\Delta+2$ & $\Delta+3$ & $\Delta+4$ & $\Delta+5$ & $\Delta+6$ & $\Delta+7$ & $\Delta+8$ \\
\hline \hline $[p+4,0]$ & --- & --- & --- & $\ell+1$ & --- & --- & --- & --- & --- \\
\hline $[p+2,2]$ & --- & --- & $\ell$, & \textcolor{red}{$\ell+1$} & $\ell$, & --- & --- & --- & ---\\
 & & & $\ell+2$ &  & $\ell+2$ &  &  &  & \\
\hline $[p,4]$ & --- & $\ell+1$ & \textcolor{red}{$\ell$}, & $\ell\pm1$, & \textcolor{red}{$\ell$}, & $\ell+1$ & --- & --- & ---  \\
&  &  & \textcolor{red}{$\ell+2$} & $\ell+3$ & \textcolor{red}{$\ell+2$} &  &  &  &  \\
\hline $[p+2,0]$ & --- & $\ell\pm1$ & \textcolor{red}{$\ell$}, & $\ell\pm1$, & \textcolor{red}{$\ell$} & $\ell\pm1$, & --- & --- & ---  \\
&  & & \textcolor{red}{$\ell+2$} & $\ell+3$ & \textcolor{red}{$\ell+2$} & $\ell+3$ &  &  &  \\
               &  &  &  &  &  &  & &  &  \\
\hline $[p,2]$ & $\ell$ & \textcolor{red}{$\ell\pm1$} & $\ell$, & \textcolor{red}{$\ell\pm1$}, & $\ell$, & \textcolor{red}{$\ell\pm1$}, & $\ell$, & --- & ---  \\
&  & & $\ell\pm2$ & \textcolor{red}{$\ell+3$} & $\ell\pm2$, & \textcolor{red}{$\ell+3$} & $\ell+2$ &  &  \\
               &  &  &  &  & $\ell+4$ &  & &  &  \\
\hline $[p,0]$ & --- & $\ell\pm1$ & \textcolor{red}{$\ell$}, & $\ell\pm1$, & \textcolor{red}{$\ell$}, & $\ell\pm1$, & \textcolor{red}{$\ell$}, & $\ell+1$ & ---  \\
&  & & \textcolor{red}{$\ell\pm2$} & $\ell\pm3$ & \textcolor{red}{$\ell\pm2$}, & $\ell+3$ & \textcolor{red}{$\ell+2$} &  &  \\
\hline
\end{tabular}
\caption{All possible conformal primaries in $\BB[p,2]_{2p+8+\ell,\ell}$ multiplets exchanged in $\EE=2$ correlators $\langle\OO_{k_1}\OO_{k_2}\OO_{k_3}\OO_{k_1+k_2+k_3-4}\rangle$.}
\label{Bp2}
\end{table}
\begin{table}[H]
\begin{tabular}{|l||c|c|c|c|c|c|c|c|c|}
\hline $\LL[p,0]_{\Delta, \ell}$: & \multicolumn{9}{|c|}{dimension $\Delta> 2p+6+\ell$} \\
\hline $\mathfrak{s o}(5)_R$ irrep & $\Delta$ & $\Delta+1$ & $\Delta+2$ & $\Delta+3$ & $\Delta+4$ & $\Delta+5$ & $\Delta+6$ & $\Delta+7$ & $\Delta+8$ \\
\hline \hline $[p+4,0]$ & --- & --- & --- & --- & $\ell$ & --- & --- & --- & --- \\
\hline $[p+2,2]$ & --- & --- & --- & $\ell\pm1$ & \textcolor{red}{$\ell$} & $\ell\pm1$ & --- & --- & \\
\hline $[p,4]$ & --- & --- & $\ell$ & \textcolor{red}{$\ell\pm1$} & $\ell$ & \textcolor{red}{$\ell\pm1$} & $\ell$ & --- & ---  \\
&  &  &  &  & $\ell\pm2$ &  &  &  &  \\
\hline $[p+2,0]$ & --- & --- & $\ell$ & \textcolor{red}{$\ell\pm1$} & $\ell$ & \textcolor{red}{$\ell\pm1$} & $\ell$ & --- & --- \\
               &  &  & $\ell\pm2$ &  & $\ell\pm2$ &  & $\ell\pm2$ &  &  \\
\hline $[p,2]$ & --- & $\ell\pm1$ & \textcolor{red}{$\ell$} & $\ell\pm1$ & \textcolor{red}{$\ell$} & $\ell\pm1$ & \textcolor{red}{$\ell$} & $\ell\pm1$ & --- \\
               &  &  & \textcolor{red}{$\ell\pm2$} & $\ell\pm3$ & \textcolor{red}{$\ell\pm2$} & $\ell\pm3$ & \textcolor{red}{$\ell\pm2$} &  &  \\
\hline $[p,0]$ & $\ell$ & \textcolor{red}{$\ell\pm1$} & $\ell$ & \textcolor{red}{$\ell\pm1$} & $\ell$ & \textcolor{red}{$\ell\pm1$} & $\ell$ & \textcolor{red}{$\ell\pm1$} & $\ell$ \\
               &  &  & $\ell\pm2$ & \textcolor{red}{$\ell\pm3$} & $\ell\pm2$ & \textcolor{red}{$\ell\pm3$} & $\ell\pm2$ &  &  \\
               &  &  &  &  & $\ell\pm4$ &  &  &  &  \\
\hline
\end{tabular}
\caption{All possible conformal primaries in $\LL[p,0]_{\Delta, \ell}$ multiplets exchanged in $\EE=2$ correlators $\langle\OO_{k_1}\OO_{k_2}\OO_{k_3}\OO_{k_1+k_2+k_3-4}\rangle$.}
\label{Lp0}
\end{table}
\end{appendix}



\printbibliography

\end{document}